\documentclass[11pt]{article}

\usepackage[letterpaper, text={6.5in, 9.0in}]{geometry}
\usepackage{authblk}

\RequirePackage{amsthm,amsmath,amsfonts,amssymb}
\RequirePackage[authoryear]{natbib}
\RequirePackage[colorlinks,citecolor=blue,urlcolor=blue]{hyperref}
\RequirePackage{graphicx}

\RequirePackage{tikz}
\usetikzlibrary{decorations.pathreplacing}

\usepackage{booktabs}

\theoremstyle{plain}
\newtheorem{lemma}{Lemma}
\newtheorem{coro}{Corollary}
\newtheorem{prop}{Proposition}
\newtheorem{theorem}{Theorem}

\theoremstyle{definition}
\newtheorem{assumption}{Assumption}
\newtheorem{fact}{Fact}

\newcommand{\indep}{\perp\!\!\!\perp}
\newcommand{\eqd}{\stackrel{d}{=}}
\newcommand{\tod}{\stackrel{D}{\longrightarrow}}
\newcommand{\toP}{\stackrel{P}{\longrightarrow}}

\newcommand{\ta}{\tilde{a}}
\newcommand{\tu}{\tilde{u}}

\newcommand{\bB}{\mathbb{B}}
\newcommand{\bE}{\mathbb{E}}
\newcommand{\bG}{\mathbb{G}}
\newcommand{\bP}{\mathbb{P}}
\newcommand{\bR}{\mathbb{R}}

\newcommand{\cB}{\mathcal{B}}
\newcommand{\cE}{\mathcal{E}}
\newcommand{\cF}{\mathcal{F}}
\newcommand{\cG}{\mathcal{G}}
\newcommand{\cJ}{\mathcal{J}}
\newcommand{\cL}{\mathcal{L}}
\newcommand{\cN}{\mathcal{N}}
\newcommand{\cR}{\mathcal{R}}
\newcommand{\cS}{\mathcal{S}_{\tilde Y\mid \tilde X}}
\newcommand{\cV}{\mathcal{V}}
\newcommand{\cW}{\mathcal{W}}

\newcommand{\hatsK}{{\hat\Psi}_{1n}}
\newcommand{\sK}{\Psi_1}
\newcommand{\sT}{\Psi_2}

\newcommand{\amin}{\underline{a}}
\newcommand{\fh}{\mathfrak{h}}

\newcommand{\col}{\text{col}}
\newcommand{\tr}{\text{tr}}
\newcommand{\diag}{\text{diag}}
\newcommand{\rank}{\text{rank}}

\begin{document}

\title{Sliced Inverse Regression Revisited: When the Number of Predictors Grows Proportionally With the Sample Size}

\author[1]{Peng Zeng}
\author[2]{Hanwen Huang}

\affil[1]{Department of Mathematics and Statistics, Auburn University}

\affil[2]{Department of Biostatistics, Data Science and Epidemiology, Medical College of Georgia, Augusta University}

\maketitle

\begin{abstract}
\cite{li1991sliced} introduced sliced inverse regression (SIR) as a dimension reduction method 
and established its theoretical properties 
under the assumption that the number of predictors $p$ remains fixed as the sample size $n$ tends to infinity. 
In this paper, we revisit the theoretical properties of SIR in the high-dimensional regime where $p/n \to \gamma \in [0,1)$, 
assuming normally distributed predictors and a fixed number of slices.
We show that the sample SIR eigenvalues exhibit an upward bias when $\gamma>0$, 
that their fluctuations are governed by the Gaussian orthogonal ensemble, 
and that the signal eigenvalues are asymptotically normal after an appropriate bias correction. 
We demonstrate that Li's chi-squared test requires a scaling adjustment to achieve the nominal asymptotic size when $\gamma>0$. 
We further propose two additional tests that achieve the nominal asymptotic size: 
an exact test based on Li's test statistic and a largest-eigenvalue test. 
In addition, we derive closed-form limits for the accuracy of the estimated SIR directions and show that their first-order expansions in $\gamma$ agree with Li's approximation. 
Finally, we establish the limiting distribution of the estimated sufficient predictors for a new observation and show that, when $\gamma>0$, they contain an additional independent Gaussian noise component. 
Simulation studies corroborate our theoretical findings.

\bigskip 
\noindent
{\bf Keywords}: 
central subspace,
dimension reduction,
Gaussian orthogonal ensemble,
matrix-variate Beta distribution,
proportional asymptotics,
sequential testing procedure. 

\bigskip 
\noindent 
{\bf MSC2020 subject classifications}: 
Primary 62H12, secondary 62H15, 62E20.
\end{abstract}

\section{Introduction}
\label{sec:intro}

Consider a regression of a scalar response $\tilde Y \in \bR$ on a predictor $\tilde X \in \bR^p$. 
\cite{li1991sliced} proposed the dimension-reduction model 
\begin{align}
 \tilde Y = f(B^T \tilde X, \epsilon),
\label{eq:model-reg}
\end{align}
where $f$ is an unknown link function, $B\in\bR^{p\times d}$ is a coefficient matrix, 
and $\epsilon$ is a random error independent of $\tilde X$. 
Under this model, the dependence of $\tilde Y$ on $\tilde X$ is fully captured by $B^T\tilde X$.
Since $f$ is arbitrary, $B$ is not identifiable. 
The inferential target is instead the column space $\col(B)$, 
referred to as the {\em effective dimension reduction subspace} by \cite{li1991sliced}.

Using an iid sample  
$\{(y_i, x_i), y_i\in\bR, x_i\in\bR^p, i = 1, \ldots, n\}$ from \eqref{eq:model-reg},
\cite{li1991sliced} proposed {\em sliced inverse regression} (SIR) to estimate $\col(B)$. 
Define the sample mean and sample covariance matrix of the predictors as 
$\bar x = \sum_i x_i/n$ and  
$\hat\Sigma = \sum_i (x_i - \bar x)(x_i - \bar x)^T/n$.  
Partition the range of the responses into $H$ slices. 
Let $n_h$ denote the number of observations in slice $h$, 
and let $\hat p_h = n_h/n$ denote the corresponding proportion. 
Denote by $\hat m_h$ the sample mean of the standardized predictors 
$\hat z_i = \hat\Sigma^{-1/2}(x_i - \bar x)$ within slice $h$.
The SIR kernel matrix is defined as 
\begin{align*}
  \hat V = \sum_{h=1}^H \hat p_h \hat m_h \hat m_h^T \in \bR^{p\times p},
\end{align*}
with eigenvalues $\hat\lambda_1 \geq \cdots \geq \hat\lambda_p \geq 0$ and 
corresponding unit eigenvectors $\hat\eta_1, \ldots, \hat\eta_{p}$. 
The $k$th {\em SIR direction} is defined as 
$\hat\beta_k = \hat\Sigma^{-1/2}\hat\eta_k$, 
which satisfies the normalization condition $\hat\beta_k^T \hat\Sigma \hat\beta_k = 1$.

Let $d_0$ denote the true dimension of $\col(B)$. 
Under the null hypothesis $H_0: d_0 = d$, 
the $p-d$ smallest eigenvalues of $\hat V$ are expected to be zero or close to zero. 
\citet[Theorem 5.1]{li1991sliced} showed that, under $H_0$ and assuming normally distributed predictors, 
\begin{align*}
    T_{d} 
    = n\sum_{i=d+1}^p \hat\lambda_i 
    \tod \chi^{2}_{f_d},    
    \quad 
    \text{ as $n\to\infty$}, 
\end{align*}
with $p$, $H$, and $d$ fixed, where $f_d = (p-d)(H-d-1)$.
This result motivates a sequential testing procedure for 
$H_{0}: d_{0} = d$ against $H_{a}: d_{0} > d$, for $d = 0, 1, 2, \ldots$.
At each step, we reject $H_0$ if $T_d > \chi^2_{f_d, \alpha}$, 
where $\chi^2_{f_d, \alpha}$ denotes the upper-$\alpha$ quantile of the $\chi^2_{f_d}$ distribution. 
We estimate $d_0$ by $\hat d_0$, the smallest value of $d$ for which $H_{0}$ is not rejected.

\cite{li1991sliced} provided a thorough analysis of SIR, 
including 
a geometric justification based on the inverse regression curve,
root-$n$ consistency of the estimated SIR directions, and 
an approximation to the expected accuracy of the estimated subspace in terms of squared trace correlation.
However, these results are derived under the assumption that $p$ is fixed. 
The main goal of this paper is to revisit these properties when $p$ grows proportionally with $n$.

\subsection{Extensions and generalizations of SIR}

Model (\ref{eq:model-reg}) implies that $\tilde Y$ and $\tilde X$ are conditionally independent
given $P_{\mathcal{S}}\tilde X$, that is, 
\begin{align}
    \tilde Y \indep \tilde X \mid P_{\mathcal{S}} \tilde X,
\label{eq:model-cs}
\end{align}
where $\indep$ denotes statistical independence and $P_{\mathcal{S}}$ is the projection matrix onto a subspace $\mathcal{S} = \col(B)$. 
This formulation does not explicitly involve the link function $f$ 
and is more commonly used in related literature. 
We refer to $\mathcal{S}$ as a {\em dimension reduction subspace} for \eqref{eq:model-cs} and 
are interested in finding the one with the smallest dimension.  
Define the {\em central subspace}, denoted by $\cS$, as 
the intersection of all dimension reduction subspaces, 
provided that the intersection itself satisfies \eqref{eq:model-cs} 
\citep{cook1994interpretation}.  
Under mild conditions, which are assumed throughout this paper, the central subspace exists. 
We refer to $d_0 = \text{dim}(\cS)$ as the {\em structural dimension} of the central subspace.

SIR has been further studied in the literature; see, for example, 
\cite{hsing1992asymptotic}, 
\cite{zhu1995asymptotics},  
\cite{chen1998can},
\cite{wu2011asymptotic},
\cite{lin2018consistency}, and
\cite{huang2026structural}.
It has also been extended to various regression settings, including 
functional data \citep{ferre2003functional}, 
high-dimensional data \citep{li2008sliced}, 
spatial point processes \citep{guan2010sufficient},
sparse models \citep{lin2019sparse}, 
metric space-valued responses \citep{virta2022sliced},
and tensor data \citep{xing2025parsimonious}.
Various methods have also been proposed to improve upon the performance of SIR; 
examples include 
SAVE \citep{cook1991sliced}, 
MAVE \citep{xia2002adaptive}, 
Fourier method \citep{zhu2006fourier}, 
directional regression \citep{li2007directional}, and 
cumulative slicing estimation \citep{zhu2010dimension}. 
The determination of structural dimension has been further studied by 
\cite{schott1994determining}, \cite{velilla1998assessing},
\cite{bura2001extending}, \cite{ye2003using}, \cite{zhu2006sliced}, \cite{zeng2008determining}, 
\cite{luo2016combining},
and \cite{huang2026structural}. 
An exhaustive review is beyond the scope of this paper; 
see \cite{cook1998book} and \cite{li2018sufficient} for further discussion.

\subsection{Asymptotic regimes}

\cite{li1991sliced} commented in Remark 5.4 that
``It is interesting to study the asymptotic behavior of the SIR estimate when both the sample size and the dimension $p$ of [$\tilde X$] increase simultaneously, but the number of components [$d$] and the number of slices $H$ are kept fixed.'' 
There are already some existing results along this line.
\cite{zhu2006sliced} studied the consistency of SIR when $p = o(\sqrt{n})$; 
\cite{wu2011asymptotic} established the convergence rate of SIR when $p = o(n/\log(n))$;
and \cite{lin2018consistency} showed 
that SIR is consistent if and only if $p/n\to 0$ and that
it is inconsistent for a single-index model when $p/n \to \gamma > 0$.
\cite{wang2021cumulative} established the same phase transition for cumulative slicing estimation.
\cite{lin2021optimality} derived the minimax rate for
estimating the central subspace.
\cite{huang2026structural} extended this rate to diverging $d$.
Sparsity-based estimators for $p\gg n$ include
\cite{lin2018consistency}, \cite{lin2019sparse}, \cite{tan2018convex}, and
\cite{tan2020sparse}.

These results suggest that SIR remains consistent when $p/n \to 0$, but may fail when $p/n \to \gamma > 0$. 
However, they do not provide a precise characterization of the properties of SIR in the latter regime. 
Consequently, it remains unclear whether SIR can be corrected or appropriately adjusted to restore its effectiveness.
To address this, we revisit the properties of SIR established in \cite{li1991sliced} under the following asymptotic regime, assuming normally distributed predictors:
\begin{align*}
    n \to \infty, \qquad 
    p/n \to \gamma \in [0, 1), \qquad 
    \text{$d$ and $H$ are fixed}, 
\end{align*}
where $d$ is the structural dimension and $H$ is the number of slices. 

Keeping $d$ and $H$ fixed in this asymptotic regime is reasonable and consistent with common practice. 
The primary goal of SIR is dimension reduction, and we typically expect $d \ll p$, as supported by many practical applications. 
When implementing SIR, it suffices to choose the number of slices $H$ with $H-1 > d$, and there is little empirical evidence that further increasing $H$ improves performance. Moreover, maintaining a sufficiently large number of observations per slice is important for obtaining reliable estimates of the slice means. In practice, $H$ is commonly chosen between 10 and 20, which is typically much smaller than $p$. 

\subsection{Overview of main results}

We provide an overview of the main results in this subsection before delving into the details.

Section~\ref{sec:review} begins by defining the relevant population quantities. 
We then introduce a matrix $K\in\bR^{(H-1)\times (H-1)}$ and show in Proposition~\ref{prop:cca} that $\hat V$ and $K$ share the same nonzero eigenvalues and that the SIR directions can be expressed in terms of the eigenvectors of $K$. 
Because $K$ has fixed dimension, analyzing SIR through $K$ substantially simplifies the analysis when $p$ grows with $n$. 
We also define a matrix $A \in \bR^{(H-1)\times (H-1)}$, analogous to $K$, obtained by applying SIR to the reduced predictor $B^Tx_i$, assuming $B$ is known. 
Although $A$ is not directly computable in practice, it provides a crucial link between $K$ and the population quantities and plays a central role in the subsequent analysis. 

Section~\ref{sec:exact} introduces the matrix $M = U^TKU$, where the columns of $U$ are eigenvectors of $A$. 
Consequently, $M$, $K$, and $\hat V$ share the same nonzero eigenvalues. 
Theorem~\ref{thm:represent} establishes the representation $M = \Lambda + D\bB D$, 
where $\Lambda$ and $D$ are diagonal matrices determined by the eigenvalues of $A$ 
and $\bB$ is independent of $\{(y_i, B^Tx_i)\}$. 
Moreover, $\bB$ follows a matrix-variate Beta distribution. 
This representation provides the foundation for all subsequent results. 

Section~\ref{sec:values} investigates the asymptotic behavior of the SIR eigenvalues. 
Theorem~\ref{thm:values1} establishes that $\hat\lambda_k \toP \rho_k^2 + \gamma(1 - \rho_k^2)$ for $k \leq d$
and $\hat\lambda_k \toP \gamma$ for $d < k \leq H - 1$,
where $\rho_k^2$ is the $k$th eigenvalue of the population SIR kernel matrix.
Thus, both the signal and null eigenvalues are asymptotically biased upward when $p/n\to \gamma > 0$. 
Theorem~\ref{thm:values2} further characterizes their fluctuations:
for $k \leq d$, the fluctuations are governed by the diagonal entries of a fixed-dimensional Gaussian orthogonal ensemble (GOE), with an explicit additional variance term,
whereas for $d < k \leq H - 1$, they are governed by the eigenvalues of a GOE. 
Finally, Corollary~\ref{coro:values} shows that, after appropriate centering and scaling, 
the bias-corrected signal eigenvalues are jointly asymptotically normal, 
with convergence at the root-$n$ rate. 

Section~\ref{sec:test} concerns sequential testing procedures for determining the structural dimension. 
Theorem~\ref{thm:Td} obtains an exact test based on an upper bound for $T_d$, 
whose distribution coincides with that of $n\,\tr(\bB_{00})$, 
where $\bB_{00}$ follows a matrix-variate Beta distribution. 
Theorem~\ref{thm:test} establishes the asymptotic normality of $T_d$, which indicates that the $\chi^2$-test proposed by \cite{li1991sliced} requires a scaling adjustment. 
Corollary~\ref{coro:regimes} shows that the asymptotic size of Li's $\chi^2$-test is 
$1 - \Phi(z_{\alpha}/\sqrt{1-\gamma})$, where $\Phi$ and $z_{\alpha}$ denote the cumulative distribution function and the upper-$\alpha$ quantile of the standard normal distribution, respectively.
Thus, the test remains asymptotically valid when $\gamma = 0$ 
but becomes conservative when $\gamma > 0$.
Theorem~\ref{thm:test} also shows that the normalized statistic $T_{d'}$ diverges when $d' < d$, 
which ensures that the sequential testing procedure asymptotically avoids underestimating the structural dimension,  
as established in Corollary~\ref{coro:test}. 
Finally, we propose an alternative Roy-type test based on $\hat\lambda_{d+1}$ and
establish its asymptotic properties in Corollary~\ref{coro:lmax}. 

Section~\ref{sec:vectors} investigates the asymptotic properties of the SIR directions. 
Since $\hat\beta_k$ is a $p$-dimensional vector and $p \to \infty$, 
we focus on its alignment with the corresponding population direction $\beta_k$ and 
on the predictive performance of $\hat\beta_k^T(x_0 - \bar x)$ for a new observation $x_0$. 
Theorem~\ref{thm:vectors-align} quantifies the alignment between the estimated and population directions by establishing that 
$\hat\beta_{1:d}^T \Sigma \beta_{1:d} \toP \sK$, where $\sK = \diag(\kappa_1, \ldots, \kappa_d)$.
This result implies that $\hat\beta_k$ is asymptotically $\Sigma$-orthogonal to $\beta_j$ for $j \neq k$
and that the $\Sigma$-inner product between $\hat\beta_k$ and $\beta_k$ converges to $\kappa_k < 1$ when $\gamma > 0$.
Two criteria, $R^2(\hat\beta_k)$ and $R^2(\hat\beta_{1:d})$, proposed by \cite{li1991sliced}, measure the accuracy of the estimated SIR directions. The former measures the alignment between a single SIR direction $\hat\beta_k$ and $\col(B)$ through the squared cosine of the angle between them, whereas the latter measures the alignment between $\col(\hat\beta_{1:d})$ and $\col(B)$ through the squared trace correlation. 
Theorem~\ref{thm:accuracy} establishes the asymptotic limits of both criteria when $\gamma \geq 0$. 
In particular, the limit of $R^2(\hat\beta_{1:d})$ agrees with the result in \cite{li1991sliced} to first order in $\gamma$. 
Theorem~\ref{thm:pred} shows that, asymptotically, $\hat\beta_k^T(x_0 - \bar x)$
behaves like $\kappa_k\beta_k^T(x_0 - \mu) + \tau_k w_{N, k}$, where $\tau_k$ is defined in Section~\ref{subsec:vectors-align} and $w_{N, k}$ is a standard normal random variable.
When $\gamma > 0$, the factor $\kappa_k < 1$ induces shrinkage towards zero, while the term $\tau_k w_{N, k}$ represents a Gaussian noise that is independent of the true reduced predictor $B^Tx_0$ and carries no additional information about the response.
This Gaussian noise vanishes only when $\gamma = 0$. 

Section~\ref{sec:simulation} assesses the theoretical findings through simulation studies. 
Section~\ref{sec:conclusion} concludes the paper with a discussion and additional remarks. 
All proofs are collected in the supplementary material. 

\subsection{Notation}

This subsection introduces concepts and notation that will be used throughout this paper. 

For a linear subspace $\cL$, $P_{\cL}$ denotes the orthogonal projection matrix onto $\cL$, and $\cL^{\perp}$ denotes its orthogonal complement. 
A {\em frame} in a subspace $\cL$ is a matrix whose columns are orthonormal and lie in $\cL$. 
For any matrix $\Sigma$, $\Sigma \cL$ denotes the linear space spanned by $\{\Sigma v \mid v \in \cL\}$. 
A random $s$-dimensional subspace of an $m$-dimensional space $\cL$ is {\em uniformly distributed} 
if its distribution is invariant under every orthogonal transformation of $\cL$ \citep{chikuse2003statistics}. 

For integers $1 \leq t \leq m$ and $0 < s < m$, the {\em matrix-variate Beta distribution} 
$\cB_t(s/2, (m-s)/2)$ is defined as the distribution of $(S_1 + S_2)^{-1/2} S_1 (S_1 + S_2)^{-1/2} \in \bR^{t\times t}$, 
where $S_1 \sim \cW_t(s, I_t)$ and $S_2\sim \cW_t(m-s, I_t)$ are independent $t\times t$ Wishart random matrices \citep{muirhead1982aspects}. 
Let $\Pi(s, t, m)$ denote the distribution of the trace of a $\cB_t(s/2, (m-s)/2)$ random matrix. 

Let $\cG_t$ denote a $t\times t$ matrix drawn from the {\em Gaussian orthogonal ensemble} (GOE), that is, a symmetric random matrix whose entries on and above the diagonal are independent, with diagonal entries distributed as $\cN(0,2)$ and off-diagonal entries as $\cN(0,1)$.

For a matrix $A$, $\col(A)$ denotes its column space. 
If $A$ has full column rank, 
$P_A = A(A^TA)^{-1}A^T$ is the orthogonal projection matrix onto $\col(A)$, and
$F_A = A(A^TA)^{-1/2}$ is a frame for $\col(A)$. 
$\|A\|_F$ and $\|A\|_{op}$ denote its Frobenius norm and operator norm, respectively.
If $A$ is symmetric, $\lambda_i(A)$ denotes the $i$th largest eigenvalue of $A$. 
For two symmetric matrices $A$ and $M$, we write $A \prec M$ (or $M \succ A$) if $M - A$ is positive definite, 
and use $\preceq$ and $\succeq$ to denote the corresponding relationship when $M - A$ is positive semidefinite. 

The Kronecker delta $\delta_{ij}$ equals 1 if $i=j$ and 0 otherwise.
$\tod$ and $\toP$ denote convergence in distribution and convergence in probability, respectively.

\section{Sliced Inverse Regression}
\label{sec:review}

This section first reviews the theoretical justification for SIR 
in terms of inverse regression curves, as proposed by \cite{li1991sliced}, in Section~\ref{subsec:irc}. 
Section~\ref{subsec:cca} then establishes a connection between SIR and canonical correlation analysis (CCA), 
which underlies the subsequent analysis. 
After formally defining the population quantities in Section~\ref{subsec:pop}, 
Section~\ref{subsec:oracle} introduces oracle quantities obtained by applying SIR to $\{(y_i, B^Tx_i)\}$, 
assuming that $B$ is known. 
These oracle quantities provide a crucial link between the sample and population quantities.

\subsection{Inverse regression curve}
\label{subsec:irc}

Let $\mu = \bE(\tilde X) \in \bR^p$ and $0 \prec \Sigma = var(\tilde X) \in \bR^{p\times p}$ denote 
the mean vector and covariance matrix of $\tilde X$, respectively. 
Define the standardized predictor as $\tilde Z = \Sigma^{-1/2}(\tilde X - \mu)$. 

\begin{assumption}
(Linearity condition)
For any $b\in\bR^p$, $\bE(b^T\tilde X\mid B^T\tilde X)$ is linear in $B^T\tilde X$. 
\label{ass:linear}
\end{assumption}

Assumption~\ref{ass:linear} holds when $\tilde X$ follows an elliptically contoured distribution, which includes the normal distribution as a special case. 
The following proposition restates Theorem 3.1 and Corollary 3.1 of \cite{li1991sliced}, with a slight extension, and provides a theoretical justification for the SIR algorithm. 
Its proof is omitted for brevity.

\begin{prop}[Inverse regression curve]
Under model (\ref{eq:model-cs}) and Assumption~\ref{ass:linear}, the centered inverse regression curve satisfies $\bE(\tilde X\mid \tilde Y) - \bE(\tilde X) \in \Sigma \cS$, 
while the standardized inverse regression curve satisfies $\bE(\tilde Z\mid \tilde Y) \in \Sigma^{1/2} \cS$. 
Moreover, for any function $\fh(\cdot)$ of $\tilde Y$, the column space of $var[\bE\{\tilde Z \mid \fh(\tilde Y)\}]$ is contained in $\Sigma^{1/2} \cS$. 
\label{prop:SIR}
\end{prop}

Proposition~\ref{prop:SIR} provides insight into estimating $\cS$ from a sample. 
The key step is to estimate the matrix $var[\bE\{\tilde Z\mid \fh(\tilde Y)\}]$ for a suitably chosen function $\fh(\cdot)$. 
\cite{li1991sliced} proposed using indicator functions for $\fh(\cdot)$ for its simplicity, which leads to the SIR algorithm. 

It is important to note that SIR may fail to recover directions in $\cS$ when $\bE(\tilde Z\mid \tilde Y)$ is degenerate in some direction of $\cS$, which can occur, 
for example, when the link is symmetric about the center in some direction of $\cS$, such as $\tilde Y = (\beta^T\tilde X)^2 + \epsilon$. 
To ensure that SIR can fully recover the central subspace, the literature commonly imposes an additional assumption that 
$\text{span}\{\bE(\tilde X\mid \tilde Y) - \mu\} = \Sigma \cS$, which is known as the coverage condition.
We assume the coverage condition throughout this paper.

\subsection{SIR as canonical correlation analysis}
\label{subsec:cca}

We establish a connection between SIR and canonical correlation analysis (CCA) in Proposition~\ref{prop:cca}, 
which plays a pivotal role in our analysis of SIR throughout this paper. 
Let $Y = (y_1, \ldots, y_n)^T \in\bR^n$ denote the response vector  
and $X = (x_1, \ldots, x_n)^T \in\bR^{n\times p}$ the design matrix. 
For each slice $h$, let $g_h \in\bR^n$ denote the corresponding indicator vector, whose $i$th element equals 1 if the $i$th observation belongs to slice $h$ and 0 otherwise.
By construction, $\sum_{h=1}^H g_h = 1_n$, where $1_n$ denotes the $n$-dimensional vector of ones.

Let $C_n = I_n - n^{-1}1_n1_n^T \in \bR^{n\times n}$ denote the centering projection matrix.
Let $\bar X = C_nX \in\bR^{n\times p}$ denote the centered predictor matrix, whose $i$th row is $(x_i - \bar x)^T$. 
Let $P_{\bar X} \in\bR^{n\times n}$ denote the orthogonal projection matrix onto $\col(\bar X)$.
Define $\bar G = C_n(g_1, \ldots, g_{H-1}) \in\bR^{n\times (H-1)}$ as the centered matrix of the indicator vectors for slices $1, \ldots, H-1$. 
Let $F_{\bar G} = \bar G(\bar G^T\bar G)^{-1/2} \in \bR^{n\times (H-1)}$ and 
define 
\begin{align*}
  K = F_{\bar G}^T P_{\bar X} F_{\bar G} \in \bR^{(H-1)\times (H-1)}, 
\end{align*}
with eigenvalues $r_1^2 \geq \cdots \geq r_{H-1}^2$. 
The eigenvalues $r_i^2$ can be interpreted as the squared sample canonical correlations between $X$ and the slice indicator vectors $(g_1, \ldots, g_{H-1})$. 

\begin{prop}[SIR and canonical correlation analysis] 
\label{prop:cca}
Assume that each slice is nonempty and $\text{rank}(\bar X) = p$.
The following statements hold:
\begin{enumerate}
\item[(a)] 
With $N = (\bar X^T \bar X)^{-1/2}\bar X^T F_{\bar G}$,
the matrices $\hat V$ and $K$ satisfy $\hat V = NN^T$ and $K = N^TN$.

\item[(b)] 
The nonzero eigenvalues of $\hat V$ and $K$ coincide.
In particular, $\hat\lambda_k = r_k^2$ for $k \leq \min (p, H-1)$,
where the eigenvalues are ordered nonincreasingly. 

\item[(c)] 
The nonzero eigenvalues $r_k^2$ are the squared sample canonical correlations between $X$ and the slice indicator vectors $g_1, \ldots, g_{H-1}$.

\item[(d)]
For $k \leq \min (p, H-1)$, if $r_k^2 > 0$ is simple, the $k$th SIR direction is given by
\begin{align}
    \hat\beta_k = \frac{\sqrt{n}}{r_k} (\bar X^T\bar X)^{-1}\bar X^TF_{\bar G} u_k,
\label{eq:betab}
\end{align}
where $u_k$ is a unit eigenvector of $K$ corresponding to the eigenvalue $r_k^2$, defined up to sign.
Equivalently, $\hat\beta_k$ is the least squares coefficient vector obtained by regressing the working response $F_{\bar G}u_k$ on $\bar X$, scaled by $\sqrt{n}/r_k$. 
\end{enumerate}  
\end{prop}

Parts (a) and (c) of Proposition~\ref{prop:cca} follow from direct matrix manipulation.  
Parts (b) and (d) follow from (a), since $NN^T$ and $N^TN$ share the same nonzero eigenvalues for any matrix $N$. 
A complete proof is provided in the supplementary material.

Proposition~\ref{prop:cca} is a nonasymptotic result that holds for any $n$. 
By Proposition~\ref{prop:cca}(b), $\hat V$ and $K$ have the same nonzero eigenvalues. 
Note that the rank of $\hat V$ is at most $H - 1$ when $H - 1 < p$, and thus $\hat V$ has at most $H - 1$ nonzero eigenvalues. 
Consequently, $\hat\lambda_{H} = \cdots = \hat\lambda_p = 0$. 
Moreover, $\hat V$ is a $p\times p$ matrix, whereas $K$ is an $(H-1)\times (H-1)$ matrix. 
Since $H$ is fixed in our asymptotic setting, we can study the spectral properties of $\hat V$ through those of the fixed-dimensional matrix $K$. This substantially simplifies the analysis. 
The SIR directions can likewise be studied through the eigenvectors of $K$ as suggested by Proposition~\ref{prop:cca}(d).

The slice indicator vectors $g_h$ depend on $y_1, \ldots, y_n$, 
and we allow any slicing rule that depends only on the responses $y_i$.  
Therefore, $g_h$, $F_{\bar G}$, and $K$ are random and depend on both the responses and the predictors. 
Our analysis typically begins by conditioning on the $\sigma$-field $\cF = \sigma(B^Tx_1, \ldots, B^Tx_n, y_1, \ldots, y_n)$, but ultimately yields a distribution that does not depend on $\cF$. 

The connection between SIR and CCA was noted as early as \cite{li1991sliced}, \cite{he1997linear}, and \cite{chen1998can}.
However, subsequent work has primarily used this connection to motivate new methods to estimate central subspaces, such as those proposed by  \cite{he1997linear} and \cite{fung2002dimension}, rather than to study the theoretical properties of SIR itself. 
The results from the CCA literature are not directly applicable to SIR for two reasons. First, CCA studies typically assume joint normality of the two sets of variables, whereas the slice indicator vectors in SIR are not normally distributed. Second, our analysis of this connection is carried out conditional on $\cF$, which differs from the conventional CCA setting. 
See \cite{gao2017sparse}, \cite{bao2019canonical} and \cite{ma2020subspace}, along with the references therein, 
for recent developments in CCA.

\subsection{Population quantities}
\label{subsec:pop}

Although the inferential target is $\col(B)$ rather than $B$ itself, it is still important to identify the population counterparts of the eigenvalues and eigenvectors of $\hat V$ for theoretical convenience. In this subsection, we define these population quantities as functionals of $\tilde Y$ and $\tilde X$. 

The population slices are defined as follows.
Let $-\infty = c_0 < c_1 < \cdots < c_H = \infty$ and define $J_h = (c_{h-1}, c_h]$ for $h = 1, \ldots, H$.
Let $\fh(y) = h$ if and only if $y\in J_h$, and define $p_h = \bP(\tilde Y \in J_h)$. We assume that $p_h > 0$ for all $h$. 
For slices with fixed boundaries, the cut-points $c_h$ are specified in advance. For slices with approximately equal counts, the $h$-th sample slice consists of the observations whose ranks fall in $(\lfloor (h-1)n/H\rfloor, \lfloor hn/H\rfloor]$. In this case, $c_h$ is the $(h/H)$-quantile of $\tilde Y$, satisfying $F_{\tilde Y}(c_h) = h/H$, where $F_{\tilde Y}$ denotes the cumulative distribution function of $\tilde Y$ and we assume that the quantile is uniquely defined. 

Let $m_h = \Sigma^{-1/2}\{\bE(\tilde X\mid \tilde Y \in J_h) - \mu\}$ denote the mean of the standardized predictor within the $h$-th slice.
The population SIR kernel matrix is defined as 
\begin{align*}
    V 
    = \sum_{h=1}^H p_h m_h m_h^T 
    = var\{\bE(\tilde Z \mid \fh(\tilde Y))\}
    \in \bR^{p\times p}. 
\end{align*}
This matrix is the population counterpart of $\hat V$. 
Its dimension is $p\times p$, so it grows with $n$ when $p$ increases with $n$. 

Assume that model (\ref{eq:model-cs}) holds and $\tilde X \sim \cN_p(\mu, \Sigma)$.  
Define the standardized sufficient predictor as $\tilde Z_S = (B^T\Sigma B)^{-1/2}B^T(\tilde X - \mu)$. 
Then $\tilde Z_S \sim \cN_d(0, I_d)$.
Let $O = \Sigma^{1/2}B(B^T\Sigma B)^{-1/2}$, so that $O \in \bR^{p\times d}$ has orthonormal columns, i.e., $O^TO = I_d$. 
Define the population SIR kernel matrix based on the standardized sufficient predictor as 
\begin{align*}
    V_S 
    = var\{\bE(\tilde Z_S \mid \fh(\tilde Y))\}
    = \sum_{h=1}^H p_h \bE(\tilde Z_S\mid \tilde Y \in J_h) \bE(\tilde Z_S\mid \tilde Y \in J_h)^T
    \in \bR^{d\times d}.
\end{align*}
Let $\rho_1^2 \geq \cdots \geq \rho_d^2$ denote the eigenvalues of $V_S$, with corresponding orthonormal eigenvectors $v_1, \ldots, v_d$. 
Let $e(\tilde Y) = (1\{\tilde Y\in J_1\}, \ldots, 1\{\tilde Y\in J_{H-1}\})^T \in \bR^{H-1}$ denote the slice indicator,
and define $\Sigma_{ZE} = cov(\tilde Z_S, e(\tilde Y)) \in \bR^{d\times (H-1)}$ and $\Sigma_{EE} = var(e(\tilde Y)) \in \mathbb{R}^{(H-1)\times (H-1)}$. 

Proposition~\ref{prop:cca-pop} is the population counterpart of Proposition~\ref{prop:cca}. 
If $\col(B) = \cS$, then $\text{rank}(V) = d$ if and only if $\rho_d^2 > 0$, in which case SIR recovers the entire central subspace (this is the coverage condition for the chosen slices). 
A complete proof of this proposition is provided in the supplementary material. 

\begin{prop}[Population SIR and canonical correlation analysis]
Suppose model (\ref{eq:model-cs}) holds, $\tilde X$ follows a normal distribution, and $p_h > 0$ for all $h = 1, \ldots, H$. 
Then the following statements hold. 
\begin{enumerate}
\item[(a)]
$V = OV_SO^T$. Consequently, $V$ and $V_S$ have the same nonzero eigenvalues. 
In particular, $\lambda_k(V) = \rho_k^2$ for all $k$, where we set $\rho_k^2 = 0$ for $k > d$.   
Moreover, the population SIR directions are $\beta_k = \Sigma^{-1/2} Ov_k = B(B^T\Sigma B)^{-1/2}v_k$ and satisfy $\beta_j^T \Sigma \beta_k = \delta_{jk}$.

\item[(b)]
The matrix $V_S$ admits the representation $V_S = \Sigma_{ZE} \Sigma_{EE}^{-1} \Sigma_{ZE}^T$. Hence $\rho_k^2$ are the squared population canonical correlations between $B^T\tilde X$ and the slice indicator vector $e(\tilde Y)$. They are the nonzero eigenvalues of 
$A^* = \Sigma_{EE}^{-1/2} \Sigma_{ZE}^T \Sigma_{ZE} \Sigma_{EE}^{-1/2} \in \bR^{(H-1)\times (H-1)}$.

\item[(c)]
The eigenvalues $\rho_k^2$ depend only on the joint distribution of $(\tilde Z_S, \tilde Y)$ and the cut points. 
They do not depend on $p$, $\mu$, or $\Sigma$, nor on the choice of $B$ within a given column space. Moreover, $0\leq \rho_k^2 < 1$. 
\end{enumerate}
\label{prop:cca-pop}
\end{prop}

\subsection{Oracle quantities}
\label{subsec:oracle}

If the true $B$ were known, we could reduce $X$ to the sufficient predictors $X_S = XB \in \bR^{n\times d}$ and apply the SIR algorithm to $X_S$. Although this is not feasible in practice, the resulting oracle procedure plays a pivotal role in our subsequent analysis of the asymptotic properties of SIR. 

Let $\bar X_S = C_nXB \in\bR^{n\times d}$ denote the centered sufficient predictor matrix, and let $P_{\bar X_S}$ denote the orthogonal projection matrix onto $\col(\bar X_S)$. 
Define 
\begin{align*}
    A = F_{\bar G}^T P_{\bar X_S} F_{\bar G} \in\mathbb{R}^{(H-1)\times (H-1)},
\end{align*} 
which is analogous to $K$, with $P_{\bar X}$ replaced by $P_{\bar X_S}$. 
Since $X_S$ has $d$ columns, the matrix $A$ has rank at most $d$. Its $d$ largest eigenvalues, denoted by $\ta_1 \geq \cdots \geq \ta_d$, are the squared sample canonical correlations between $X_S$ and the slice indicator vectors. 
By Proposition~\ref{prop:cca}, applied to the $d$-dimensional data $\{(y_i, B^Tx_i)\}$, these eigenvalues are the nonzero eigenvalues of the oracle SIR kernel matrix computed using the sufficient predictors $X_S$. 
Although they cannot be computed in practice because $B$ is unknown, they serve as natural sample counterparts of the population eigenvalues $\rho_k^2$ defined in Section~\ref{subsec:pop}. 
Moreover, they depend on the data only through $\{(y_i, B^Tx_i)\}$, and hence do not depend on $p$. 
Let $\hat v_k \in \bR^d$ denote the $k$th oracle SIR direction, normalized as $\hat v_k^T (B^T\hat\Sigma B) \hat v_k = 1$. 

\begin{assumption}
The joint distribution of $(\tilde Y, \tilde Z_S)$ does not depend on $n$, all slice probabilities $p_h$ are positive, and for slices with approximately equal counts, $F_{\tilde Y}(c_h) = h/H$ for $h < H$.
\label{ass:link}
\end{assumption}

\begin{prop}
Under model (\ref{eq:model-cs}), suppose Assumption~\ref{ass:link} holds, $\tilde X$ follows a normal distribution, and $n\to\infty$. Then $\ta_k \toP \rho_k^2$ for $k \leq d$, even when $p$ grows with $n$. Moreover, if $\rho_k^2$ is a simple eigenvalue of $V_S$, then $(B^T\Sigma B)^{1/2} \hat v_k \toP v_k$ up to sign. 
\label{prop:oracle}
\end{prop}

It is important to note that $\hat V$ is not a consistent estimator of $V$ when $p/n\to\gamma > 0$, as implied by Theorem~\ref{thm:values1}. In contrast, the oracle quantities $\ta_k$ and $\hat v_k$ converge to their respective population counterparts because they are defined in the fixed dimension $d$, rather than the increasing dimension $p$. 
Some of the asymptotic results established later require $\ta_k \toP \rho_k^2$ as an assumption.  
Proposition~\ref{prop:oracle} demonstrates that this assumption is valid. 
A complete proof of this proposition is provided in the supplementary material. 
The proof essentially follows the standard argument for establishing the consistency of SIR in the fixed-dimensional setting.

\section{An Exact Representation}
\label{sec:exact}

This section establishes an exact representation in Theorem~\ref{thm:represent}, 
which serves as the foundation for all subsequent analysis of the eigenvalues of $K$, or equivalently, those of $\hat V$.
The motivation is as follows.
The first $d$ eigenvalues of $K$ correspond to signals and are expected to be related to the eigenvalues of $A$.
The definitions of $K$ and $A$ in Section~\ref{sec:review} yield 
\begin{align*}
    K = A + F_{\bar G}^T(P_{\bar X} - P_{\bar X_S})F_{\bar G}.
\end{align*}
Let $A = U\Lambda U^T$ be the spectral decomposition of $A$, 
where $U$ and $\Lambda$ are defined as in Theorem~\ref{thm:represent}.
It follows that 
\begin{align*}
    U^TKU = \Lambda + U^TF_{\bar G}^T(P_{\bar X} - P_{\bar X_S})F_{\bar G}U.
\end{align*}
A careful analysis of the second term shows that it can be expressed as $D\bB D$. 
A key result we use is that, if $P$ is an orthogonal projection matrix onto a uniformly distributed subspace
and $F$ is a fixed frame, then $F^TPF$ follows a matrix-variate Beta distribution and this distribution does not depend on $F$. 
We apply this result with $P = P_{\bar X} - P_{\bar X_S}$ and a matrix $F$ constructed from $F_{\bar G}U$, which is $\cF$-measurable. 
This yields $\bB$, which is independent of $\cF$. 

Theorem~\ref{thm:represent} introduces the matrix $M = \Lambda + D\bB D$, 
where $\Lambda$ and $D$ are diagonal matrices depending only on eigenvalues of $A$, 
and $\bB$ is independent of $\{(y_i, B^Tx_i)\}$.
The preceding discussion implies that the eigenvalues of $M$ coincide with those of $K$.
This structure facilitates the analysis of the eigenvalues of $K$ in subsequent sections. 
A complete proof of Theorem~\ref{thm:represent} is provided in the supplementary material. 

\begin{assumption}
The observations are iid.
The predictor $\tilde X$ follows a normal distribution $\cN_p(\mu, \Sigma)$, where $\Sigma \succ 0$.
The model (\ref{eq:model-cs}) holds with $P_{\col(B)}$ for a $p\times d$ matrix $B$ of rank $d$.
The slice labels are a function of $(y_1, \ldots, y_n)$ and every slice is nonempty. 
Furthermore, assume that $d < H - 1 < p < n - 1$, and $H + d < n$. 
\label{ass:null}
\end{assumption}

Assumption~\ref{ass:null} holds when the true structural dimension is $d_0 = d$ and $\col(B) = \cS$. 
It also holds when $d_0 < d$ and $B$ is constructed by augmenting a basis of $\cS$ with additional linearly independent vectors.

\begin{theorem} \label{thm:represent}
Suppose Assumption~\ref{ass:null} holds. 
Then there exist an $\cF$-measurable $(H-1)\times (H-1)$ orthogonal matrix $U$ and 
a random matrix $\bB\sim \cB_{H-1}((p-d)/2, (n-p-1)/2)$, independent of $\cF$, such that 
\begin{align*}
    U^TKU = M := \Lambda + D\bB D,
\end{align*}
where $\Lambda = \diag(\ta_1, \ldots, \ta_d, 0, \ldots, 0)$ and
$D = \diag(\sqrt{1 - \ta_1}, \ldots, \sqrt{1 - \ta_d}, 1, \ldots, 1)$.
In particular, the eigenvalues of $M$ are $r_1^2, \ldots, r_{H-1}^2$.
\end{theorem}

Recall that the $\sigma$-field $\cF$ is defined as $\cF = \sigma(B^Tx_1, \ldots, B^Tx_n, y_1, \ldots, y_n)$. 
Theorem~\ref{thm:represent} expresses $r_i^2$ in terms of $\ta_i$ and a random matrix that is independent of $\{(y_i, B^Tx_i)\}$. 
Theorem~\ref{thm:represent} is an exact finite-sample result and does not rely on asymptotic approximations. 
Moreover, all matrices involved have fixed dimensions. As $n, p\to\infty$, only the parameters governing the distribution of $\bB$ diverge. 
Since $A$ has rank at most $d$, $\Lambda$ has at most $d$ nonzero diagonal entries. 

When proving Theorem~\ref{thm:represent}, the normality assumption is used only for the noise coordinates. Specifically, it suffices that, after
an affine transformation, $X = (X_S, X_N)$, where $X_N$ has independent $\cN(0, 1)$ entries, $X_N$ is independent of $(Y, X_S)$, and $\text{rank}(C_n X_S) = d$ almost surely. 

\section{SIR Eigenvalues}
\label{sec:values}

This section presents our results on the SIR eigenvalues. 
These eigenvalues provide information about the number of directions in $\cS$ that SIR can recover.
Section~\ref{subsec:values1} derives the first-order asymptotic behavior of the SIR eigenvalues,
while Section~\ref{subsec:values2} characterizes their fluctuations.  

\subsection{The first-order behavior of the SIR eigenvalues}
\label{subsec:values1}

Let $\gamma_n = (p-d)/(n-d-1)$. 
Theorem~\ref{thm:values1} characterizes the first-order asymptotic behavior of the sample SIR eigenvalues as $n\to\infty$. 
We briefly explain the intuition behind Theorem~\ref{thm:values1}.
By Proposition~\ref{prop:cca} and Theorem~\ref{thm:represent}, $\hat V$, $K$, and $M$ share the same nonzero eigenvalues.
Although $M$ has a simpler form than $\hat V$ and $K$, it is not diagonal. 
To facilitate the analysis, define 
\begin{align*}
    M^* 
    = \Lambda + \gamma_n D^2 
    = \diag\{\gamma_n + (1 - \gamma_n)\ta_1, \dots, \gamma_n + (1 - \gamma_n)\ta_d,
             \gamma_n, \dots, \gamma_n\}, 
\end{align*}
which is obtained by replacing $\bB$ by its mean in the expression of $M$.
Note that $M^*$ is diagonal, with eigenvalues in nonincreasing order. 
For $k \leq d$, we have 
$\lambda_k(M) - \lambda_k(M^*) = \hat\lambda_k - \{\gamma_n + (1 - \gamma_n)\ta_k\}$,
whereas for $k > d$, 
$\lambda_k(M) - \lambda_k(M^*) = \hat\lambda_k - \gamma_n$.
By Weyl's inequality, $|\lambda_k(M) - \lambda_k(M^*)| \leq \|M - M^*\|_F$.
Thus, Theorem~\ref{thm:values1} follows by bounding $\|M - M^*\|_F$. 
The complete proof is provided in the supplementary material. 

\begin{theorem}[First-order behavior of the sample SIR eigenvalues]
Suppose Assumption~\ref{ass:null} holds, $H$ and $d$ are fixed, and $n\to\infty$. 
Then 
\begin{align}
    \max_{k\leq d} |\hat\lambda_k - \{\gamma_n + (1 - \gamma_n)\ta_k\}| 
    + 
    \max_{d < k \leq H-1} |\hat\lambda_k - \gamma_n|
    = O_p(\sqrt{p}/n).
\label{eq:value-rate}
\end{align}
In particular, if $p/n \to \gamma \in [0, 1)$ and $\ta_k \toP \rho_k^2$ for $k \leq d$, then 
$\hat\lambda_k \toP \rho_k^2 + \gamma (1 - \rho_k^2)$ for $k \leq d$, whereas $\hat\lambda_k \toP \gamma$ for $d < k \leq H - 1$.
\label{thm:values1}
\end{theorem}

Theorem~\ref{thm:values1} implies that $\hat\lambda_k \toP \rho^2_k$ for $k \leq d$ and $\hat\lambda_k \toP 0$ for $k > d$ when $p/n \to 0$.
These results suggest the consistency of the SIR algorithm and include the asymptotic regime $p = o(n^{1/2})$ considered in \cite{zhu2006sliced}. 
In this setting, $p$ grows sufficiently slowly relative to $n$, and the increasing noise component in $X$ does not alter the asymptotic properties of SIR in the fixed-dimensional regime.

The sample signal eigenvalues are biased upward when $p/n\to\gamma > 0$, in the sense that
the limiting squared sample canonical correlation is $\rho_k^2 + \gamma(1 - \rho_k^2)$, rather than the population squared canonical correlation $\rho_k^2$.
The null eigenvalues are similarly inflated from 0 to $\gamma$. Since $H - 1$ is fixed, there is no phase transition: every positive population squared canonical correlation $\rho_k^2 > 0$ separates from the null eigenvalues, in contrast to the setting where both dimensions grow \citep{bao2019canonical}. 
Formally, letting the second dimension ratio tend to zero in Theorem~1.7 of \cite{bao2019canonical} yields the same limit $\rho_k^2 + \gamma (1 - \rho_k^2)$.

The condition $\ta_k \toP \rho_k^2$ is reasonable even when $p/n\to\gamma > 0$, because the structural dimension $d$ remains fixed; see Proposition~\ref{prop:oracle}. 
We state this condition explicitly rather than Assumption~\ref{ass:link}, because it is implied by Assumption~\ref{ass:link} but also holds under more general slicing rules.

\subsection{Fluctuations of the SIR eigenvalues}
\label{subsec:values2}

First-order limits alone are insufficient for statistical inference. 
Testing hypotheses and constructing confidence intervals require characterizing 
the magnitude and distribution of fluctuations around these limits.
To characterize the fluctuations of the SIR eigenvalues, we need the following additional assumption,
which is satisfied under Assumption~\ref{ass:link} whenever $\rho_d^2 > 0$, by Proposition~\ref{prop:oracle}.

\begin{assumption}
There exists a constant $\amin > 0$ such that $\bP(\ta_d \geq \amin) \to 1$ as $n\to\infty$. 
\label{ass:signal}
\end{assumption}

\begin{theorem}[Fluctuations of the sample SIR eigenvalues]
Suppose Assumptions~\ref{ass:null} and~\ref{ass:signal} hold, $H$ and $d$ are fixed, $p\to\infty$, and $p/n\to\gamma\in[0, 1)$. 
Let $\cG_{H-1}$ be a GOE matrix independent of $\cF$, partitioned into blocks $\cG_{11} \in \bR^{d\times d}$ and $\cG_{00} \in \bR^{(H-d-1)\times (H-d-1)}$. Then the following statements hold. 
\begin{enumerate}
\item[(a)]
{\em Null eigenvalues}. Jointly for $j = 1, \ldots, H - d - 1$,
\begin{align*}
    \frac{n}{\sqrt{p}} (\hat \lambda_{d + j} - \gamma_n) 
    \tod 
    \sqrt{1 - \gamma} \, \lambda_j(\cG_{00}),
\end{align*}
where $\lambda_j(\cG_{00})$ is the $j$th largest eigenvalue of $\cG_{00}$. 

\item[(b)] 
{\em Signal eigenvalues}. Suppose, in addition, that $\ta_k \toP \rho_k^2$ for $k \leq d$, where $\rho_1^2 > \cdots > \rho_d^2 > 0$. Then, jointly with the convergence in (a),
\begin{align*}
    \frac{n}{\sqrt{p}}\{\hat \lambda_k - \gamma_n - (1 - \gamma_n)\ta_k\}
    \tod 
    \sqrt{1-\gamma}(1 - \rho_k^2)(\cG_{11})_{kk},
    \quad k \leq d,
\end{align*}
where $(\cG_{11})_{kk}$ denotes the $k$th diagonal entry of $\cG_{11}$.

\item[(c)]
{\em Joint convergence and independence}. 
The convergences in (a) and (b) hold jointly with any $\cF$-measurable sequence $\Xi_n \tod \Xi$, where the limit matrix $\cG_{H-1}$ is independent of $\Xi$. 
\end{enumerate}
\label{thm:values2}
\end{theorem}

The null eigenvalues therefore exhibit fluctuations governed by the eigenvalues of a fixed-size GOE matrix, centered at $\gamma_n$ and scaled by $\sqrt{p(1 - \gamma)}/n$. Summing over $j$ recovers the scale of the central limit theorem
for Li's test statistic in Theorem~\ref{thm:test}, since $\tr(\cG_{00}) \sim \cN(0, 2(H-1-d))$. The signal eigenvalues fluctuate around their
inflated oracle values with asymptotically independent Gaussian errors.
Part (c) of Theorem~\ref{thm:values2} emphasizes that all these fluctuations are independent of $\{(y_i, B^Tx_i)\}$.

The bias in the SIR eigenvalues can be removed using the known constants $\gamma_n$, at the cost of
an additional asymptotic variance of $2\gamma (1 - \rho_k^2)^2/(1-\gamma)$ on the root-$n$ scale. 
The following corollary formalizes this result. 

\begin{coro}[Bias-corrected eigenvalues]
Let $\hat\rho_k^2 = (\hat\lambda_k - \gamma_n)/(1 - \gamma_n)$ for $k \leq d$.
Under the assumptions of Theorem~\ref{thm:values2}(b), the following statements hold. 
\begin{enumerate}
\item[(a)] 
For each $k\leq d$, $\hat\rho_k^2 - \ta_k = O_p(\sqrt{p}/n)$.
Consequently, $\hat\rho_k^2 \toP \rho_k^2$.

\item[(b)]
We have
$\sqrt{n}(\hat\rho_k^2 - \ta_k)_{k\leq d} \tod 
\cN_d(0, \frac{2\gamma}{1-\gamma}\diag((1-\rho_1^2)^2, \ldots, (1 - \rho_d^2)^2))$, 
where the limiting distribution is independent of $\cF$.

\item[(c)]
Suppose that the oracle eigenvalues satisfy $\sqrt{n}(\ta_k - \rho_k^2)_{k\leq d} \tod \cN_d(0, \Sigma_{or})$. Then 
\begin{align*}
    \sqrt{n}(\hat\rho_k^2 - \rho_k^2)_{k\leq d}
    \tod \cN_d\left(0, \Sigma_{or} + \frac{2\gamma}{1-\gamma}\diag((1 - \rho_k^2)^2)_{k\leq d}\right).
\end{align*}
\end{enumerate}
\label{coro:values}
\end{coro}

The fixed-dimensional asymptotic variance $\Sigma_{or}$ is the variance of SIR eigenvalues based on the signal predictor $XB$; see \cite{hsing1992asymptotic} and \cite{zhu1995asymptotics}.
The additional variance in Corollary~\ref{coro:values}(c) arises from the $p - d$ noise coordinates. 
It vanishes when $\gamma = 0$, in which case the bias-corrected SIR eigenvalues have the same asymptotic distribution as in the fixed-dimensional setting.

\section{Sequential Test for the Structural Dimension}
\label{sec:test}

This section revisits Li's sequential testing procedure for determining the structural dimension $d_0$. 
We develop a corrected test and two additional new tests to correct Li's $\chi^2$-test in the proportional regime:
the exact test based on an upper bound for $T_d$, 
the corrected test based on a normal approximation,
and the largest-eigenvalue test based on the largest null eigenvalue $\hat\lambda_{d+1}$.

\subsection{The exact test}

Proposition~\ref{prop:cca} implies that 
\begin{align*}
    T_d 
    = n\sum_{i=d+1}^p \hat\lambda_i 
    = n\sum_{i=d+1}^{H-1} r_i^2. 
\end{align*}
In the exact representation in Theorem~\ref{thm:represent}, $\Lambda$ approximates the first $d$ eigenvalues, while $D\bB D$ is expected to capture the behavior of the remaining eigenvalues. 
Let $\bB_{00}$ denote the bottom-right $(H-d-1)\times (H-d-1)$ block of $\bB$, and define
\begin{align*}
    T_d^* = n\, \tr(\bB_{00}). 
\end{align*}
Since $\bB$ is independent of $\cF = \sigma(B^Tx_1, \ldots, B^Tx_n, y_1, \ldots, y_n)$, so is $\bB_{00}$. 
As $T_d^*$ approximates $T_d$, we can derive the asymptotic properties of $T_d$ by studying $T_d^*$. 

The following theorem shows that $T^*_d$ provides an upper bound for $T_d$
and characterizes several properties of $T_d^*$.
 
\begin{theorem} \label{thm:Td}
Under Assumption~\ref{ass:null}, the following statements hold.
\begin{enumerate}
\item[(a)]
$T_d^*/n \sim \Pi(p-d, H-d-1, n-d-1)$ and 
$T_d^*$ is independent of $\cF$. 
Moreover, $T_d \leq T_d^*$ almost surely, with equality when $d = 0$.

\item[(b)]
The mean and variance of $T_d^*$ are 
\begin{align*}
   \bE(T_d^*) 
   &= \frac{n(p-d)(H-d-1)}{(n-d-1)}, \\
   var(T_d^*) 
   &= \frac{2n^2(p-d)(H-d-1)(n-p-1)(n - H)}{(n-d-1)^2(n-d+1)(n-d-2)}.
\end{align*}

\item[(c)]
$\bP(T_d > c_{n, \alpha}(d)) \leq \alpha$ for any $n$, $p$, and $H$ satisfying Assumption~\ref{ass:null}, 
where $c_{n, \alpha}(d)$ denotes the upper-$\alpha$ quantile of $n\Pi(p-d, H-d-1, n-d-1)$. 
\end{enumerate}
\end{theorem}

Theorem~\ref{thm:Td} suggests rejecting $H_0: d_0 = d$ at significance level $\alpha$ if $T_d > c_{n, \alpha}(d)$. 
We refer to this procedure as the {\em exact test}.  
The critical value $c_{n,\alpha}(d)$ can be readily computed via Monte Carlo simulation,
as its calculation involves only $(H-d-1)\times (H-d-1)$ Wishart matrices. 
This test is conservative in the sense that its probability of type I error is bounded above by $\alpha$. 
However, as shown in Corollary~\ref{coro:regimes}(c), the asymptotic size converges to $\alpha$ as $n\to\infty$.

\subsection{The corrected test}

Theorem~\ref{thm:values1} shows that the null eigenvalues are inflated by a common bias. 
Thus, $T_d$ can be corrected by an appropriate rescaling. 
The following theorem formalizes this correction through a normal approximation. 

\begin{theorem}
Suppose Assumptions~\ref{ass:null} and~\ref{ass:signal} hold, 
$H$ and $d$ are fixed, $p\to\infty$ and $p/n\to\gamma\in[0,1)$.
Then, the following statements hold.
\begin{enumerate}
\item[(a)] 
The difference between $T_d^*$ and $T_d$ satisfies
$0\leq T_d^* - T_d = O_p(p/n)$ and
\begin{equation*}
  \frac{T_d - (p-d)(H-d-1)}{\sqrt{2(p-d)(H-d-1)(1-p/n)}}
  \tod \cN(0,1).
\end{equation*}

\item[(b)] 
For every $d'<d$,
\begin{equation*}
  \frac{T_{d'} - (p-d')(H-d'-1)}{\sqrt{2(p-d')(H-d'-1)(1-p/n)}}
  \toP\infty.
\end{equation*}
\end{enumerate}
\label{thm:test}
\end{theorem}

Assumption~\ref{ass:signal} is void when testing $H_0: d_0 = 0$. 
In this case, the test statistic $T_0 = n\tr(\hat V)$, which is the sum of all SIR eigenvalues, is equal to $n$ times Pillai's trace statistic from a MANOVA in which the slices are treated as groups \citep{muirhead1982aspects}.

The sampling distribution $\chi^2_{f_d}$ in \cite{li1991sliced} has mean $f_d$ and variance $2f_d$,
where $f_d = (p-d)(H-d-1)$.
Thus, its mean is correct to first order, but its variance fails to account for the
factor $(1-p/n)$. After removing the $d$ signal eigenvalues, the remaining $H-d-1$
eigenvalues behave like the null eigenvalues of a problem with $p-d$ predictors and $H-d-1$
slice contrasts, up to a nonnegative correction of order $O_p(p/n)$.

Theorem~\ref{thm:test} suggests rejecting $H_0: d_0 = d$ at significance level $\alpha$ if  
\begin{align*}
    \frac{T_d - (p-d)(H-d-1)}{\sqrt{2(p-d)(H-d-1)(1-p/n)}}
    > z_{\alpha},
\end{align*}  
where $z_{\alpha}$ denotes the upper-$\alpha$ quantile of $\cN(0, 1)$. 
We refer to this procedure as the {\em corrected test}. 
We use a one-sided test because only large values of $T_d$ provide evidence in favor of $H_a$. 

Theorem~\ref{thm:test} also provides insight into the power of the test. 
In practice, a scree plot can be constructed by plotting the normalized $T_d$ against $d$, 
with the structural dimension selected at the point where a sharp drop occurs.
Theorem~\ref{thm:test} provides a theoretical justification for this procedure:
the normalized $T_{d'}$ diverges for $d' < d$, but is bounded in probability at $d$.

\subsection{Asymptotic behavior of the test}

The following corollary characterizes the asymptotic behavior of Li's $\chi^2$-test, as well as the properties of the exact and corrected tests. 

\begin{coro}
Under the assumptions of Theorem~\ref{thm:test}, with the true dimension $d = d_0$,
the following statements hold.
\begin{enumerate}
\item[(a)] 
If $p/n \to 0$, Li's $\chi^2$-test has asymptotic size $\alpha$.
\item[(b)] 
If $p/n \to \gamma \in (0,1)$, Li's $\chi^2$-test has asymptotic size
$1 - \Phi(z_{\alpha}/\sqrt{1-\gamma}) < \alpha$.
\item[(c)] 
Both the exact and corrected tests have asymptotic size $\alpha$. 
Moreover, the exact test has size at most $\alpha$ for every $n$.
\end{enumerate}
\label{coro:regimes}
\end{coro}

Li's $\chi^2$-test is asymptotically valid when $p$ is fixed \citep{li1991sliced}, and remains asymptotically valid when $p\to\infty$ and $p/n \to 0$. This robustness stems from the rotational invariance of the noise,
which allows the test to extend naturally to diverging $p$. 
When $\gamma > 0$ and $\alpha = 0.05$, the asymptotic size $1 - \Phi(z_{\alpha}/\sqrt{1-\gamma})$ is approximately $0.029$ at $\gamma = 0.25$, $0.010$ at $\gamma = 0.5$, and $0.0005$ at $\gamma = 0.75$. 
Therefore, Li's $\chi^2$-test becomes conservative when $\gamma > 0$
and rejects a true $H_0$ far less often than the nominal level. 
As demonstrated by the simulation studies in Section~\ref{sec:simulation}, Li's $\chi^2$-test loses power against weak signals in finite samples.

When applying the sequential testing procedure to determine the structural dimension, 
failure to reject $H_0: d_0 = d$ for $d < d_0$ leads to an underestimate of the structural dimension. 
Corollary~\ref{coro:test} shows that this does not occur asymptotically, allowing the structural dimension to be estimated at a prespecified significance level. 
This is noteworthy because, when $\gamma > 0$, the SIR directions are inconsistent (Theorem~\ref{thm:accuracy}), 
yet determining the structural dimension remains feasible. In other words, the sequential testing procedure is valid even when $\cS$ cannot be consistently estimated. 

\begin{coro} 
\label{coro:test}
Suppose Assumption~\ref{ass:null} holds with $d = d_0$ and $\col(B) = \cS$
and Assumption~\ref{ass:signal} holds.
Assume that 
$H$ is fixed, $p\to\infty$, and $p/n\to\gamma\in[0,1)$.
The following statements hold.
\begin{enumerate}
\item[(a)]
If the corrected test is used at each step of the sequential testing procedure, 
then $\bP(\hat d_0 = d_0)\to 1-\alpha$. 
\item[(b)]
If the exact test is used at each step of the sequential testing procedure, 
then $\bP(\hat d_0 > d_0)\leq \alpha$ for every $n$.
Moreover, $\bP(\hat d_0 = d_0)\to 1-\alpha$.
\end{enumerate}
\end{coro}

In fixed dimension, the trace statistic $T_d$ and Bartlett's likelihood-ratio statistic
\begin{align*}
 \tilde T_{d} = -\left\{n - 1 - \frac{p+H}{2}\right\}\sum_{i>d}\log(1 - \hat\lambda_i) 
\end{align*}
are asymptotically equivalent. 
However, this equivalence breaks down when $p/n \to \gamma > 0$.
By Theorem~\ref{thm:values1}, the log-form statistic $\tilde T_d$ is centered around $n(H-d-1)(1-\gamma/2)\{-\log(1-\gamma)\}$, which exceeds $f_d$ by a term of order $n$, far larger than its standard deviation of order
$\sqrt{p}$. Consequently, calibrating $\tilde T_d$ using a $\chi^2_{f_d}$ distribution yields a test whose size tends to one; see also \cite{bai2013testing} and \cite{he2021likelihood}. In contrast, the trace statistic $T_d$ is correctly centered and requires only a variance correction.

\subsection{The largest-eigenvalue test}

Theorem~\ref{thm:values2}(a) characterizes the distribution of the individual null eigenvalues,
thereby yielding a Roy-type test based on the largest null eigenvalue as an alternative to the $\chi^2$-test.
This test is also asymptotically valid and is referred to as the {\em largest-eigenvalue test}. 

\begin{coro}[Largest-eigenvalue test] 
Under the assumptions of Theorem~\ref{thm:values2} with $d = d_0$, 
\begin{align*}
  L_d 
  := \frac{n}{\sqrt{p(1-p/n)}} (\hat\lambda_{d+1}-\gamma_n)
  \tod
  \lambda_{\max}(\cG_{H-d-1}).
\end{align*}
Thus, rejecting $H_0: d_0 = d$ when $L_d$ exceeds the upper-$\alpha$ quantile of
$\lambda_{\max}(\cG_{H-d-1})$ yields a test with asymptotic size $\alpha$. 
Moreover,  
if this test is used at each step of the sequential testing procedure,
$\bP(\hat d_0 = d_0) \to 1 - \alpha$. 
\label{coro:lmax}
\end{coro}

The critical value for the largest-eigenvalue test can be obtained via Monte Carlo simulation, 
since the asymptotic distribution of $L_d$ coincides with that of the largest eigenvalue of 
a fixed-dimensional GOE matrix.

\section{SIR Directions}
\label{sec:vectors}

\citet[Section~5]{li1991sliced} established the root-$n$ consistency of the sample SIR directions for fixed $p$ and also quantified their accuracy. 
We revisit these results in the proportional regime. 
Since the sample SIR directions are vectors in a space whose dimension grows with $n$, we focus primarily on their accuracy and the resulting sufficient predictors. 
Specifically, 
Section~\ref{subsec:vectors-align} studies the alignment between the estimated and population SIR directions,
Section~\ref{subsec:vectors-accu} examines the accuracy of the sample SIR directions, and
Section~\ref{subsec:vectors-pred} investigates the estimated sufficient predictors for a new observation. 

Recall that $\hat\beta_k$ denotes the $k$th estimated SIR direction.  
Define $\hat \beta_{1:d} = (\hat\beta_1,\dots, \hat\beta_d)$. 
We normalize the population SIR directions as in Proposition~\ref{prop:cca-pop}(a): 
$\beta_k = B(B^T\Sigma B)^{-1/2}v_k$, 
so that $\beta_{1:d}^T \Sigma \beta_{1:d} = I_d$.
We choose the sign of each sample SIR direction $\hat\beta_k$ such that $\hat\beta_k^T \Sigma \beta_k \geq 0$. 

\subsection{Alignment with the population SIR directions}
\label{subsec:vectors-align}

For $x\in(0,1]$ and $\gamma\in[0,1)$, define 
\begin{equation*}
  \varphi_{\gamma}(x)
  = \frac{(1-\gamma)x}{(1-\gamma)x + \gamma(1-x)}.
\end{equation*}
For $k\leq d$, let 
\begin{equation*}
  \kappa_k^2 = \frac{\rho_k^2}{\rho_k^2 + \gamma(1 - \rho_k^2)},
  \qquad
  \tau_k^2 = \frac{\gamma(1-\rho_k^2)}{(1-\gamma)\{\rho_k^2 + \gamma(1 - \rho_k^2)\}},
\end{equation*}
so that $\kappa_k^2/(\kappa_k^2+\tau_k^2)=\varphi_\gamma(\rho_k^2)$. 
Define the diagonal matrices 
\begin{align*}
\sK = \diag(\kappa_1,\dots,\kappa_d), \qquad 
\sT= \diag(\tau_1,\dots,\tau_d). 
\end{align*}

\begin{theorem} \label{thm:vectors-align}
Suppose Assumption~\ref{ass:null} holds.
Assume that 
$H$ and $d$ are fixed, 
$p/n \to \gamma\in[0,1)$, 
$\rho_1^2 > \cdots >\rho_d^2 > 0$.
Furthermore, suppose that
$\ta_k\toP \rho_k^2$, and 
the oracle directions are consistent up to sign, in the sense that
$(B^T \Sigma B)^{1/2}\hat v_k \toP v_k$, for $k\leq d$.
Then,  
\begin{align*}
  \hat\beta_{1:d}^T \Sigma \beta_{1:d} \toP \sK, \qquad 
  \hat \beta_{1:d}^T \Sigma \hat \beta_{1:d} \toP \sK^2 + \sT^2.
\end{align*}
\end{theorem}

Theorem~\ref{thm:vectors-align} implies that 
$\hat\beta_k$ is asymptotically $\Sigma$-orthogonal to $\beta_j$, for $j\neq k$, 
and its $\Sigma$-inner product with $\beta_k$ converges to $\kappa_k$,
which is strictly less than one when $\gamma > 0$. 
Since $\hat\beta_k^T \hat\Sigma \hat\beta_k = 1$, 
one might expect $\hat\beta_k^T \Sigma \hat\beta_k \approx 1$. 
However, 
$\hat\beta_k^T\Sigma\hat\beta_k \toP \kappa_k^2 + \tau_k^2 > 1$ when $\gamma > 0$, 
where $\kappa_k^2$ represents the contribution of the component of $\hat\beta_k$ lying in $\col(B)$ and $\tau_k^2$ represents that of the component $\Sigma$-orthogonal to $\col(B)$.

It is worth noting that, since the primary inferential target is $\col(B)$ rather than $\beta_{1:d}$, 
a more appropriate assessment of estimation performance is to evaluate the alignment of $\col(\hat\beta_{1:d})$ with $\col(B)$, as discussed in Section~\ref{subsec:vectors-accu}. 
Nevertheless, Theorem~\ref{thm:vectors-align} establishes results that facilitate the analysis in Section~\ref{subsec:vectors-accu}. 

\subsection{Accuracy of the SIR directions}
\label{subsec:vectors-accu}

\cite{li1991sliced} introduced two criteria for assessing the accuracy of estimated SIR directions: 
one measures the alignment of an individual estimated SIR direction with $\col(B)$,
while the other assesses the accuracy of the estimated SIR directions collectively.  
\begin{itemize}
\item 
For any $\hat b\in\bR^p$, the alignment between $\hat b$ and $\col(B)$ can be measured by the squared cosine of the angle between $\hat b$ and the subspace $\col(B)$, defined as 
\begin{align*}
  R^2(\hat b)
  = \cos_{\Sigma}^2(\hat b, \col(B))  
  := \max_{\beta \in \col(B)} \text{corr}^2(\hat b^T\tilde X, \beta^T\tilde X) 
  = \frac{\hat b^T\Sigma B(B^T\Sigma B)^{-1} B^T\Sigma \hat b}{\hat b^T\Sigma \hat b}. 
\end{align*}

\item 
For a rank-$d$ matrix $\hat B = (\hat b_1, \ldots, \hat b_d)$, the agreement between $\col(\hat B)$ and $\col(B)$ can be measured by the squared trace correlation, defined as the average of the squared canonical correlations between $\hat B^T\tilde X$ and $B^T\tilde X$:
\begin{align*}
    R^2(\hat B) = \frac{1}{d}\tr 
    \{(\hat B^T\Sigma\hat B)^{-1}\hat B^T\Sigma B(B^T\Sigma B)^{-1}B^T\Sigma\hat B\}.
\end{align*}
\end{itemize}
Both criteria are affine invariant and depend on $\hat B$ and $B$ only through $\col(\hat B)$ and $\col(B)$. 
Moreover, both take values in $[0, 1]$, with values of 0 and 1 corresponding to orthogonality and perfect alignment, respectively.

The following theorem characterizes the accuracy of the estimated SIR directions. 
We emphasize that the $k$th population eigenvalue is assumed to be simple for (a) and 
all population eigenvalues are assumed to be distinct for (b). 

\begin{theorem}[Accuracy of the estimated SIR directions]
Suppose Assumption~\ref{ass:null} holds, $H$ and $d$ are fixed,
$p/n \to \gamma \in [0,1)$ and $\ta_k \toP \rho_k^2$ for $k\leq d$.
The following statements hold.
\begin{enumerate}
\item[(a)] 
Suppose $\rho_k^2 > 0$ and $\rho_k^2 \neq \rho_j^2$ for all $j\neq k$. 
Then
\begin{equation*} 
  R^2(\hat\beta_k)
  = \cos^2_{\Sigma} (\hat\beta_k, \col(B))
  \toP
  \varphi_{\gamma}(\rho_k^2)
  = \frac{(1-\gamma)\rho_k^2}{(1-\gamma)\rho_k^2 + \gamma(1-\rho_k^2)}.
\end{equation*}

\item[(b)] 
Suppose $\rho_1^2 > \cdots > \rho_d^2 > 0$.
Then 
\begin{align*}
  R^2(\hat \beta_{1:d}) \toP 
  \frac{1}{d} \sum_{k=1}^d \varphi_{\gamma}(\rho_k^2).
\end{align*}
\end{enumerate}
\label{thm:accuracy}
\end{theorem}

The limit in Theorem~\ref{thm:accuracy}(a) equals $1$ when $\gamma = 0$, 
decreases monotonically with $\gamma$, 
approaches $0$ as $\gamma \to 1$, 
and equals $\rho_k^2$ when $\gamma = 1/2$. 
These results indicate that the estimated SIR directions asymptotically align with $\col(B)$ when $p/n\to 0$. 
As $\gamma$ increases, their alignment deteriorates, 
and they become asymptotically orthogonal to $\col(B)$ as $\gamma\to 1$.
\cite{lin2018consistency} established the inconsistency of SIR for a single index model when $p/n\to \gamma > 0$, but did not derive an explicit limit as in Theorem~\ref{thm:accuracy}.

For fixed $p$, \citet[equation~(5.1)]{li1991sliced} approximated the expected squared trace correlation by
\begin{equation}\label{eq:li51}
  \bE[R^2(\hat \beta_{1:d})]
  =
  1-\frac{p-d}{nd}\sum_{k=1}^d \left(\frac{1}{\rho_k^2} - 1 \right)
  + o(n^{-1}).
\end{equation}
Since $1 - \varphi_{\gamma}(\rho^2) = \gamma(\rho^{-2}-1)/\{1+\gamma(\rho^{-2}-2)\}$,
the limit in Theorem~\ref{thm:accuracy}(b) can be written as 
\begin{align*}
  1 - \frac{1}{d}\sum_{k=1}^d\frac{\gamma(\rho_k^{-2}-1)}{1+\gamma(\rho_k^{-2}-2)}
  =
  1 - \frac{\gamma}{d}\sum_{k=1}^d (\rho_k^{-2} - 1) + O(\gamma^2).
\end{align*}
To first order in $\gamma = \lim(p-d)/n$, this agrees with \eqref{eq:li51}.
Thus, Theorem~\ref{thm:accuracy} confirms the formula proposed by \cite{li1991sliced} and provides its exact counterpart in the proportional regime. 
The first-order approximation overstates the accuracy when $\rho_k^2 > 1/2$ and 
understates it when $\rho_k^2 < 1/2$. Moreover, it can be substantially inaccurate;  
see Section~\ref{subsec:sim-vectors} for details. 

Theorem~\ref{thm:accuracy} also sharpens the minimax analysis of
\cite{lin2021optimality}. When $\Sigma = I_p$, we have 
$\|P_{\hat\beta_{1:d}}-P_B\|_F^2=2d\{1-R^2(\hat\beta_{1:d})\}$, 
whose limit value,
$2\sum_k\{1-\varphi_\gamma(\rho_k^2)\}$, 
is of the same order as the minimax rate $d(p-d)/(n\rho_d^2)$ established therein,
but with an explicit asymptotic constant.

\subsection{Estimated sufficient predictors on new data}
\label{subsec:vectors-pred}

Let $(y_0, x_0)$ be a new observation independent of the training sample.
We expect the estimated sufficient predictor $\hat\beta_{1:d}^T (x_0 - \bar x)$ to capture all regression-related information. It is therefore of interest to examine its asymptotic properties.

\begin{theorem}[Estimated sufficient predictors on new data]\label{thm:pred}
Under the assumptions of Theorem~\ref{thm:vectors-align},
the following statements hold.
\begin{enumerate}
\item[(a)] 
Let $w_S = \beta_{1:d}^T (x_0 - \mu)$. Then, $w_S \sim \cN_d (0, I_d)$ and 
\begin{align*}
  (\hat\beta_{1:d}^T (x_0 - \bar x),\ w_S,\ y_0)
  \tod
  (\sK w_S + \sT w_N,\ w_S,\ y_0),   
\end{align*}
where $w_N\sim \cN_d(0, I_d)$ is independent of $(w_S, y_0)$.

\item[(b)] 
For each $k\leq d$, 
\begin{align*}
\hat\beta_k^T(x_0 - \bar x) - \beta_k^T (x_0 - \mu)
\tod
\cN(0,(1-\kappa_k)^2 + \tau_k^2). 
\end{align*}
The limiting distribution is degenerate if and only if $\gamma = 0$.
\end{enumerate}
\end{theorem}

\emph{Attenuation and noise.} 
Asymptotically, $\hat\beta_k^T (x_0-\bar x) \approx \kappa_k\beta_k^T (x_0-\mu)+\tau_kw_{N, k}$.
When $\gamma > 0$, the factor $\kappa_k < 1$ induces shrinkage towards zero, while the term
$\tau_kw_{N, k}$ represents Gaussian noise independent of the true predictors
and the response, and hence carries no additional information about $y_0$. 
The components of the estimated sufficient predictor are asymptotically independent, 
with weaker directions (those associated with smaller $\rho_k^2$) experiencing greater shrinkage and larger noise.

\emph{No rescaling restores consistency.} 
Can the consistency of $\hat\beta_k^T (x_0 - \bar x)$ be restored through rescaling, as in the case of Li's $\chi^2$-test or the bias-corrected eigenvalues? The answer is negative. 
The minimum mean squared error after optimal rescaling is
\begin{align*}
  \min_c
  \bE[\{c\, \hat\beta_k^T (x_0 - \bar x) - \beta_k^T(x_0 - \mu)\}^2
  \mid \text{sample}]
  \toP
  \frac{\tau_k^2}{\kappa_k^2 + \tau_k^2}
  = 1 - \varphi_{\gamma}(\rho_k^2),
\end{align*}
where the minimum is attained at $c = \kappa_k/(\kappa_k^2 + \tau_k^2)$. 
This limit coincides with the loss $1 - R^2(\hat\beta_k)$ in Theorem~\ref{thm:accuracy}. 

\emph{In-sample versus out-of-sample.} 
In the training sample,
$\hat\beta_k^T (x_i-\bar x)$ has variance exactly $1$ due to the normalization constraint. 
For a new observation, however, its limiting variance is, when $\gamma > 0$, 
\begin{align*}
\kappa_k^2+\tau_k^2 
= 1 + \frac{\gamma^2(1-\rho_k^2)}{(1-\gamma)\{\gamma + (1-\gamma)\rho_k^2\}}
> 1,
\end{align*}
reflecting the fact that the estimated directions are fitted partly to the noise predictors in the sample. 
The additional variability vanishes when $\gamma = 0$, indicating that the effect of noise predictors becomes negligible only when $p$ grows sufficiently slowly relative to $n$.

\emph{Estimating the constants.} 
By Corollary~\ref{coro:values}, $\hat\rho_k^2$ is consistent. 
Thus, substituting $\hat\rho_k^2$ and $p/n$ into
the expressions of 
the accuracy $\varphi_{\gamma}(\rho_k^2)$, 
the attenuation factor $\kappa_k$, 
and the noise variance $\tau_k^2$
yields consistent estimators of these quantities.
In particular, $\kappa_k^2$ and $\tau_k^2$ can be estimated by
\begin{align*}
\hat\kappa_k^2 = (\hat\rho_k^2)_+/\hat\lambda_k, \qquad 
\hat\tau_k^2 = (p/n)(1 - \hat\rho_k^2) / \{(1-p/n) \hat\lambda_k\},
\end{align*}
where $(x)_+ = \max(x, 0)$. 
We truncate $\hat\rho_k^2$ at zero because it can take negative values, particularly for weak directions when $\gamma_n$ is large. 
The rescaled predictor $\hat\beta_k^T(x_0-\bar x)/\hat\kappa_k$ is asymptotically conditionally
unbiased for $\beta_k^T(x_0-\mu)$, 
whereas $\hat\kappa_k\hat\beta_k^T(x_0-\bar x)/(\hat\kappa_k^2 + \hat\tau_k^2)$ 
is the best linear predictor of $\beta_k^T (x_0 - \mu)$.
Neither estimator is consistent when $\gamma > 0$.

\section{Simulation Studies}
\label{sec:simulation}
This section assesses the finite-sample accuracy of the theoretical results.
Section~\ref{subsec:sim-test} studies the sequential tests for the structural dimension,
Section~\ref{subsec:sim-values} the SIR eigenvalues, and
Section~\ref{subsec:sim-vectors} the SIR directions.
The R code that reproduces all results is available in the supplementary material.

We generate $\tilde X \sim \cN_p(0, I_p)$ and $\epsilon\sim\cN(0, 1)$ independent of $\tilde X$, and consider three models:
\begin{align*}
  &\text{Model 0:} \quad \tilde Y = \epsilon, && d_0 = 0, \\
  &\text{Model 1:} \quad \tilde Y = \tilde X_1 + 0.5\epsilon, && d_0 = 1, \\
  &\text{Model 2:} \quad \tilde Y = \frac{\tilde X_1}{0.5 + (\tilde X_2 + 1.5)^2} + 0.2\epsilon, && d_0 = 2.
\end{align*}
Model 2 is the rational model in equation (6.3) of \cite{li1991sliced} with noise level 0.2. Since SIR is affine invariant, the choice $\mu = 0$ and $\Sigma = I_p$ is made without loss of generality.
Slices have approximately equal counts and are formed from the ranks of the responses, as described in Section~\ref{subsec:pop}.
The eigenvalues and directions of SIR are computed through Proposition~\ref{prop:cca}.
The population quantities $\rho_k^2$ and $v_k$ are computed from the eigen-decomposition of $V_S$, using a sample of size $4\times 10^6$ from the joint distribution of $(\tilde Z_S, \tilde Y)$.
For Model 1, $\rho_1^2 = 0.716$ for $H = 5$ and $\rho_1^2 = 0.769$ for $H = 10$.
For Model 2, $(\rho_1^2, \rho_2^2) = (0.572, 0.241)$ for $H = 5$ and $(0.596, 0.309)$ for $H = 10$.
Thus, the second direction in Model 2 is much weaker than the first.
The nominal level of all tests is $\alpha = 0.05$.

\subsection{Sequential tests for the structural dimension}
\label{subsec:sim-test}

We compare four tests of $H_0: d_0 = d$:
Li's $\chi^2$-test, which rejects $H_0$ when $T_d > \chi^2_{f_d, \alpha}$;
the corrected test based on Theorem~\ref{thm:test};
the exact test based on Theorem~\ref{thm:Td}; and
the largest-eigenvalue test based on Corollary~\ref{coro:lmax}.
The critical value $c_{n, \alpha}(d)$ of the exact test is computed from $10^5$ draws of $n\,\tr\{(S_1 + S_2)^{-1}S_1\}$,
where $S_1\sim\cW_{H-d-1}(p - d, I_{H-d-1})$ and $S_2\sim\cW_{H-d-1}(n - p - 1, I_{H-d-1})$ are independent,
and the critical value of the largest-eigenvalue test is computed from $2\times 10^5$ draws of $\lambda_{\max}(\cG_{H-d-1})$.
We set $n = 1000$, $H\in\{5, 10\}$ and $p = \gamma n$ with $\gamma = 0.05, 0.10, \ldots, 0.90$,
and use 2000 replications for each setting,
so the Monte Carlo standard error of an empirical size near $0.05$ is about $0.005$.

\begin{table}[t]
\centering
\caption{Null behavior of Li's statistic $T_d$ at the true dimension $d = d_0$ and the empirical sizes of the four tests based on 2000 Monte Carlo replications with $n = 1000$.
The ``MC'' columns report Monte Carlo estimates.
The values of $\bE(T_d^*)$ and the exact variance ratio $var(T_d^*)/(2f_d)$ are obtained from Theorem~\ref{thm:Td}(b).
The asymptotic size $1 - \Phi(z_{\alpha}/\sqrt{1-\gamma})$ of Li's $\chi^2$-test in Corollary~\ref{coro:regimes} is in parentheses.}
\label{tab:size}
\footnotesize\setlength{\tabcolsep}{3.5pt}
\begin{tabular}{rrrr rr rrr cccc}
\toprule
& & & & \multicolumn{2}{c}{mean of $T_d$} & \multicolumn{3}{c}{$var(T_d)/(2f_d)$} & \multicolumn{4}{c}{empirical size}\\
\cmidrule(lr){5-6}\cmidrule(lr){7-9}\cmidrule(lr){10-13}
Model & $H$ & $\gamma$ & $f_d$ & MC & $\bE(T_d^*)$ & MC & exact & $1-\gamma$ & $\chi^2$ & corrected & exact & largest\\
\midrule
0 & 5 & 0.10 & 400 & 400.1 & 400.4 & 0.901 & 0.897 & 0.90 & 0.042 (0.041) & 0.048 & 0.047 & 0.082\\
 &  & 0.25 & 1000 & 1000.3 & 1001.0 & 0.806 & 0.747 & 0.75 & 0.037 (0.029) & 0.062 & 0.056 & 0.060\\
 &  & 0.50 & 2000 & 2000.9 & 2002.0 & 0.519 & 0.498 & 0.50 & 0.010 (0.010) & 0.055 & 0.051 & 0.045\\
 &  & 0.75 & 3000 & 3004.4 & 3003.0 & 0.237 & 0.249 & 0.25 & 0.000 (0.001) & 0.053 & 0.044 & 0.034\\
 & 10 & 0.10 & 900 & 901.5 & 900.9 & 0.867 & 0.893 & 0.90 & 0.045 (0.041) & 0.056 & 0.052 & 0.146\\
 &  & 0.25 & 2250 & 2253.2 & 2252.3 & 0.722 & 0.744 & 0.75 & 0.032 (0.029) & 0.057 & 0.055 & 0.089\\
 &  & 0.50 & 4500 & 4504.9 & 4504.5 & 0.491 & 0.495 & 0.50 & 0.010 (0.010) & 0.064 & 0.057 & 0.054\\
 &  & 0.75 & 6750 & 6757.3 & 6756.8 & 0.252 & 0.247 & 0.25 & 0.000 (0.001) & 0.064 & 0.056 & 0.013\\
\midrule
1 & 5 & 0.10 & 297 & 297.0 & 297.6 & 0.912 & 0.901 & 0.90 & 0.038 (0.041) & 0.053 & 0.045 & 0.066\\
 &  & 0.25 & 747 & 749.3 & 748.5 & 0.745 & 0.750 & 0.75 & 0.031 (0.029) & 0.060 & 0.051 & 0.066\\
 &  & 0.50 & 1497 & 1498.0 & 1500.0 & 0.507 & 0.500 & 0.50 & 0.008 (0.010) & 0.061 & 0.049 & 0.048\\
 &  & 0.75 & 2247 & 2251.4 & 2251.5 & 0.243 & 0.249 & 0.25 & 0.001 (0.001) & 0.053 & 0.042 & 0.040\\
 & 10 & 0.10 & 792 & 793.4 & 793.6 & 0.890 & 0.896 & 0.90 & 0.046 (0.041) & 0.057 & 0.050 & 0.138\\
 &  & 0.25 & 1992 & 1996.0 & 1996.0 & 0.769 & 0.747 & 0.75 & 0.034 (0.029) & 0.057 & 0.051 & 0.091\\
 &  & 0.50 & 3992 & 3997.2 & 4000.0 & 0.539 & 0.497 & 0.50 & 0.018 (0.010) & 0.060 & 0.048 & 0.051\\
 &  & 0.75 & 5992 & 6001.9 & 6004.0 & 0.244 & 0.248 & 0.25 & 0.000 (0.001) & 0.072 & 0.048 & 0.018\\
\midrule
2 & 5 & 0.10 & 196 & 195.5 & 196.6 & 0.891 & 0.904 & 0.90 & 0.037 (0.041) & 0.050 & 0.043 & 0.064\\
 &  & 0.25 & 496 & 497.0 & 497.5 & 0.776 & 0.754 & 0.75 & 0.029 (0.029) & 0.056 & 0.049 & 0.058\\
 &  & 0.50 & 996 & 994.8 & 999.0 & 0.497 & 0.502 & 0.50 & 0.007 (0.010) & 0.045 & 0.040 & 0.040\\
 &  & 0.75 & 1496 & 1494.4 & 1500.5 & 0.242 & 0.250 & 0.25 & 0.000 (0.001) & 0.044 & 0.030 & 0.021\\
 & 10 & 0.10 & 686 & 686.5 & 688.1 & 0.863 & 0.900 & 0.90 & 0.038 (0.041) & 0.056 & 0.045 & 0.101\\
 &  & 0.25 & 1736 & 1735.9 & 1741.2 & 0.727 & 0.750 & 0.75 & 0.025 (0.029) & 0.056 & 0.041 & 0.071\\
 &  & 0.50 & 3486 & 3487.1 & 3496.5 & 0.470 & 0.499 & 0.50 & 0.009 (0.010) & 0.049 & 0.032 & 0.040\\
 &  & 0.75 & 5236 & 5233.6 & 5251.8 & 0.256 & 0.249 & 0.25 & 0.000 (0.001) & 0.046 & 0.022 & 0.009\\
\bottomrule
\end{tabular}
\end{table}

\begin{figure}[t]
\centering
\includegraphics[width=\textwidth]{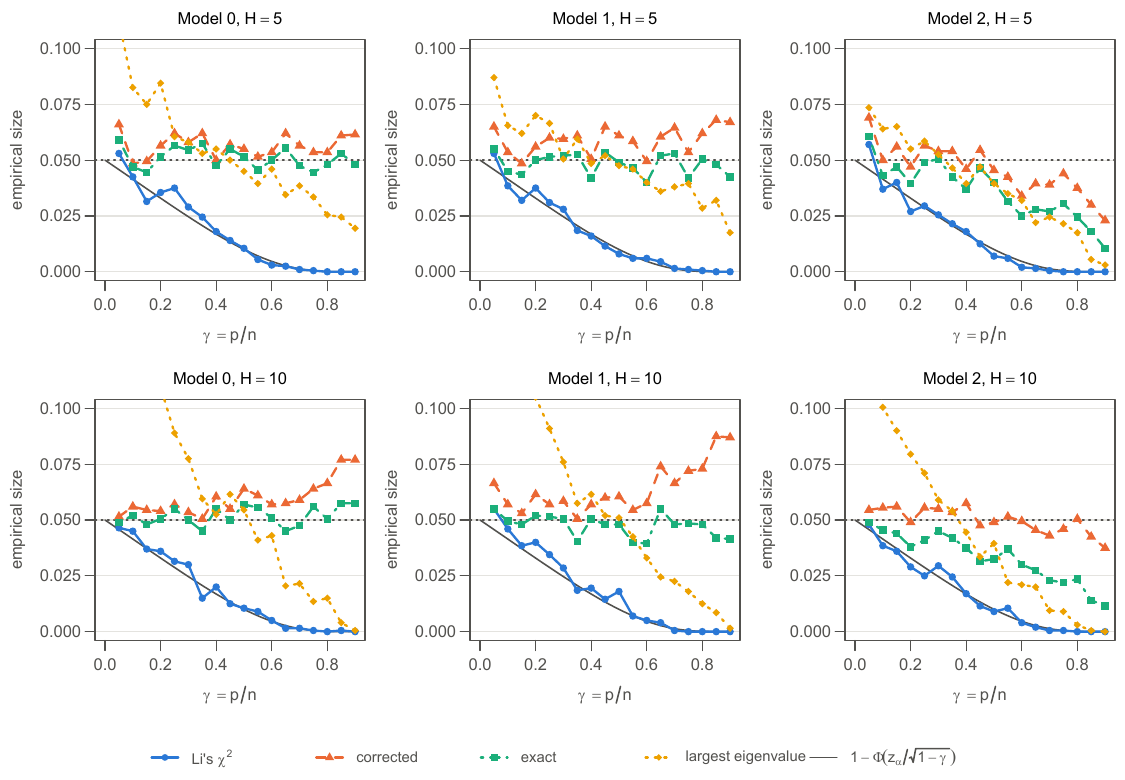}
\caption{Empirical sizes of the four tests at the true dimension against $\gamma = p/n$ based on 2000 Monte Carlo replications with $n = 1000$. The solid gray curve is the asymptotic size $1 - \Phi(z_{\alpha}/\sqrt{1-\gamma})$ of Li's $\chi^2$-test, and the dotted line marks the nominal level $0.05$.}
\label{fig:size}
\end{figure}

Table~\ref{tab:size} and Figure~\ref{fig:size} summarize the null behavior of $T_d$ at the true dimension.
The Monte Carlo mean of $T_d$ agrees with $\bE(T_d^*)$ to within $0.6\%$ in all settings.
For $d_0 > 0$ it lies slightly below $\bE(T_d^*)$, consistent with $T_d\leq T_d^*$ in Theorem~\ref{thm:Td}(a) up to Monte Carlo error.
The ratio $var(T_d)/(2f_d)$ is close to its exact value and to its limit $1-\gamma$,
confirming that the $\chi^2_{f_d}$ approximation of \cite{li1991sliced} overstates the variance of $T_d$ by the factor $1/(1-\gamma)$.
As a consequence, the empirical size of Li's $\chi^2$-test follows the curve $1 - \Phi(z_{\alpha}/\sqrt{1-\gamma})$ of Corollary~\ref{coro:regimes}(b) closely,
decreasing from about $0.05$ at $\gamma = 0.05$ to $0.010$ at $\gamma = 0.5$ and to essentially zero for $\gamma\geq 0.7$.

The exact and corrected tests remove this conservativeness.
When $d_0 = 0$, $T_0 = T_0^*$, and the exact test has sizes between $0.044$ and $0.059$, which is within Monte Carlo error of $0.05$.
When $d_0 > 0$, the exact test is conservative because $T_d \leq T_d^*$,
and it becomes more conservative as $\gamma$ increases in Model 2,
where the gap $T_d^* - T_d$ is larger because the second signal eigenvalue is small;
for example, its size is $0.022$ for Model 2 with $H = 10$ and $\gamma = 0.75$.
The corrected test has sizes between $0.044$ and $0.072$ in Table~\ref{tab:size}.
It is slightly liberal in Models 0 and 1, with average sizes of $0.057$--$0.064$ over the $\gamma$ grid, and the excess grows for $\gamma$ close to one.
This is because the statistic is centered at $f_d$ rather than at $\bE(T_d^*) = nf_d/(n-d-1)$;
the resulting shift, relative to the standard deviation, is of order $\{\gamma(H-1)/(n(1-\gamma))\}^{1/2}$,
which is asymptotically negligible but grows as $\gamma\to 1$.
The largest-eigenvalue test relies on the GOE approximation of Theorem~\ref{thm:values2}(a), which requires $p$ to be large relative to $H$.
It is liberal for small $p$,
with sizes of $0.10$--$0.15$ when $p = 100$ and $0.15$--$0.23$ when $p = 50$ for $H = 10$,
and conservative for $\gamma$ close to one.
For $H = 5$ and $0.3\leq\gamma\leq 0.6$, its sizes are between $0.032$ and $0.060$.
Overall, the exact test is the most reliable of the four.
Its largest empirical size over all 108 settings is $0.061$, which is consistent with the guarantee $\bP(T_d > c_{n,\alpha}(d))\leq\alpha$ in Theorem~\ref{thm:Td}(c) up to Monte Carlo error.

\begin{table}[t]
\centering
\caption{Distribution of the estimated structural dimension $\hat d_0$ from the sequential testing procedure based on each of the four tests: proportions of $\hat d_0 < d_0$, $\hat d_0 = d_0$ and $\hat d_0 > d_0$. Results are based on 2000 Monte Carlo replications with $n = 1000$.}
\label{tab:dhat}
\footnotesize\setlength{\tabcolsep}{3pt}
\begin{tabular}{rrr ccc ccc ccc ccc}
\toprule
& & & \multicolumn{3}{c}{Li's $\chi^2$} & \multicolumn{3}{c}{corrected} & \multicolumn{3}{c}{exact} & \multicolumn{3}{c}{largest}\\
\cmidrule(lr){4-6}\cmidrule(lr){7-9}\cmidrule(lr){10-12}\cmidrule(lr){13-15}
Model & $H$ & $\gamma$ & $<$ & $=$ & $>$ & $<$ & $=$ & $>$ & $<$ & $=$ & $>$ & $<$ & $=$ & $>$\\
\midrule
1 & 5 & 0.10 & 0.000 & 0.962 & 0.038 & 0.000 & 0.947 & 0.053 & 0.000 & 0.955 & 0.045 & 0.000 & 0.934 & 0.066\\
 &  & 0.25 & 0.000 & 0.969 & 0.031 & 0.000 & 0.940 & 0.060 & 0.000 & 0.949 & 0.051 & 0.000 & 0.933 & 0.066\\
 &  & 0.50 & 0.000 & 0.992 & 0.008 & 0.000 & 0.939 & 0.061 & 0.000 & 0.951 & 0.049 & 0.000 & 0.953 & 0.047\\
 &  & 0.75 & 0.059 & 0.940 & 0.001 & 0.000 & 0.946 & 0.054 & 0.001 & 0.957 & 0.042 & 0.000 & 0.961 & 0.039\\
 & 10 & 0.10 & 0.000 & 0.954 & 0.046 & 0.000 & 0.943 & 0.057 & 0.000 & 0.951 & 0.049 & 0.000 & 0.862 & 0.138\\
 &  & 0.25 & 0.000 & 0.966 & 0.034 & 0.000 & 0.943 & 0.057 & 0.000 & 0.949 & 0.051 & 0.000 & 0.909 & 0.091\\
 &  & 0.50 & 0.000 & 0.982 & 0.018 & 0.000 & 0.940 & 0.060 & 0.000 & 0.952 & 0.048 & 0.000 & 0.949 & 0.051\\
 &  & 0.75 & 0.476 & 0.524 & 0.000 & 0.035 & 0.892 & 0.072 & 0.044 & 0.907 & 0.048 & 0.000 & 0.982 & 0.018\\
\midrule
2 & 5 & 0.10 & 0.000 & 0.963 & 0.037 & 0.000 & 0.950 & 0.050 & 0.000 & 0.957 & 0.043 & 0.000 & 0.936 & 0.064\\
 &  & 0.25 & 0.001 & 0.970 & 0.029 & 0.000 & 0.944 & 0.056 & 0.000 & 0.951 & 0.049 & 0.000 & 0.942 & 0.058\\
 &  & 0.50 & 0.208 & 0.784 & 0.007 & 0.066 & 0.888 & 0.045 & 0.080 & 0.880 & 0.040 & 0.002 & 0.959 & 0.039\\
 &  & 0.75 & 0.935 & 0.064 & 0.000 & 0.404 & 0.552 & 0.044 & 0.447 & 0.523 & 0.030 & 0.237 & 0.742 & 0.021\\
 & 10 & 0.10 & 0.000 & 0.962 & 0.038 & 0.000 & 0.945 & 0.055 & 0.000 & 0.955 & 0.045 & 0.000 & 0.899 & 0.101\\
 &  & 0.25 & 0.009 & 0.966 & 0.025 & 0.006 & 0.938 & 0.055 & 0.006 & 0.953 & 0.041 & 0.000 & 0.929 & 0.071\\
 &  & 0.50 & 0.435 & 0.556 & 0.009 & 0.180 & 0.771 & 0.049 & 0.218 & 0.750 & 0.032 & 0.000 & 0.961 & 0.039\\
 &  & 0.75 & 0.966 & 0.034 & 0.000 & 0.559 & 0.395 & 0.046 & 0.642 & 0.336 & 0.022 & 0.330 & 0.661 & 0.008\\
\bottomrule
\end{tabular}
\end{table}

\begin{figure}[t]
\centering
\includegraphics[width=\textwidth]{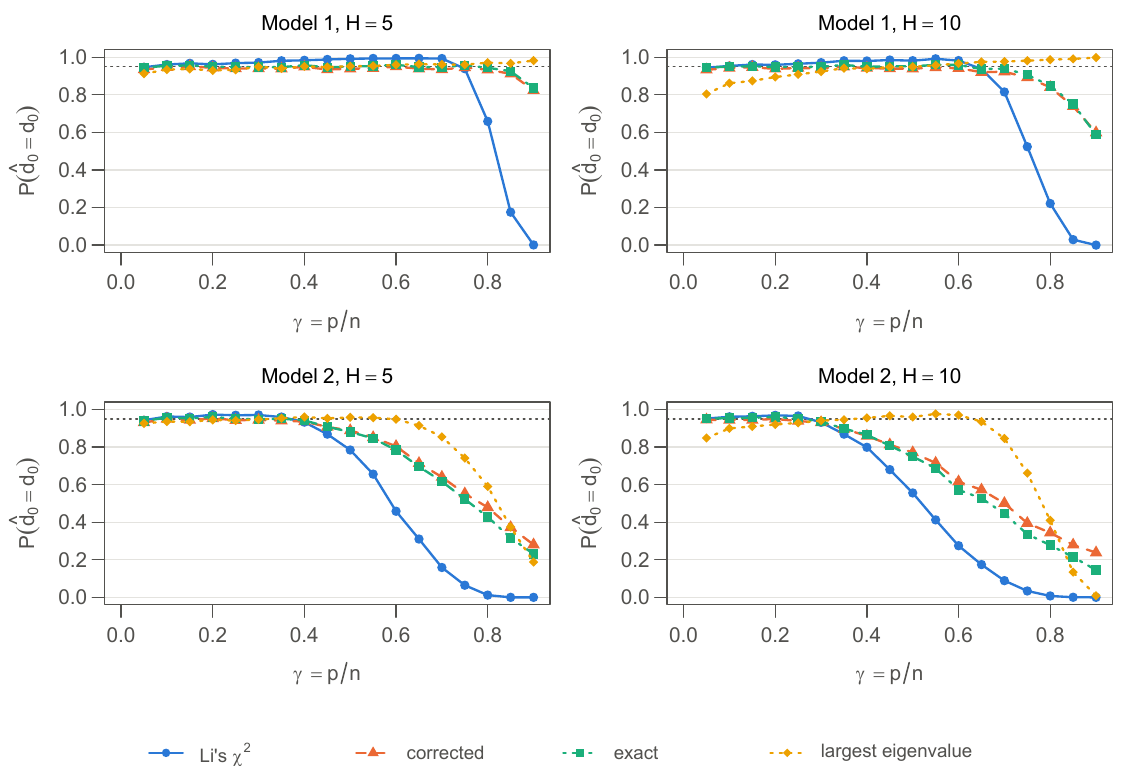}
\caption{Proportion of replications with $\hat d_0 = d_0$ against $\gamma = p/n$ for the sequential testing procedure based on each of the four tests. The results are based on 2000 Monte Carlo replications with $n = 1000$. The dotted line marks $1 - \alpha = 0.95$.}
\label{fig:dhat}
\end{figure}

Table~\ref{tab:dhat} and Figure~\ref{fig:dhat} report the distribution of $\hat d_0$.
In Model 1, the sequential procedures based on the corrected and exact tests select $\hat d_0 = d_0$ in about $95\%$ of the replications for $\gamma\leq 0.75$ when $H = 5$ and for $\gamma\leq 0.6$ when $H = 10$,
and overestimate $d_0$ at about the nominal rate, as Corollary~\ref{coro:test} predicts.
For larger $\gamma$, the proportion of $\hat d_0 = d_0$ decreases because of underestimation; for example, it is $0.89$ and $0.91$ for $H = 10$ and $\gamma = 0.75$.
Li's procedure selects $\hat d_0 = d_0$ more often for moderate $\gamma$, but only because its size is far below $\alpha$.
This conservativeness makes it underestimate $d_0$ for larger $\gamma$:
in Model 1 with $H = 10$ and $\gamma = 0.75$, Li's procedure selects $\hat d_0 = 0$ in $47.6\%$ of the replications,
compared with $3.5\%$ and $4.4\%$ for the corrected and exact procedures.

In Model 2, all procedures underestimate $d_0$ for large $\gamma$, because the second direction is weak and $n = 1000$ is moderate.
Theorem~\ref{thm:test}(b) guarantees that underestimation vanishes asymptotically,
but the finite-sample power at the step $d' = d_0 - 1$ is governed by the ratio of the signal
$n(1-\gamma)\rho_{d_0}^2$ in $T_{d'}$ to its null standard deviation $\{2(p-d')(H-d'-1)(1-\gamma)\}^{1/2}$.
For $H = 10$ and $\gamma = 0.5$, this ratio is about $2.5$, compared with $z_\alpha = 1.64$,
so the corrected test rejects $H_0: d_0 = 1$ in only about $80\%$ of the replications.
Li's procedure is again the worst, with $43.5\%$ underestimation in this setting, compared with $18.0\%$ and $21.8\%$ for the corrected and exact procedures.
The largest-eigenvalue test is much more powerful here:
it underestimates $d_0$ in none of the replications for $H = 10$ and $\gamma = 0.5$.
The reason is that the signal is concentrated in a single eigenvalue,
whereas $T_{d'}$ accumulates the fluctuations of all $H - d' - 1$ null eigenvalues;
the signal-to-noise ratio of $L_{d'}$ is larger than that of $T_{d'}$ by a factor of about $\{2(H - d' - 1)\}^{1/2}$.
The price is the size distortion of the largest-eigenvalue test for small $p$ discussed above.
In practice, we recommend the exact test as the default,
and the largest-eigenvalue test as a complement when $p$ is large and the signal is weak.

\subsection{SIR eigenvalues}
\label{subsec:sim-values}

To assess the results in Section~\ref{sec:values}, we consider six settings:
Model 1 with $n = 1000$, $H = 5$ and $p\in\{250, 500, 750\}$;
Model 2 with $n = 1000$, $H = 10$ and $p\in\{250, 500\}$; and
Model 2 with $n = 2000$, $H = 5$ and $p = 1000$.
Each setting uses 1000 replications.
The oracle eigenvalues $\ta_k$ are computed by applying SIR to the sufficient predictors $B^Tx_i$, using the same slices.

\begin{table}[t]
\centering
\caption{Signal eigenvalues and the largest null eigenvalue; 1000 replications.
The columns $\hat\lambda_k$, $\ta_k$ and $\hat\rho_k^2$ are Monte Carlo means, and ``corr'' is the correlation between $\hat\rho_k^2 - \ta_k$ and $\ta_k$.
Theoretical values are in parentheses:
$\gamma_n + (1-\gamma_n)\rho_k^2$ for $\hat\lambda_k$ (Theorem~\ref{thm:values1});
$2\gamma(1-\rho_k^2)^2/(1-\gamma)$ for $n\,var(\hat\rho_k^2 - \ta_k)$ (Corollary~\ref{coro:values});
and the mean, standard deviation and $95\%$ quantile of $\lambda_{\max}(\cG_t)$, $t = H - d_0 - 1$, for $L_d$ at $d = d_0$ (Corollary~\ref{coro:lmax}).
The last column is the empirical size of the largest-eigenvalue test.}
\label{tab:eigen}
\footnotesize\setlength{\tabcolsep}{3pt}
\begin{tabular}{rrrrr r c c c c c c}
\toprule
Model & $n$ & $p$ & $H$ & $\gamma$ & $k$ & $\rho_k^2$ & $\hat\lambda_k$ & $\ta_k$ & $\hat\rho_k^2$ & $n\,var(\hat\rho_k^2-\ta_k)$ & corr\\
\midrule
1 & 1000 & 250 & 5 & 0.25 & 1 & 0.716 & 0.789 (0.787) & 0.718 & 0.719 & 0.052 (0.054) & $\phantom{-}0.034$\\
1 & 1000 & 500 & 5 & 0.50 & 1 & 0.716 & 0.860 (0.858) & 0.719 & 0.720 & 0.161 (0.161) & $-0.029$\\
1 & 1000 & 750 & 5 & 0.75 & 1 & 0.716 & 0.931 (0.929) & 0.718 & 0.722 & 0.479 (0.483) & $\phantom{-}0.024$\\
2 & 1000 & 250 & 10 & 0.25 & 1 & 0.596 & 0.702 (0.696) & 0.601 & 0.603 & 0.109 (0.109) & $-0.021$\\
 &  &  &  &  & 2 & 0.309 & 0.486 (0.481) & 0.311 & 0.316 & 0.305 (0.319) & $-0.004$\\
2 & 1000 & 500 & 10 & 0.50 & 1 & 0.596 & 0.803 (0.798) & 0.602 & 0.607 & 0.306 (0.327) & $\phantom{-}0.006$\\
 &  &  &  &  & 2 & 0.309 & 0.664 (0.654) & 0.312 & 0.328 & 0.883 (0.956) & $-0.044$\\
2 & 2000 & 1000 & 5 & 0.50 & 1 & 0.572 & 0.787 (0.786) & 0.573 & 0.574 & 0.384 (0.366) & $\phantom{-}0.003$\\
 &  &  &  &  & 2 & 0.241 & 0.621 (0.621) & 0.240 & 0.243 & 1.188 (1.151) & $-0.042$\\
\midrule
& & & & & $t$ & & mean $L_d$ & sd $L_d$ & $q_{0.95}(L_d)$ & size & \\
\midrule
1 & 1000 & 250 & 5 & 0.25 & 3 &  & 2.09 (2.07) & 1.16 (1.09) & 4.04 (3.92) & 0.060 & \\
1 & 1000 & 500 & 5 & 0.50 & 3 &  & 2.06 (2.07) & 1.07 (1.09) & 3.80 (3.92) & 0.041 & \\
1 & 1000 & 750 & 5 & 0.75 & 3 &  & 2.03 (2.07) & 1.03 (1.09) & 3.78 (3.92) & 0.042 & \\
2 & 1000 & 250 & 10 & 0.25 & 7 &  & 4.27 (4.19) & 1.02 (0.93) & 6.04 (5.78) & 0.078 & \\
2 & 1000 & 500 & 10 & 0.50 & 7 &  & 4.03 (4.19) & 0.89 (0.93) & 5.61 (5.78) & 0.029 & \\
2 & 2000 & 1000 & 5 & 0.50 & 2 &  & 1.15 (1.25) & 1.20 (1.19) & 3.22 (3.25) & 0.048 & \\
\bottomrule
\end{tabular}
\end{table}

Table~\ref{tab:eigen} confirms Theorem~\ref{thm:values1}:
the signal eigenvalues $\hat\lambda_k$ are biased upward by $\gamma(1-\rho_k^2)$, and their means agree with $\gamma_n + (1-\gamma_n)\rho_k^2$ within $0.01$.
The bias is substantial.
For example, in Model 2 with $p = 1000$ and $n = 2000$, the second eigenvalue $\hat\lambda_2 \approx 0.62$ is more than twice $\rho_2^2 = 0.241$.
The bias-corrected eigenvalues $\hat\rho_k^2$ are centered at $\rho_k^2$, as stated in Corollary~\ref{coro:values}(a).
The largest deviation is $0.019$, which occurs for the weak direction in Model 2 with $\gamma = 0.5$ and $H = 10$.
The variance of $\sqrt n(\hat\rho_k^2 - \ta_k)$ agrees with $2\gamma(1-\rho_k^2)^2/(1-\gamma)$ to within about $8\%$.
The correlation between $\hat\rho_k^2 - \ta_k$ and $\ta_k$ is negligible, as the independence of the limit from $\cF$ in Corollary~\ref{coro:values}(b) predicts.
Consequently, the variance of $\hat\rho_k^2$ is approximately the sum of the oracle variance and the additional variance due to the noise predictors, as in Corollary~\ref{coro:values}(c).
For example, in Model 1 with $\gamma = 0.75$, $n\,var(\hat\rho_1^2) = 0.698$, whereas $n\,var(\ta_1) + n\,var(\hat\rho_1^2 - \ta_1) = 0.204 + 0.479 = 0.683$.
The additional variance is comparable to the oracle variance at $\gamma = 0.5$ and more than twice as large at $\gamma = 0.75$.

The lower part of Table~\ref{tab:eigen} concerns the largest null eigenvalue.
The mean, standard deviation and $95\%$ quantile of the standardized statistic $L_d$ are close to those of $\lambda_{\max}(\cG_{H-d-1})$,
as Theorem~\ref{thm:values2}(a) and Corollary~\ref{coro:lmax} predict,
and the largest-eigenvalue test has empirical sizes between $0.029$ and $0.078$.

\subsection{SIR directions}
\label{subsec:sim-vectors}

\begin{table}[t]
\centering
\caption{Alignment, accuracy and out-of-sample behavior of the SIR directions; 1000 replications; settings as in Table~\ref{tab:eigen}.
Monte Carlo means are reported, with the limits in parentheses:
$\kappa_k$ for $\hat\beta_k^T\Sigma\beta_k$ (Theorem~\ref{thm:vectors-align});
$\kappa_k^2 + \tau_k^2$ for $\hat\beta_k^T\Sigma\hat\beta_k$ (Theorem~\ref{thm:vectors-align});
$(1-\kappa_k)^2 + \tau_k^2$ for the prediction error $\text{PE}_k$ (Theorem~\ref{thm:pred}(b));
and $\varphi_{\gamma}(\rho_k^2)$ and $d^{-1}\sum_k\varphi_{\gamma}(\rho_k^2)$ for $R^2(\hat\beta_k)$ and $R^2(\hat\beta_{1:d})$ (Theorem~\ref{thm:accuracy}).
``Plug-in'' is the mean of the estimate $\varphi_{p/n}(\hat\rho_k^2)$ of $\varphi_\gamma(\rho_k^2)$, and the last column is Li's approximation~\eqref{eq:li51}.}
\label{tab:directions}
\footnotesize\setlength{\tabcolsep}{3pt}
\begin{tabular}{rrr r c ccc cc cc}
\toprule
Model & $p$ & $\gamma$ & $k$ & $\rho_k^2$ & $\hat\beta_k^T\Sigma\beta_k$ & $\hat\beta_k^T\Sigma\hat\beta_k$ & PE$_k$ & $R^2(\hat\beta_k)$ & plug-in & $R^2(\hat\beta_{1:d})$ & \eqref{eq:li51}\\
\midrule
1 & 250 & 0.25 & 1 & 0.716 & .954 (.954) & 1.03 (1.03) & .124 (.122) & .883 (.883) & .885 & .883 (.883) & .901\\
1 & 500 & 0.50 & 1 & 0.716 & .916 (.914) & 1.17 (1.17) & .341 (.338) & .716 (.716) & .720 & .716 (.716) & .802\\
1 & 750 & 0.75 & 1 & 0.716 & .879 (.878) & 1.71 (1.69) & .956 (.931) & .452 (.457) & .465 & .452 (.457) & .703\\
2 & 250 & 0.25 & 1 & 0.596 & .923 (.925) & 1.05 (1.05) & .206 (.199) & .814 (.816) & .820 & .688 (.694) & .638\\
 &  &  & 2 & 0.309 & .793 (.801) & 1.13 (1.12) & .545 (.518) & .562 (.572) & .580 &  & \\
2 & 500 & 0.50 & 1 & 0.596 & .859 (.864) & 1.26 (1.25) & .545 (.525) & .589 (.596) & .607 & .439 (.452) & .273\\
 &  &  & 2 & 0.309 & .663 (.687) & 1.56 (1.53) & 1.235 (1.155) & .288 (.309) & .328 &  & \\
2 & 1000 & 0.50 & 1 & 0.572 & .852 (.853) & 1.27 (1.27) & .571 (.566) & .571 (.572) & .574 & .404 (.407) & .029\\
 &  &  & 2 & 0.241 & .616 (.624) & 1.62 (1.61) & 1.393 (1.364) & .236 (.241) & .243 &  & \\
\bottomrule
\end{tabular}
\end{table}

\begin{figure}[t]
\centering
\includegraphics[width=\textwidth]{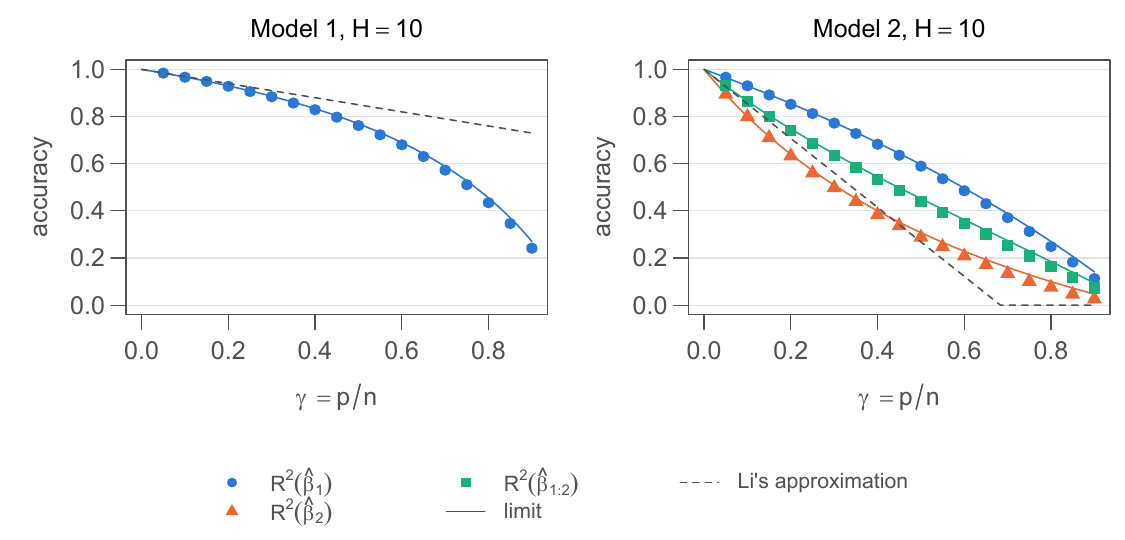}
\caption{Accuracy of the SIR directions against $\gamma = p/n$; $n = 1000$, $H = 10$, 500 replications. Points are Monte Carlo means of $R^2(\hat\beta_k)$ and $R^2(\hat\beta_{1:d})$, solid curves are their limits in Theorem~\ref{thm:accuracy}, and the dashed curve is Li's approximation~\eqref{eq:li51} for $R^2(\hat\beta_{1:d})$, truncated at zero.}
\label{fig:directions}
\end{figure}

Table~\ref{tab:directions} uses the same replications as Table~\ref{tab:eigen},
and Figure~\ref{fig:directions} reports $R^2(\hat\beta_k)$ and $R^2(\hat\beta_{1:d})$ for $n = 1000$, $H = 10$ and $\gamma = 0.05, 0.10, \ldots, 0.90$, with 500 replications for each $\gamma$.
For a new observation $x_0$ independent of the sample, the prediction error
\begin{align*}
  \text{PE}_k = \bE[\{\hat\beta_k^T(x_0 - \bar x) - \beta_k^T(x_0 - \mu)\}^2 \mid \text{sample}]
  = \|\hat\beta_k - \beta_k\|^2 + (\hat\beta_k^T\bar x)^2
\end{align*}
is computed exactly, without generating new data.

The diagonal entries of $\hatsK = \hat\beta_{1:d}^T\Sigma\beta_{1:d}$ agree with $\kappa_k$ to within $0.025$.
The off-diagonal entries are small; for Model 2, the average of the largest absolute off-diagonal entry is $0.05$--$0.07$,
so $\hat\beta_k$ is nearly $\Sigma$-orthogonal to $\beta_j$ for $j\neq k$, as stated in Theorem~\ref{thm:vectors-align}.
Although $\hat\beta_k^T\hat\Sigma\hat\beta_k = 1$ by construction, $\hat\beta_k^T\Sigma\hat\beta_k$ is close to $\kappa_k^2 + \tau_k^2 > 1$ and reaches $1.71$ in Model 1 with $\gamma = 0.75$.
Thus, the estimated sufficient predictor has a larger variance on new data than in the training sample.
The prediction error $\text{PE}_k$ agrees with $(1-\kappa_k)^2 + \tau_k^2$.
It exceeds $1$, the variance of $\beta_k^T(x_0 - \mu)$ itself, for the weak direction in Model 2 with $\gamma = 0.5$.

The accuracy criteria $R^2(\hat\beta_k)$ and $R^2(\hat\beta_{1:d})$ agree with the limits in Theorem~\ref{thm:accuracy} to within about $0.02$ in Table~\ref{tab:directions},
and Figure~\ref{fig:directions} shows that the agreement holds over the whole range of $\gamma$.
The accuracy decreases steadily as $\gamma$ increases, and it is close to $\rho_k^2$ at $\gamma = 0.5$.
The Monte Carlo means are slightly below the limits, by at most $0.03$ over the grid in Figure~\ref{fig:directions}.
The difference is largest for $\gamma$ close to one, and for the weak direction in Model 2.
The plug-in estimate $\varphi_{p/n}(\hat\rho_k^2)$ is close to the true accuracy, so the accuracy of SIR can be assessed from the data.
In contrast, Li's approximation \eqref{eq:li51} is accurate only for small $\gamma$.
It overstates the accuracy in Model 1, where $\rho_1^2 > 1/2$,
and in Model 2 it can be far too pessimistic:
with $n = 2000$, $p = 1000$ and $H = 5$, it gives $0.03$, whereas the actual value of $R^2(\hat\beta_{1:2})$ is $0.40$.
For $\gamma \geq 0.69$ in Model 2 with $H = 10$, it is even negative.

\section{Discussion and Conclusion}
\label{sec:conclusion}

We revisit the theoretical properties of SIR in the high-dimensional regime 
where $p/n\to\gamma \in [0, 1)$, assuming normally distributed predictors and a fixed number of slices. 
A key step is the exact representation established in Theorem~\ref{thm:represent}, which shows that the nonzero SIR eigenvalues coincide with the eigenvalues of $M = \Lambda + D\bB D$.
Here, $\Lambda$ and $D$ depend only on the oracle SIR eigenvalues, 
while $\bB$ is independent of the response and the sufficient predictors. 

The main findings are as follows:  
(i) The sample SIR eigenvalues exhibit an upward bias when $\gamma > 0$.
Their fluctuations are governed by the GOE, and 
the signal eigenvalues are asymptotically normal after an appropriate bias correction. 
(ii) Li's chi-squared test requires a scaling adjustment to achieve the nominal asymptotic size when $\gamma > 0$.
We further propose two additional tests that achieve the nominal asymptotic size:
an exact test based on Li's test statistic $T_d$ and a largest-eigenvalue test. 
(iii) We derive closed-form limits for the accuracy of the estimated SIR directions
and show that their first-order expansions in $\gamma$ agree with Li's approximation. 
(iv) We establish the limiting distribution of the estimated sufficient predictors for a new observation
and show that, when $\gamma > 0$, they contain an additional independent Gaussian noise component. 

The results in this paper are derived under the assumption that the predictors are normally distributed. 
However, \cite{bai2018limiting} showed that the distributions of MANOVA eigenvalues and invariant test statistics are asymptotically robust to departures from normality, suggesting that a similar phenomenon may hold for SIR.
Moreover, results from random matrix theory suggest that some of our asymptotic results may extend beyond the Gaussian setting. In particular, many results for high-dimensional sample covariance matrices hold under weaker distributional assumptions, such as independent components with finite fourth moments. 
However, extending our theory to non-Gaussian predictors requires additional consideration of the linearity condition, which is essential for the validity of SIR. 
In particular, independence of the predictor components alone does not guarantee this condition. 
Identifying suitable non-Gaussian distributions under which both the random matrix results and the linearity condition hold is an interesting direction for future research.

We focus on the behavior of Li's sequential testing procedure under normally distributed predictors in the proportional asymptotic regime. 
Although the three tests proposed in Section~\ref{sec:test} have the same asymptotic size,
their empirical performance differs, as demonstrated in Section~\ref{subsec:sim-test}. 
Further investigation is warranted to better understand their respective properties. 
In addition, other extensions of SIR have explored nonnormal predictor distributions and alternative estimation algorithms, as reviewed in Section~\ref{sec:intro}. 
Extending these lines of research to the development of testing procedures is an interesting direction for future work.

We restrict our attention to the case $n > p$ throughout this paper. 
When $p > n$, corresponding to the asymptotic regime $p/n \to \gamma \geq 1$, 
the sample covariance matrix $\hat\Sigma$ is singular, so $\hat\Sigma^{-1/2}$ is not well-defined and all $r_i^2 = 1$ since $\col(\bar X) = 1_n^{\perp}$. 
The standard SIR algorithm therefore needs to be modified to accommodate this setting.
Addressing this regime would require additional structure, such as regularization or sparsity assumptions,  
and is an important direction for further research. 

Many other methods have been proposed in the literature to improve the performance of SIR in estimating the central subspace. Most of these methods have been studied only under the conventional asymptotic regime in which $p$ remains fixed as $n\to\infty$. 
It is therefore important to revisit these methods and investigate their theoretical properties in the proportional asymptotic regime.

\bibliographystyle{chicago} 
\bibliography{references}


\newpage 
\clearpage
\setcounter{page}{1}
\setcounter{equation}{0}
\setcounter{theorem}{0}
\setcounter{lemma}{0}
\setcounter{section}{0}
\setcounter{figure}{0}
\setcounter{table}{0}
\setcounter{fact}{0}

\renewcommand{\thepage}{S\arabic{page}}
\renewcommand{\theequation}{S\arabic{equation}}
\renewcommand{\thetheorem}{S\arabic{theorem}}
\renewcommand{\thelemma}{S\arabic{lemma}}
\renewcommand{\thesection}{S\arabic{section}}
\renewcommand{\thefact}{S\arabic{fact}}
\renewcommand{\thefigure}{S\arabic{figure}}
\renewcommand{\thetable}{S\arabic{table}}

\renewcommand{\theHequation}{S\arabic{equation}}
\renewcommand{\theHlemma}{S\arabic{lemma}}
\renewcommand{\theHsection}{S\arabic{section}}
\renewcommand{\theHfact}{S\arabic{fact}}
\renewcommand{\theHfigure}{S\arabic{figure}}
\renewcommand{\theHtable}{S\arabic{table}}

\quad 
\vspace{1em}

\begin{center}
{\LARGE Supplementary Material} \\[1em]
{\Large Sliced Inverse Regression Revisited: When the Number of Predictors Grows Proportionally With the Sample Size} \\[1em]
Peng Zeng and Hanwen Huang \\[1em]
Department of Mathematics and Statistics, Auburn University and
Department of Biostatistics, Data Science and Epidemiology, Medical College of Georgia, Augusta University
\end{center}

\vspace{1em}

This document contains the proofs for this paper.
Section~\ref{sec:facts} presents several results that will be used in the subsequent proofs. 
Section~\ref{sec:reduce} shows that, without loss of generality, we may assume $\mu = 0$, $\Sigma = I_p$ and $B = (I_d, 0)^T$ in all subsequent proofs. 
Sections~\ref{sec:proofs-sec2}--\ref{sec:proofs-sec6} provide proofs of the results in Sections~\ref{sec:review}--\ref{sec:vectors}, respectively. 

\bigskip 
To keep the proofs compact, we write
\begin{align*}
  q = H-1, \quad 
  s = p-d, \quad 
  t = H - d - 1, \quad 
  m = n - d - 1, \quad 
  \gamma_n = s/m, \quad 
  f_d = st.
\end{align*}
So 
$\bB\sim\cB_q(s/2,(m-s)/2)$ in Theorem~\ref{thm:represent}, 
$\cG_{00}$ is $t\times t$ in Theorem~\ref{thm:values2}, and
$T_d^*/n \sim \Pi(s, t, m)$ in Theorem~\ref{thm:Td}. 
Assumption~\ref{ass:null} implies $s < m$ and $m > q$.

In addition to the notation introduced in the paper, we introduce the following notation.
$E_s = (I_s, 0)^T$ denotes the matrix of the first $s$ columns of an identity matrix.
$e_s$ denotes the vector whose $s$th component is 1 and all other components are 0.
The symbol $\eqd$ denotes equality in distribution. 

\section{Auxiliary Facts}
\label{sec:facts}

This section presents several results that will be used in the subsequent proofs.

\begin{fact}[Fan's minimum principle, Corollary 4.3.39 in \cite{horn2013matrix}]\label{fact:kyfan}
For a symmetric $q\times q$ matrix $M$ and $0\le d<q$,
$\sum_{i=d+1}^q\lambda_i(M)=\min\{\tr(F^TMF):F\in\bR^{q\times(q-d)},\
F^TF=I_{q-d}\}$.
\end{fact}

\begin{fact}[Weyl's inequality, \cite{horn2013matrix}]\label{fact:weyl}
For symmetric matrices $M$ and $M'$ of the same dimension, 
(a) $|\lambda_i(M)-\lambda_i(M')|\le \|M-M'\|_{op}\le\|M-M'\|_F$ for every $i$; 
(b) if $M\preceq M'$, then $\lambda_i(M)\le\lambda_i(M')$ for every $i$.
\end{fact}

\begin{fact}[Davis-Kahan theorem, \cite{yu2015useful}]\label{fact:dk}
Let $M$ and $M'$ be symmetric matrices. 
Suppose that $\lambda_k(M')$ has eigengap
$g = \min_{j \neq k} |\lambda_{j}(M')-\lambda_k(M')| > 0$.
Let $u$ and $u'$ be unit eigenvectors corresponding to $\lambda_k(M)$ and $\lambda_k(M')$, respectively. 
If $u^Tu' \geq 0$, then $\|u - u'\|\le 2^{3/2}\|M - M'\|_{op}/g$.
\end{fact}

\begin{fact}[Theorem 2.2.1 in \cite{chikuse2003statistics}]\label{fact:unif}
A random $m\times t$ frame whose distribution is invariant under left multiplication by orthogonal
matrices is uniformly distributed on the Stiefel manifold.
This distribution is unique and coincides with that of 
$Z(Z^TZ)^{-1/2}$, 
where $Z$ is an $m\times t$ matrix with iid $\cN(0, 1)$ entries. 
\end{fact}

\begin{fact}[Extreme singular values, Theorem~II.13 in \cite{davidson2001local}]\label{fact:rmt}
Let $Z \in \bR^{m\times s}$, with $s\le m$, have iid\ $\cN(0,1)$ entries.
Then, for every $x > 0$, 
\begin{align*}
    \bP\{\sqrt{m}- \sqrt{s} - x 
    \le 
    \sigma_{\min}(Z)
    \le
    \sigma_{\max}(Z)
    \le
    \sqrt{m} + \sqrt{s} + x\}
    \ge
    1 - 2e^{-x^2/2},
\end{align*} 
where $\sigma_{\max}(Z)$ and $\sigma_{\min}(Z)$ are the largest and smallest singular values of $Z$, respectively. 
In particular, if $s/m\to\gamma\in[0,1)$, then for every $\varepsilon>0$, 
all eigenvalues of $Z^TZ/m$ lie in
$[(1-\sqrt\gamma)^2-\varepsilon,(1+\sqrt\gamma)^2+\varepsilon]$ 
with probability tending to one. 
\end{fact}

\begin{fact}[Marchenko-Pastur, \cite{bai2010spectral}]\label{fact:mp}
Let $Z \in \bR^{m\times s}$, with $s\le m$, have iid\ $\cN(0,1)$ entries. 
If $s\to\infty$ and $s/m \to \gamma \in (0,1)$, the empirical spectral distribution of $Z^TZ/m$
converges weakly in probability to the Marchenko-Pastur distribution $F_{\gamma}$, which satisfies
$\int x^{-1}dF_{\gamma}(x) = 1/(1-\gamma)$. 
\end{fact}

\section{Reduce to Standard Coordinates}
\label{sec:reduce}

All results are invariant under affine transformation $x_i \mapsto Lx_i + b$, where $L$ is invertible. The model (\ref{eq:model-cs}) remains unchanged under $B \mapsto BJ$, where $J$ is invertible.  Thus, without loss of generality, we may assume $\mu = 0$, $\Sigma = I_p$, and $B = (I_d, 0)^T$. 
We formalize this statement in the following lemma and provide its proof.

\begin{lemma}[Affine invariance]\label{lem:reduction}
Let $x_i' = Lx_i + b$ with $L$ invertible, and define $B' = L^{-T}B$ and $\Sigma' = L\Sigma L^T$. 
Under this transformation, $K$, $\hat\lambda_k$ for $k\le q$, $T_d$, $A$, and $\ta_k$ for $k\leq q$ remain unchanged.
Moreover,
$B'^Tx'_i = B^Tx_i + B^TL^{-1}b$, so the model in \eqref{eq:model-cs} and the $\sigma$-field $\cF$ remain unchanged.
The estimated and population directions transform as 
$\hat\beta_k' = L^{-T}\hat\beta_k$
and $\beta_k'=L^{-T}\beta_k$. 
Consequently, $R^2(\cdot)$, $\hat\beta_{1:d}^T\Sigma\beta_{1:d}$ and
$\hat\beta_{1:d}^T\Sigma\hat\beta_{1:d}$, 
$\hat\beta_k^T(x_0-\bar x)$ and $\beta_k^T(x_0-\mu)$, 
and $\rho_k^2$ are all invariant under the transformation. 
The same invariance properties hold when $B$ is replaced by $BJ$ for any invertible matrix $J$. 
Consequently, without loss of generality, we may assume that $\mu=0$, $\Sigma=I_p$ and $B=(I_d,0)^T$.
Under this parameterization, when the population directions are needed, 
we may further take $\beta_k = e_k$ for $k\le d$.
\end{lemma}

\begin{proof}
Since $\bar X'=\bar XL^T$, the matrices $\bar X'$ and $\bar X$ have the same column space. 
Hence, $P_{\bar X}$, $K$, and the eigenvalues of $K$ remain unchanged.
Moreover, $\bar X'B' = \bar XB$, 
so $P_{\bar X_S}$, $A$, and the eigenvalues $\ta_k$ of $A$ are also unchanged. 
By Proposition~\ref{prop:cca}(d), 
\begin{align*}
    \hat\beta_k'
    = \frac{\sqrt{n}}{r_k} (L\bar X^T\bar XL^T)^{-1}L\bar X^TF_{\bar G}u_k 
    = L^{-T}\hat\beta_k.
\end{align*} 
Consequently,
\begin{align*}
\hat\beta_k'^T\Sigma'B'=\hat\beta_k^T\Sigma B,  \quad 
B'^T\Sigma'B'= B^T\Sigma B, \\
\hat\beta_k'^T\Sigma'\hat\beta_k'=\hat\beta_k^T\Sigma\hat\beta_k, \quad  
\hat\beta_k'^T(x'_0 - \bar x') = \hat\beta_k^T(x_0 - \bar x). 
\end{align*}
The distribution of $(\tilde Z_S, \tilde Y)$ is also unchanged. 
Therefore, the squared population canonical correlations $\rho_k^2$ remain unchanged, 
and the population SIR direction $\beta_k$ transforms as stated. 

To establish the claimed parameterization, take 
\begin{align*}
   L = Q^T\Sigma^{-1/2}, \qquad 
   b = -L\mu,
\end{align*}
where $Q$ is a $p\times p$ orthogonal matrix. 
Then $\mu'=0$ and $\Sigma' = I_p$. 

Next, choose an invertible matrix $J$ such that $\Sigma^{1/2}BJ = (Ov_1,\dots,Ov_d)$, 
where $v_1, \ldots, v_d$ are orthonormal eigenvectors of $V_S$. 
Choose $Q$ so that its first $d$ columns are $\Sigma^{1/2}BJ$, with the remaining columns forming an orthogonal completion. 
Then $B'J=(I_d,0)^T$ and the corresponding population directions satisfy $\beta_k'=e_k$ for $k \leq d$.
\end{proof}

The proof of this lemma relies on Proposition~\ref{prop:cca} and its population counterpart Proposition~\ref{prop:cca-pop}, whose proofs do not use Lemma~\ref{lem:reduction}.

\bigskip 

In the remaining proofs, without loss of generality, we assume that $\mu = 0$, $\Sigma = I_p$, and $B = (I_d, 0)^T$. Under these assumptions, Assumption~\ref{ass:null} can be partially restated as follows.
\begin{itemize}
\item 
The entries of $X$ are iid $\cN(0, 1)$. 
\item 
Partition $X = (X_S, X_N)$, where $X_S \in\bR^{n\times d}$ contains the signal predictors and $X_N \in\bR^{n\times (p-d)}$ contains the noise predictors. 
\item 
The response $Y$ depends on $X$ only through $X_S$. 
\item 
$X_N$ is independent of $(Y, X_S)$. 
\end{itemize}
Additionally, 
$\tilde Z_S$ is the first $d$ coordinates of $\tilde X$ and $V=\diag(V_S,0)$.

\section{Proofs for Section~\ref{sec:review}}
\label{sec:proofs-sec2}

This section provides proofs of the results in Section~\ref{sec:review}.

\begin{proof}[Proof of Proposition~\ref{prop:cca}]
We first establish some properties of $\bar G$ and $F_{\bar G}$ that will be used later. 
Let $G = (g_1, \ldots, g_H) \in \bR^{n\times H}$ and let $P_G$ denote the orthogonal projection matrix onto $\col(G)$.
Since the slices are nonempty and disjoint, $\dim\{\col(G)\} = H$, $1_n \in \col(G)$, and 
\begin{align*}
    P_G 
    = \sum_{h=1}^H g_h(g_h^Tg_h)^{-1}g_h^T 
    = \sum_{h=1}^H \frac{1}{n_h}g_hg_h^T. 
\end{align*}
Moreover, since $C_n g_H = -\sum_{h<H} C_ng_h$,  we have
\begin{align*}
   \col(\bar G) = C_n\col(G) = \col(G) \cap 1_n^\perp,
\end{align*}
which has dimension $q = H - 1$.
Consequently, $\bar G^T\bar G$ is invertible, and $F_{\bar G}$ satisfies
\begin{align*}
F_{\bar G}^TF_{\bar G} = I_q, \quad 
F_{\bar G}F_{\bar G}^T = P_{G} - n^{-1}1_n1_n^T = P_{\bar G}, \quad 
C_nF_{\bar G} = F_{\bar G}.
\end{align*}
We are now ready to prove the proposition.
\begin{enumerate}
\item[(a)]
First, we have 
\begin{align*}
  N^TN 
  = F_{\bar G}^T\bar X(\bar X^T\bar X)^{-1} \bar X^T F_{\bar G}
  = F_{\bar G}^T P_{\bar X} F_{\bar G}
  = K,
\end{align*}
where the second equality follows from the definition of the projection matrix $P_{\bar X}$. 
Next, let $\bar x_h$ denote the sample mean of $x_i$ in slice $h$. We have
\begin{align*}
   NN^T 
   &= (\bar X^T\bar X)^{-1/2} \bar X^T F_{\bar G} F_{\bar G}^T\bar X (\bar X^T\bar X)^{-1/2} \\
   &= (\bar X^T\bar X)^{-1/2} X^T F_{\bar G} F_{\bar G}^T X (\bar X^T\bar X)^{-1/2} \\
   &= (n\hat\Sigma)^{-1/2} X^T(P_G - n^{-1}1_n1_n^T)X (n\hat\Sigma)^{-1/2} \\
   &= (n\hat\Sigma)^{-1/2} 
      \left\{\sum_{h=1}^H n_h\bar x_h\bar x_h^T - n\bar x\bar x^T \right\} 
      (n\hat\Sigma)^{-1/2} \\
   &= \hat V,
\end{align*}
where the second equality follows from $\bar X = C_n X$ and $C_n F_{\bar G} = F_{\bar G}$ and 
the third equality follows from $n\hat\Sigma = \bar X^T\bar X$ and $F_{\bar G} F_{\bar G}^T = P_G - n^{-1}1_n1_n^T$. 
Therefore, part (a) follows. 

\item[(b)]
The nonzero eigenvalues of $\hat V$ and $K$ coincide, because $NN^T$ and $N^TN$ share the same nonzero eigenvalues for any matrix $N$.

\item[(c)]
The squared sample canonical correlations between $X$ and $G$ are the eigenvalues of 
\begin{align*}
  (\bar X^T\bar X)^{-1} \bar X^T \bar G(\bar G^T\bar G)^{-1}\bar G^T \bar X 
 = \{(\bar X^T\bar X)^{-1} \bar X^T F_{\bar G}\} \{F_{\bar G}^T\bar X\},
\end{align*}
where we use the definition of $P_{\bar G}$ and $F_{\bar G} F_{\bar G}^T = P_{\bar G}$. 
Equivalently, the nonzero eigenvalues of this matrix are the nonzero eigenvalues of 
$\{F_{\bar G}^T\bar X \}\{(\bar X^T\bar X)^{-1} \bar X^T F_{\bar G}\} = F_{\bar G}^T P_{\bar X} F_{\bar G} = K$, 
since $AB$ and $BA$ have the same nonzero eigenvalues for any two matrices $A$ and $B$. 

\item[(d)]
If $(r_k^2, u_k)$ is an eigenvalue-eigenvector pair of $K$ with $r_k^2 > 0$,
then 
\begin{align*}
     N^TNu_k = Ku_k = r_k^2u_k,
\end{align*} 
which implies that 
$NN^T(Nu_k)=r_k^2Nu_k$.
Thus, $Nu_k$ is an eigenvector of $\hat V = NN^T$ corresponding to eigenvalue $\hat\lambda_k = r_k^2$.
Since $r_k^2$ is simple and $\hat\eta_k$ is the corresponding unit eigenvector of $\hat V$,
we have $\hat\eta_k = Nu_k/\|Nu_k\| = Nu_k/r_k$, where the last equality follows from 
\begin{align*}
    \|Nu_k\|^2 = u_k^TN^TNu_k = u_k^T r_k^2 u_k = r_k^2. 
\end{align*} 
Finally, the $k$th SIR direction is 
\begin{align*}
 \hat\beta_k 
 = \hat\Sigma^{-1/2}\hat\eta_k
 = \sqrt n(\bar X^T\bar X)^{-1/2}Nu_k/r_k
 = \frac{\sqrt{n}}{r_k} (\bar X^T\bar X)^{-1}\bar X^TF_{\bar G}u_k.
\end{align*}
This establishes part (d). 
\end{enumerate}
\end{proof}

\begin{proof}[Proof of Proposition~\ref{prop:cca-pop}]
This proposition is the population counterpart of Proposition~\ref{prop:cca}.
\begin{enumerate}
\item[(a)]
Since $\tilde Y\indep\tilde X\mid B^T\tilde X$ and $\tilde X$ is normally distributed,
we have 
\begin{align*}
   \bE(\tilde X\mid B^T\tilde X,\tilde Y)
   = \bE(\tilde X\mid B^T\tilde X)
   = \mu+\Sigma B(B^T\Sigma B)^{-1}B^T(\tilde X-\mu). 
\end{align*}
Taking the conditional expectation of both sides given $\tilde Y\in J_h$ 
and substituting the result into 
$m_h = \Sigma^{-1/2}\{\bE(\tilde X \mid \tilde Y \in J_h) - \mu\}$ yields 
\begin{align*}
   m_h
    = \Sigma^{1/2}B(B^T\Sigma B)^{-1/2}\bE(\tilde Z_S\mid\tilde Y\in J_h)
    = O\,\bE(\tilde Z_S\mid\tilde Y\in J_h).
\end{align*} 
Consequently, $V=OV_SO^T$. 
Since $O^TO=I_d$,
the nonzero eigenvalues of $V$ coincide with those of $V_S$, 
and the corresponding eigenvectors are $Ov_k$.
Therefore, 
\begin{align*}
    \beta_k=\Sigma^{-1/2}Ov_k = B(B^T\Sigma B)^{-1/2}v_k
\end{align*} 
and $\beta_j^T\Sigma\beta_k=v_j^Tv_k = \delta_{jk}$.

\item[(b)]
Every function of $\fh(\tilde Y)$ is an affine function of $e(\tilde Y)$, 
since $1\{\tilde Y\in J_H\} = 1-1_q^T e(\tilde Y)$. 
Consequently, $\bE(\tilde Z_S\mid\fh(\tilde Y))$ is the linear regression of $\tilde Z_S$ on $e(\tilde Y)$, 
given by
\begin{align*}
 \bE(\tilde Z_S\mid\fh(\tilde Y))
 = \Sigma_{ZE}\Sigma_{EE}^{-1}\{e(\tilde Y) - \bE[e(\tilde Y)]\}, 
\end{align*}
and hence $V_S=\Sigma_{ZE}\Sigma_{EE}^{-1}\Sigma_{ZE}^T$. 
Here, 
$\Sigma_{EE}=\diag(p_1,\dots,p_q)-(p_hp_{h'})_{h,h'\le q}$ 
is positive definite because $p_h>0$ for every $h$.

The matrices $V_S$ and $A^*$ have the same nonzero eigenvalues, because 
\begin{align*}
    V_S = \{\Sigma_{ZE}\Sigma_{EE}^{-1/2}\}\{\Sigma_{EE}^{-1/2}\Sigma_{ZE}^T\}, \qquad 
    A^* = \{\Sigma_{EE}^{-1/2}\Sigma_{ZE}^T\}\{\Sigma_{ZE}\Sigma_{EE}^{-1/2}\}.
\end{align*}
Thus, the nonzero eigenvalues of $V_S$ coincide with those of $A^*$.
Moreover, since $\Sigma_{ZZ} = var(\tilde Z_S) = I_d$, 
$A^* = \Sigma_{EE}^{-1/2}\Sigma_{ZE}^T \Sigma_{ZZ}^{-1} \Sigma_{ZE}\Sigma_{EE}^{-1/2}$.
Therefore, these eigenvalues are the squared population canonical correlations between 
$\tilde Z_S$ and $e(\tilde Y)$, or equivalently, 
between $B^T\tilde X$ and $e(\tilde Y)$. 

\item[(c)]
The first claim follows directly from the definition of $V_S$.

For any invertible matrix $J \in \bR^{d\times d}$, define 
\begin{align*}
    R = (J^TB^T\Sigma BJ)^{-1/2} J^T (B^T\Sigma B)^{1/2}.
\end{align*}
Then $R$ is orthogonal, since $R^TR = I_d$. 
Replacing $B$ by $BJ$ transforms $\tilde Z_S$ into $R\tilde Z_S$  
and $V_S$ into $RV_SR^T$. 
Because $V_S$ and $RV_SR^T$ have the same nonzero eigenvalues, 
$\rho_k^2$ do not depend on the choice of $J$ and hence depend only on $\col(B)$.  

Suppose, for contradiction, that $\rho_1^2 = 1$.
Then there exists a unit vector $u\in\bR^d$ such that $u^T\tilde Z_S$ is almost surely 
a function of $\fh(\tilde Y)$. 
Consequently, $u^T\tilde Z_S$ takes at most $H$ values.
However, since $\tilde Z_S$ is normally distributed, $u^T\tilde Z_S$ is a nondegenerate normal random variable and cannot take only finitely many values. 
This contradiction establishes that $\rho_k^2 < 1$ for all $k$.
\end{enumerate}
\end{proof}

\begin{proof}[Proof of Proposition~\ref{prop:oracle}]
By Lemma~\ref{lem:reduction}, 
we may work with the iid\ rows $(z_i, y_i)$ of $(X_S,Y)$, 
where $z_i\sim\cN(0,I_d)$. 
For each observation, 
let $e_i = e(y_i)$ denote the slice indicator vector defined using the population cut points, and 
let $\hat e_i$ denote the indicator vector corresponding to its sample slice.
Write $\hat\Sigma_{ZZ}$, $\hat\Sigma_{ZE}$, $\hat\Sigma_{EE}$
for the sample covariances of $(z_i,\hat e_i)$, computed with divisor $n$. 
Then
$\hat\Sigma_{ZZ} = \bar X_S^T\bar X_S/n$,
$\hat\Sigma_{EE} = \bar G^T\bar G/n$ 
and $\hat\Sigma_{ZE}^T = \bar G^T\bar X_S/n$, 
so
\begin{align*}
  A
  &= F_{\bar G}^TP_{\bar X_S}F_{\bar G} \\
  &= \{\bar G (\bar G^T\bar G)^{-1/2}\}^T \{\bar X_S(\bar X_S^T\bar X_S)^{-1}\bar X_S^T\} \{\bar G (\bar G^T\bar G)^{-1/2}\} \\
  &= \hat\Sigma_{EE}^{-1/2}\hat\Sigma_{ZE}^T \hat\Sigma_{ZZ}^{-1}\hat\Sigma_{ZE}\hat\Sigma_{EE}^{-1/2}.
\end{align*}
The corresponding oracle SIR kernel matrix is 
$\hat V_S=\hat\Sigma_{ZZ}^{-1/2}\hat\Sigma_{ZE}\hat\Sigma_{EE}^{-1}\hat\Sigma_{ZE}^T\hat\Sigma_{ZZ}^{-1/2}$. 
Hence $A$ and $\hat V_S$ have the same nonzero eigenvalues $\ta_1, \dots, \ta_d$.

The law of large numbers implies that $\hat\Sigma_{ZZ}$ converges in probability to $I_d$.
When fixed cut points are used, $\hat e_i = e_i$.
In this case, the law of large numbers implies that 
$\hat\Sigma_{ZE}$ and $\hat\Sigma_{EE}$ converge in probability 
to $\Sigma_{ZE}$ and $\Sigma_{EE}$, respectively.
For other slicing schemes, we must account for the differences between $\hat e_i$ and $e_i$ 
to establish the convergence results for $\hat\Sigma_{ZE}$ and $\hat\Sigma_{EE}$. 

For equal-count slicing, the observations satisfying $y_i\le c_h$
are exactly the $nF_n(c_h)$ smallest observations,
where $F_n$ denotes the empirical distribution function. 
When ties occur, a tie-breaking rule may be used,
and the choice of rule does not affect the asymptotic result. 
The number of observations for which $\hat e_i\ne e_i$ is bounded by  
$N_n = \sum_{h<H}|nF_n(c_h)-\lfloor hn/H\rfloor|$.
Since $F_n(c_h) \to F_{\tilde Y}(c_h)=h/H$ almost surely, we have $N_n/n \to 0$ almost surely. 
Each difference $\hat e_i - e_i$ has entries in $\{-1,0,1\}$, 
with at most two nonzero entries. 

Let $\tilde\Sigma_{ZE}$ denote the counterpart of $\hat\Sigma_{ZE}$ obtained by 
replacing $\hat e_i$ with $e_i$. 
Then 
\begin{align*}
    \hat\Sigma_{ZE} - \tilde\Sigma_{ZE}
    = \frac{1}{n}\sum_iz_i(\hat e_i-e_i)^T - \bar z (\bar{\hat e} - \bar e)^T,
\end{align*}
where $\bar{\hat e}$ and $\bar e$ denote the sample means of $\hat e_i$ and $e_i$, respectively. 
By the Cauchy-Schwarz inequality, 
\begin{align*}
  \Bigl\|\frac{1}{n}\sum_iz_i(\hat e_i-e_i)^T\Bigr\|_F
  \le \Bigl(\frac1n\sum_i\|z_i\|^2\Bigr)^{1/2} \Bigl(\frac{2N_n}{n}\Bigr)^{1/2}
  \toP 0.
\end{align*}
Analogous arguments show that $(\bar{\hat e} - \bar e) \toP 0$. 
Since $\bar z = O_p(1)$, it follows that $\hat\Sigma_{ZE} - \tilde\Sigma_{ZE} \toP 0$.
By the law of large numbers, $\tilde\Sigma_{ZE} \toP \Sigma_{ZE}$, 
and hence $\hat\Sigma_{ZE} \toP \Sigma_{ZE}$. 
Similar arguments establish the convergence of $\hat\Sigma_{EE}$ to $\Sigma_{EE}$.
Consequently, $\hat V_S \toP \Sigma_{ZE}\Sigma_{EE}^{-1}\Sigma_{ZE}^T = V_S$.  
It then follows from Fact~\ref{fact:weyl} that $\ta_k = \lambda_k(\hat V_S) \toP \rho_k^2$. 
Notably, this argument does not depend on $p$.

If $\rho_k^2$ is a simple eigenvalue, Fact~\ref{fact:dk} implies that there exists 
a unit eigenvector $\hat v_k^S$ of $\hat V_S$ such that $\hat v_k^S \toP v_k$ up to sign.
Applying the SIR algorithm to $\{(y_i, z_i)\}$, 
the $k$th oracle SIR direction is given by $\hat v_k = \hat\Sigma_{ZZ}^{-1/2}\hat v_k^S$. 
Since $\hat\Sigma_{ZZ} \toP I_d$, it follows that $\hat v_k \toP v_k$ up to sign.
Hence, the result follows from the fact that $B^T\Sigma B = I_d$ in the standard coordinates.
\end{proof}

\section{Proofs for Section~\ref{sec:exact}}
\label{sec:proofs-sec3}

This section provides proofs of the results in Section~\ref{sec:exact}.
The proof of Theorem~\ref{thm:represent} is divided into several lemmas, 
with Lemma~\ref{lem:rep} essentially restating the theorem. 
Throughout this section, Assumption~\ref{ass:null} holds, and we work in the
coordinates specified by Lemma~\ref{lem:reduction}. 
Let $\cF = \sigma(X_S, Y)$. 
The slice labels, $\bar G$, $F_{\bar G}$, $\bar X_S = C_nX_S$, $P_{\bar X_S}$, 
$A$ and $\ta_1,\dots,\ta_d$ are all $\cF$-measurable.

\begin{figure}[htb]
\begin{center}
\begin{tikzpicture}[scale = 1]
\draw[thick] (0, 0) -- (10, 0);

\foreach \x in {0, 1, 4, 8, 10} 
    \draw[thick] (\x, -0.15) -- (\x, 0.15); 

\draw[decorate, decoration = {brace, amplitude = 6pt}]
(4, 0.2) -- (10, 0.2) node [midway, above = 8pt] {$\cL$};

\draw[decorate, decoration = {brace, amplitude = 6pt, mirror}]
(0, -0.2) -- (1, -0.2) node [midway, below = 8pt] {$\col(1_n)$}; 

\draw[decorate, decoration = {brace, amplitude = 6pt, mirror}]
(1, -0.2) -- (4, -0.2) node [midway, below = 8pt] {$\col(\bar X_S)$}; 

\draw[decorate, decoration = {brace, amplitude = 6pt, mirror}]
(4, -0.2) -- (8, -0.2) node [midway, below = 8pt] {$\cV = \col(\check X_N)$}; 

\draw[decorate, decoration = {brace, amplitude = 6pt, mirror}]
(1, -1.2) -- (8, -1.2) node [midway, below = 8pt] {$\col(\bar X)$}; 
\end{tikzpicture}
\end{center}
\caption{Illustration of the relationships among the subspaces involved in the proof.}
\label{fig:subspaces}
\end{figure}
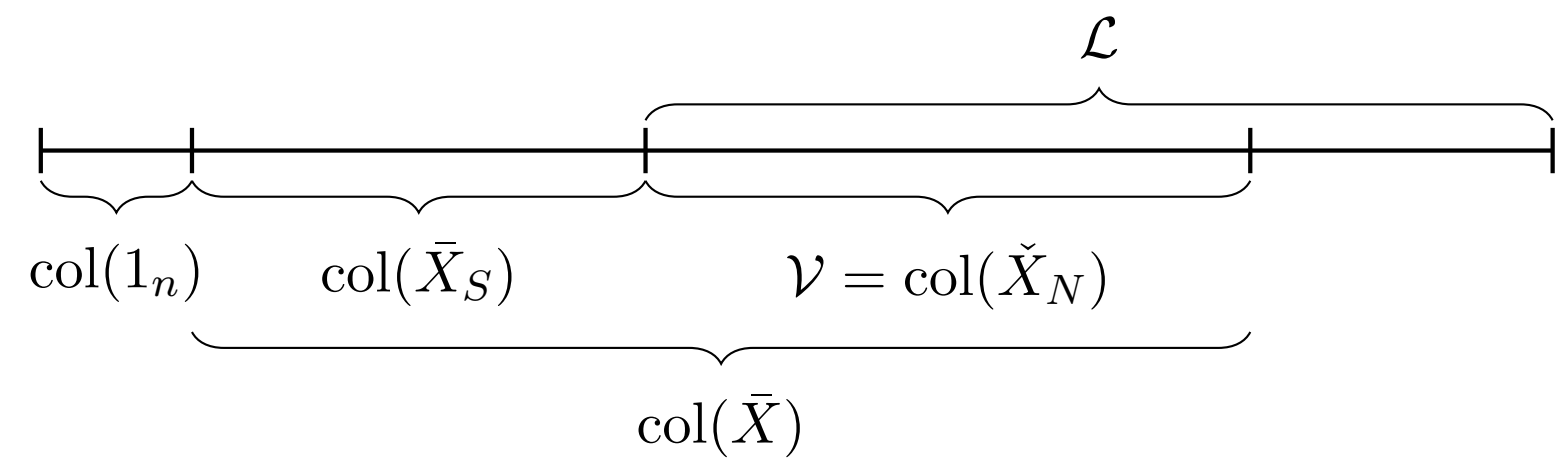

\begin{table}[hbt] 
\caption{Summary of the subspaces used in the proof, together with their corresponding projection matrices, frames, and dimensions.}
\begin{center}
\begin{tabular}{lccc}
\hline
Subspace & Projection matrix & Frame & Dimension \\
\hline
$\col(1_n)$ & & & 1  \\
$\col(\bar X_S)$ & $P_{\bar X_S}$ & $F_S$ & $d$ \\
$\cV = \col(\check X_N)$ & $P_{\cV}$ & & $s = p-d$ \\
$\col(\bar X)$ & $P_{\bar X}$ & & $p$ \\
$\cL$ & $P_{\cL}$ & $F_{\cL}$ & $m = n - d - 1$ \\
$1_n^{\perp}$ & $C_n$ &  & $n - 1$ \\
\hline 
\end{tabular}
\end{center}
\label{table:subspaces}
\end{table}

Lemma~\ref{lem:geometry} introduces several subspaces, together with their corresponding projection matrices and frames. Figure~\ref{fig:subspaces} illustrates the relationships among these subspaces, while  Table~\ref{table:subspaces} provides a summary for quick reference.

\begin{lemma}[Geometry]\label{lem:geometry}
\begin{enumerate}
\item[(a)] 
$X_N$ is independent of $\cF$ and has iid\ $\cN(0,1)$ entries.

\item[(b)] 
Let $\cL = 1_n^\perp \cap \col(\bar X_S)^\perp$. 
Almost surely, $\rank(\bar X_S) = d$, so $\dim(\cL) = m$ and $P_{\cL} = C_n - P_{\bar X_S}$. 
Define $\check X_N = P_{\cL}X_N$ and $\cV = \col(\check X_N)$. 
Then $P_{\bar X} = P_{\bar X_S} + P_{\cV}$ and $P_{\cV}P_{\bar X_S} = 0$.

\item[(c)] 
Let $F_{\cL}\in\bR^{n\times m}$ be an $\cF$-measurable frame
spanning $\cL$ and define $Z = F_{\cL}^TX_N \in \bR^{m\times s}$. 
Then $\check X_N = F_{\cL}Z$, and, conditionally on $\cF$, 
the entries of $Z$ are iid\ $\cN(0,1)$. 
Consequently, $\rank(\check X_N) = s$ almost surely, 
and, conditionally on $\cF$, $\cV$ is uniformly distributed over the $s$-dimensional subspaces of $\cL$.
\end{enumerate}
\end{lemma}

\begin{proof}
\begin{enumerate}
\item[(a)] 
A single observation satisfies  
$\tilde X = (\tilde X_S^T, \tilde X_N^T)^T \sim \cN_p(0,I_p)$, 
so $\tilde X_N \indep \tilde X_S$ and $\tilde X_N$ has iid $\cN(0, 1)$ entries.
Moreover, by \eqref{eq:model-cs}, $\tilde Y \indep \tilde X_N \mid\tilde X_S$.
Thus, the joint density factorizes as $f(x_S)f(x_N)f(y\mid x_S)$, 
implying that $\tilde X_N \indep (\tilde X_S,\tilde Y)$.
By independence across observations, the matrix $X_N$ is independent of $\cF$.

\item[(b)] 
Because $\bar X_S = C_nX_S$, we have $\col(\bar X_S) \subset 1_n^\perp$, 
and hence $P_{\bar X_S} = C_nP_{\bar X_S} = P_{\bar X_S}C_n$.
Therefore, $C_n - P_{\bar X_S}$ is the orthogonal projection matrix onto $\cL$, 
which has dimension $n-1-d = m$. 

Moreover, $C_nX_N = P_{\bar X_S}X_N + \check X_N$, where 
$P_{\bar X_S}X_N$ has columns in $\col(\bar X_S)$ and 
$\check X_N$ has columns in $\cL$. 
Thus, $\col(\bar X) = \col(\bar X_S) \oplus \col(\check X_N)$, 
an orthogonal sum.
Consequently, $P_{\bar X} = P_{\bar X_S} + P_{\cV}$ and $P_{\cV}P_{\bar X_S} = 0$.

\item[(c)] 
Since $P_{\cL} = F_{\cL}F_{\cL}^T$, we have $\check X_N = F_{\cL}F_{\cL}^TX_N = F_{\cL}Z$. 
Conditionally on $\cF$, the columns of $X_N$ are iid\ $\cN_n(0,I_n)$, 
so the columns of $Z$ are iid\ $\cN_m(0, I_m)$. 
Since $s < m$, $\rank(Z) = s$ almost surely. 
Moreover, for any orthogonal $Q \in \bR^{m\times m}$, 
$QZ \eqd Z$ conditional on $\cF$. 
Thus, $\col(Z)$ is uniformly distributed over the $s$-dimensional subspaces of $\bR^m$,
which implies that $\cV = F_{\cL}\col(Z)$ is uniformly distributed over the $s$-dimensional subspaces of $\cL$.
\end{enumerate}
\end{proof}

Lemma~\ref{lem:proj} summarizes several results on matrix-variate Beta distributions.

\begin{lemma}[Uniform projections]\label{lem:proj}
Let $P$ be the orthogonal projection matrix onto a uniformly distributed $s$-dimensional subspace of $\bR^m$,
where $0 < s < m$, and let $F$ be a fixed $m\times t$ frame, with $t\le m$.
The following properties hold. 
\begin{enumerate}
\item[(a)] 
The matrix $F^TPF$ follows a matrix-variate Beta distribution, 
$F^TPF \sim \cB_t(s/2,(m-s)/2)$.
In particular, its distribution does not depend on $F$.

\item[(b)] 
The expectation is $\bE(F^TPF) = (s/m)I_t$, and 
\begin{align*}
  var\{\tr(F^TPF)\} = \frac{2ts(m-s)(m-t)}{m^2(m+2)(m-1)},
  \qquad
  \bE\|F^TPF-(s/m)I_t\|_F^2 \le \frac{2t^2s(m-s)}{m^2(m-1)}.
\end{align*}

\item[(c)] 
If $t$ is fixed, $s\to\infty$, and $s/m \to \gamma \in [0,1)$, then
\begin{align*}
  \frac{m}{\sqrt s}\Bigl(F^TPF-\frac smI_t\Bigr)
  \tod
  (1-\gamma)^{1/2} \cG_t.
\end{align*}

\item[(d)] 
If $t$ is fixed, $s\to\infty$, and $s/m \to \gamma \in [0,1)$, then
\begin{align*}
  \frac{\tr(F^TPF)-ts/m}{\sqrt{2ts(m-s)/m^3}}
  \tod \cN(0,1).
\end{align*}

\end{enumerate}
\end{lemma}

\begin{proof}
\begin{enumerate}
\item[(a)] 
Recall that $E_s = (I_s, 0)^T \in\bR^{m\times s}$. By Fact~\ref{fact:unif}, a uniformly distributed $s$-dimensional subspace of $\bR^m$ can be represented as $\col(OE_s)$, where $O\in\bR^{m\times m}$ is uniformly distributed on the orthogonal group $O(m)$. 
Therefore   
\begin{align*}
  P \eqd OE_sE_s^TO^T \quad\text{ and }\quad
  F^TPF \eqd F^TOE_sE_s^TO^TF = \tilde F^T E_sE_s^T \tilde F, 
\end{align*}
where $\tilde F = O^TF \in\bR^{m\times t}$.
Here, $O$ and $\tilde F$ are random matrices, whereas $E_s$ and $F$ are fixed. 
By Fact~\ref{fact:unif}, $\tilde F$ is uniformly distributed on $\mathcal{V}_{m, t}$, the Stiefel manifold of $m\times t$ frames. 
Furthermore, $\tilde F \eqd Z(Z^TZ)^{-1/2}$, where $Z\in\bR^{m\times t}$ has iid $\cN(0, 1)$ entries. 

Partition $Z$ as $Z = (Z_1^T, Z_2^T)^T$, where $Z_1\in\bR^{s\times t}$ consists of the first $s$ rows and $Z_2\in\bR^{(m-s)\times t}$ consists of the remaining $m-s$ rows. 
Define $S_j = Z_j^TZ_j$ for $j = 1, 2$. 
Then $S_1 \sim \cW_t(s, I_t)$ and $S_2\sim \cW_t(m-s, I_t)$ are independent, and $Z^TZ = S_1 + S_2$. Consequently, 
\begin{align*}
  \tilde F^TE_sE_s^T\tilde F
  &\eqd (S_1 + S_2)^{-1/2} Z^TE_sE_s^TZ (S_1 + S_2)^{-1/2} \\
  &= (S_1 + S_2)^{-1/2} S_1 (S_1 + S_2)^{-1/2}.
\end{align*}
Hence, by the definition of the matrix-variate Beta distribution, $F^TPF \sim \cB_t(s/2, (m-s)/2)$. 
The parameters of this distribution depend only on $m$ and $s$, but are independent of $F$. 

\item[(b)] 
The distribution of $F^TPF$ does not depend on $F$ due to (a). 
Without loss of generality, we may take $F = (I_t, 0)^T$, and 
hence, $F^TPF = (P_{ij})_{i, j \leq t}$. 

Because the distribution of $P$ is uniform, it is invariant under the transformation $P \mapsto OPO^T$ for any orthogonal matrix $O$. Hence, it is also invariant under the transformation $P\mapsto Q P Q^T$ for any permutation matrix $Q$, because $Q$ is always orthogonal. 
This implies that $P_{ii}$ are identically distributed, and so are $P_{ij}$ for $i\neq j$, and the pairs $(P_{ii}, P_{jj})$ for $i \neq j$. We compute them separately. 
\begin{itemize}
\item 
{\em mean $\bE(P_{ii})$ and $\bE(P_{ij})$}. 
Due to the invariance of the distribution of $P$, we have $\bE(P) = O\bE(P)O^T$ for any orthogonal matrix $O$. Therefore, we must have $\bE(P) = cI_m$ for a constant $c$.
It implies that $\bE(P_{ij}) = 0$ for $i \neq j$. Because $mc = \tr(cI_m) = \bE[\tr(P)] = s$, we have $c = s/m$. Therefore, $\bE(P_{ii}) = s/m$. 
Combining the results yields $\bE(F^TPF) = (s/m)I_t$.

\item
{\em diagonal variance $var(P_{ii})$}. 
Applying (a) with $F = e_1 = (1, 0, \ldots, 0)^T$ yields $P_{11} = e_1^TPe_1 \sim \cB_1(s/2, (m-s)/2)$, which is exactly a univariate Beta($s/2$, $(m-s)/2$) distribution. 
Hence, 
\begin{align*}
 var(P_{ii}) = var(P_{11}) 
 = \frac{2s(m-s)}{m^2(m + 2)}.
\end{align*}

\item 
{\em off-diagonal second moment $\bE(P_{ij}^2)$.} 
Because $P$ is idempotent $P^2 = P$, $\sum_{j=1}^m P_{1j}^2 = P_{11}$.
It implies that $\bE(P_{11}^2) + (m-1) \bE(P_{12}^2) = \bE(P_{11})$.  Hence, 
\begin{align*}
  \bE(P_{12}^2) 
  = \frac{\bE(P_{11}) - \bE(P_{11}^2)}{m-1}
  = \frac{\bE(P_{11}) - \bE(P_{11})^2 - var(P_{11})}{m-1}
  = \frac{s(m-s)}{m(m-1)(m+2)}. 
\end{align*}

\item 
{\em diagonal covariance $cov(P_{ii}, P_{jj})$}. 
Because $\tr(P) = \sum_{i=1}^m P_{ii} = s$ is a constant, 
\begin{align*}
 0 
 = var\left(\sum_{i=1}^m P_{ii}\right) 
 = m\, var(P_{11}) + m(m-1) \, cov(P_{11}, P_{22}),
\end{align*}
which yields $cov(P_{11}, P_{22}) = -var(P_{11})/(m-1)$.
\end{itemize}

The trace variance is 
\begin{align*}
  var[\tr(F^TPF)]
  &= var\left(\sum_{i=1}^{t} P_{ii}\right)
   = t \, var(P_{11}) + t(t-1)\, cov(P_{11}, P_{22}) \\
  &= \frac{t(m-t)}{m-1} var(P_{11})
  = \frac{2ts(m-s)(m-t)}{m^2(m-1)(m + 2)}.
\end{align*}

Finally, the expected Frobenius bound is 
\begin{align*}
 \bE\|F^TPF - (s/m)I_t\|_F^2
 &= t \, var(P_{11}) + t(t-1) \bE(P_{12}^2) \\
 &= \frac{2ts(m-s)}{m^2(m+2)} + \frac{t(t-1)s(m-s)}{m(m-1)(m+2)} \\
 &\leq \frac{2ts(m-s)}{m^2(m-1)} + \frac{t(t-1)s(m-s)}{m^2(m-1)} \\
 &= \frac{t(t+1)s(m-s)}{m^2(m-1)} \\
 &\leq \frac{2t^2s(m-s)}{m^2(m-1)}.  
\end{align*}

\item[(c)] 
We have $F^TPF \eqd  (S_1 + S_2)^{-1/2}S_1(S_1 + S_2)^{-1/2}$, due to the definition of matrix-variate Beta distribution.
Write $S_1 = sI_t + s^{1/2}G_1$ and $S_2 = (m-s)I_{t} + (m-s)^{1/2}G_2$. 
Since $S_1$ is a sum of $s$ iid\ matrices $Z_iZ_i^T$, where $Z_i\sim\cN_t(0, I_t)$, 
the distinct entries of $G_1$ are uncorrelated, with variance $2$ for diagonal entries and $1$ for off-diagonal entries.  
Thus, the multivariate central limit theorem yields $G_1 \tod \cG_t$ as $s\to\infty$. 
Similarly, as $m-s\to\infty$, $G_2\tod\cG'_t$, where $\cG'_t$ is independent of $G_1$.
Since $t$ is fixed, each entry of $G_1$ and $G_2$ has bounded variance, so $G_1 = O_p(1)$ and $G_2 = O_p(1)$. 

Let $\Delta = S_1 + S_2 - mI_t$, so that $\Delta = s^{1/2}G_1 + (m-s)^{1/2}G_2 = O_p(m^{1/2})$.
Notice that $P(\|\Delta/m\|_{op} < 1/2) \to 1$.
On the event $\|\Delta/m\|_{op} < 1/2$, the spectral calculus yields 
\begin{align*}
  (S_1 + S_2)^{-1/2} 
  = m^{-1/2} (I_t + \Delta/m)^{-1/2}
  = m^{-1/2} (I_t - \Delta/(2m) + \cR), 
\end{align*}
where the remainder $\cR$ is symmetric and satisfies $\|\cR\|_{op} \leq C_0 \|\Delta/m\|_{op}^2 = O_p(m^{-1})$ for a universal constant $C_0$. 
Substituting the preceding expansion yields 
\begin{align*}
  &m(S_1+S_2)^{-1/2}S_1(S_1+S_2)^{-1/2} \\
  &= (I_t - \Delta/(2m) + \cR)(sI_t + s^{1/2}G_1)(I_t - \Delta/(2m) + \cR)  \\
  &= sI_t + s^{1/2}G_1 - \frac{s}{m}\Delta + \cE,
\end{align*}
where $\cE$ collects the remaining terms and 
\begin{align*}
    \cE 
    = -\frac{s^{1/2}}{2m}(\Delta G_1 + G_1\Delta) 
      + 2s\cR 
      + \frac{s}{4m^2}\Delta^2 
      - \frac{s}{2m}(\cR\Delta + \Delta\cR)
      + o_p(1). 
\end{align*}
Hence
\begin{align}
  \frac m{\sqrt s}\Bigl(F^TPF-\frac smI_t\Bigr)
  \eqd
    \Bigl(1-\frac{s}{m}\Bigr) G_1 
  - \Bigl\{\frac{s}{m}\Bigl(1-\frac{s}{m}\Bigr)\Bigr\}^{1/2}G_2
  + O_p(m^{-1/2}).
\label{eq:FPF-dist}
\end{align}
The right-hand side converges in distribution to 
$(1-\gamma)\cG_t-\{\gamma(1-\gamma)\}^{1/2}\cG'_t$.
By the independence of $\cG_t$ and $\cG'_t$, this limit has the same distribution 
as $(1-\gamma)^{1/2}\cG_t$, 
because $(1-\gamma)^2+\gamma(1-\gamma) = 1-\gamma$. 
This establishes (c). 

\item[(d)]
Taking traces on both sides of (\ref{eq:FPF-dist}) and dividing by $\{2t(m-s)/m\}^{1/2}$ yields
\begin{align*}
    \frac{\tr(F^TPF) - ts/m}{\sqrt{2ts(m-s)/m^3}}
    \eqd 
    \frac{a_n}{\sqrt{2t}} \tr(G_1) 
    - \frac{b_n}{\sqrt{2t}} \tr(G_2)
    + O_p\left(\frac{1}{\sqrt{2t(m-s)}}\right),
\end{align*}
where $a_n = \sqrt{(m-s)/m}$ and $b_n = \sqrt{s/m}$ are deterministic weights.
Let $\chi_v^2$ denote a chi-squared random variable with $v$ degrees of freedom. 
Then 
\begin{align*}
  \frac{1}{\sqrt{2t}} \tr(G_1) \eqd \frac{\chi_{ts}^2 - ts}{\sqrt{2ts}},
  \qquad 
  \frac{1}{\sqrt{2t}} \tr(G_2) \eqd \frac{\chi_{t(m-s)}^2 - t(m-s)}{\sqrt{2t(m-s)}}
\end{align*} 
are independent and are asymptotically $\cN(0, 1)$ as $s\to\infty$ and $m-s\to \infty$ by the CLT. 
Moreover, since $a_n^2 + b_n^2 = 1$, the leading term converges to $\cN(0, 1)$. 
The remainder approaches 0 because $(m-s)\to\infty$. 
The result holds due to Slutsky's Theorem. 
\end{enumerate}
\end{proof}

Partition $M$ and $\bB$, defined in Theorem~\ref{thm:represent}, into blocks of sizes $d$ and $t$ as follows.
\begin{align*}
  M = \begin{pmatrix} M_{11} & M_{10} \\ M_{01} & M_{00} \end{pmatrix},
  \qquad 
  \bB = \begin{pmatrix} \bB_{11} & \bB_{10} \\ \bB_{01} & \bB_{00} \end{pmatrix},
\end{align*}
where $M_{00} = \bB_{00} \in \bR^{t\times t}$ and 
$M_{01} = \bB_{01}D_1 \in \bR^{t\times d}$, 
where $D_1 = \diag\{(1 - \ta_1)^{1/2}, \ldots, (1 - \ta_d)^{1/2}\}$.

\begin{lemma}[Exact representation]\label{lem:rep}
Define $W = (I_n-P_{\bar X_S})F_{\bar G}$.
Then $W = P_{\cL}F_{\bar G}$, $W^TW = I_q-A$, and
\begin{equation}\label{eq:KA}
  K=A+W^TP_{\cV}W.
\end{equation}
There exist an $\cF$-measurable orthogonal $q\times q$ matrix
$U = (\tu_1, \dots, \tu_q)$, an $\cF$-measurable $n\times q$ frame
$F_{\phi} = (\phi_1, \dots, \phi_q)$ with columns in $\cL$, 
and a random matrix $\bB\sim\cB_q(s/2,(m-s)/2)$, independent of $\cF$, 
such that
\begin{align*}
  WU = F_{\phi} D, \qquad
  \bB = F_{\phi}^T P_{\cV} F_{\phi}, \qquad
  U^TKU = M = \Lambda + D \bB D,
\end{align*}
where $\Lambda$ and $D$ are defined in Theorem~\ref{thm:represent}. 
We have $M_{00} = \bB_{00}\sim\cB_t(s/2,(m-s)/2)$ and 
$\bB_{00}$ is independent of $\cF$. 
Moreover, $0\preceq M\preceq I_q$.
\end{lemma}

\begin{proof}
Since $F_{\bar G} = C_nF_{\bar G}$ and $P_{\bar X_S}C_n = P_{\bar X_S}$, 
we have 
\begin{align*}
    W = (C_n-P_{\bar X_S})F_{\bar G}=P_{\cL}F_{\bar G},
    \qquad 
    W^TW=F_{\bar G}^T(I_n-P_{\bar X_S})F_{\bar G}=I_q-A. 
\end{align*} 
By Lemma~\ref{lem:geometry}(b), $K = A + F_{\bar G}^T (P_{\bar X} - P_{\bar X_S}) F_{\bar G} = A + F_{\bar G}^TP_{\cV}F_{\bar G}$. 
Furthermore, since $P_{\cV} = P_{\cL}P_{\cV}P_{\cL}$, it follows that
$F_{\bar G}^TP_{\cV}F_{\bar G} = W^TP_{\cV}W$.
Hence, \eqref{eq:KA} follows.  

Let $U = (\tu_1, \ldots, \tu_{H-1})$ be an orthogonal matrix whose columns are eigenvectors of $A$ 
satisfying $A\tu_j = \ta_j \tu_j$. Since $A$ has rank at most $d$, we have $\ta_j = 0$ for $j > d$. 
Then $U^T A U = \Lambda$. 
Moreover, $0 \le \ta_j \le 1$, since $0\preceq A\preceq F_{\bar G}^TF_{\bar G} = I_q$. 
Because $A$ is $\cF$-measurable and $U$ can be chosen as a Borel-measurable function of $A$, $U$ is also $\cF$-measurable. 

Recall that $W = (I_n - P_{\bar X_S})F_{\bar G}$ defined above
and $W^TW = I_{q} - A$. For any $i$ and $j$,
\begin{align*}
  (W\tu_i)^T(W\tu_j) = \tu_i^T(I_{q} - A)\tu_j = (1 - \ta_i)\delta_{ij},
\end{align*}
where $\delta_{ij} = 1$ if $i = j$ and $= 0$ otherwise. 
Moreover, $W\tu_j \in \cL$. 
If $\ta_j < 1$, set $\phi_j = W\tu_j/\sqrt{1 - \ta_j}$.
If $\ta_j = 1$, then $W\tu_j = 0$ and let $\phi_j$ be a unit vector in $\cL$ orthogonal to the other $\phi$'s.
Such a choice is possible, because $\text{dim}(\cL) = m > q$. 
Then $F_{\phi} = (\phi_1, \ldots, \phi_{H-1})$ is a frame with columns in $\cL$ and $WU = F_{\phi} D$. 

By \eqref{eq:KA}, 
\begin{align*}
    U^TKU   
    = U^TAU + U^T W^TP_{\cV} WU 
    = \Lambda + DF_{\phi}^T P_{\cV} F_{\phi} D. 
\end{align*}
By Lemma~\ref{lem:geometry}(b)(c), $\cV = \col(\check X_N) = F_{\cL}\col(Z)$ and hence 
$P_{\cV} = F_{\cL}P_{\col(Z)}F_{\cL}^T$. 
Define $F = F_{\cL}^TF_{\phi}$, which is an $\cF$-measurable $m\times q$ frame. 
Then 
\begin{align*}
    \bB
    = F_{\phi}^T P_{\cV} F_{\phi}
    = F_{\phi}^T F_{\cL}P_{\col(Z)}F_{\cL}^T F_{\phi}
    = F^TP_{\col(Z)}F. 
\end{align*}
Conditionally on $\cF$, $F$ is fixed and $\col(Z)$ is uniformly distributed over the $s$-dimensional subspaces of $\bR^m$ by Lemma~\ref{lem:geometry}(c).  
Therefore, by Lemma~\ref{lem:proj}(a), the conditional distribution of $\bB$ is
$\cB_q(s/2,(m-s)/2)$. 
Since this conditional distribution does not depend on $\cF$, $\bB$ is independent of $\cF$. 

The bottom-right block of $\bB$ is $\bB_{00} = F_0^T P_{\col(Z)} F_0$,
where $F_0 = F_{\cL}^T(\phi_{d+1}, \ldots, \phi_{H-1})$ is $\cF$-measurable and has orthonormal columns. 
Following the same argument as above, $\bB_{00} \sim \cB_{q-d}(s/2, (m-s)/2)$ 
and $\bB_{00}$ is independent of $\mathcal{F}$. 

Finally, $M$ is orthogonally similar to $K = F_{\bar G}^TP_{\bar X}F_{\bar G}$.
Since $0\preceq K\preceq F_{\bar G}^TF_{\bar G} = I_q$, 
it follows that $0\preceq M\preceq I_q$.
\end{proof}

\begin{proof}[Proof of Theorem~\ref{thm:represent}]
This is Lemma~\ref{lem:rep}.
\end{proof}

\section{Proofs for Section~\ref{sec:values}}
\label{sec:proofs-sec4}

This section provides proofs of the results in Section~\ref{sec:values}.
We use the coordinates of Lemma~\ref{lem:reduction} and the notation of
Section~\ref{sec:proofs-sec3}. By Theorem~\ref{thm:represent}, $r_1^2,\dots,r_q^2$ are the
eigenvalues of $M$. 
As discussed in Section~\ref{sec:values}, the key step is to bound the differences between eigenvalues by Weyl's inequality. We first present two lemmas that will be used in the subsequent proofs. 

\begin{lemma}[Block perturbation]\label{lem:block}
Let $M = \begin{pmatrix}M_{11}&M_{10}\\M_{01}&M_{00}\end{pmatrix}$ be a symmetric
$q\times q$ matrix, where $M_{11}$ is $d\times d$, and define
$\delta = \lambda_{\min}(M_{11}) - \lambda_{\max}(M_{00})$.
Then the following results hold: 
\begin{enumerate}
\item[(a)] 
$\sum_{i>d}\lambda_i(M) \le \tr(M_{00})$.
Moreover, if $\delta > 0$, then
\begin{align*}
\tr(M_{00}) - \sum_{i>d}\lambda_i(M) \le \frac{2\|M_{01}\|_F^2}{\delta}.
\end{align*}

\item[(b)] 
If $\delta > 0$, then for $j = 1,\dots, q-d$,
\begin{align*}
0\le 
\lambda_j(M_{00}) - \lambda_{d+j}(M)
\le
\frac{2\|M_{01}\|_{op}^2}{\delta}. 
\end{align*}
\end{enumerate}
\end{lemma}

\begin{proof}
For $\tau > 0$, define two matrices $N_{\tau} \in\bR^{q\times q}$ 
and $N'_{\tau} \in \bR^{q\times q}$ below.
\begin{align*}
  N_{\tau}
  &= \begin{pmatrix} \tau^{-1/2} M_{10} \\ -\tau^{1/2} I_{q-d} \end{pmatrix}
    \begin{pmatrix} \tau^{-1/2} M_{01} & -\tau^{1/2} I_{q-d} \end{pmatrix}
  = \begin{pmatrix}
  \tau^{-1} M_{10}M_{01} & -M_{10} \\ -M_{01} & \tau I_{q-d}
  \end{pmatrix}, \\ 
  N'_{\tau}
  &= \begin{pmatrix} \tau^{1/2} I_{d} \\ \tau^{-1/2} M_{01} \end{pmatrix}
    \begin{pmatrix} \tau^{1/2} I_{d} & \tau^{-1/2} M_{10}  \end{pmatrix}
  = \begin{pmatrix}
  \tau I_{d}  & M_{10} \\ M_{01} &  \tau^{-1}  M_{01}M_{10}
  \end{pmatrix}. 
\end{align*}
Since $M_{10} = M_{01}^T$, $N_{\tau}$ and $N'_{\tau}$ are positive semidefinite. 
Therefore, 
\begin{align*}
    M - N'_{\tau} \preceq M \preceq M + N_{\tau},
\end{align*}
where
\begin{align*}
    M + N_{\tau}  &= \diag(M_{11}+\tau^{-1}M_{10}M_{01},\ M_{00}+\tau I_{q-d}), \\
    M - N'_{\tau} &= \diag(M_{11}-\tau I_{d},\ M_{00}-\tau^{-1}M_{01}M_{10}).
\end{align*}

\begin{enumerate}
\item[(a)]
The first claim follows from Fact~\ref{fact:kyfan} by taking $F = (0,I_{q-d})^T$. 

Consider $M + N_{\tau}$. 
The eigenvalues of its top-left block are at least $\lambda_{\min}(M_{11})$, 
while those of the bottom-right block are at most $\lambda_{\max}(M_{00}) + \tau$. 
Thus, if $0 < \tau < \delta$, the $d$ largest eigenvalues of $M + N_{\tau}$ are exactly those of its top-left block. 
By Fact~\ref{fact:weyl}(b), we know $\lambda_i(M) \leq \lambda_i(M + N_{\tau})$ for $i \leq d$. 
Consequently, 
\begin{align*}
    \sum_{i\le d}\lambda_i(M)
    \le
    \sum_{i\le d}\lambda_i(M + N_{\tau})
    = 
    \tr(M_{11}) + \tau^{-1}\|M_{01}\|_F^2. 
\end{align*}
Subtracting this inequality from $\tr(M) = \tr(M_{11}) + \tr(M_{00})$ yields 
\begin{align*}
  \sum_{i=d+1}^q \lambda_i(M) \geq \tr(M_{00}) - \tau^{-1} \|M_{01}\|_F^2. 
\end{align*}
Substituting $\tau = \delta/2$ establishes the desired bound and completes the proof.

\item[(b)]
Following the same argument as in (a), 
if $0 < \tau < \delta$, the $q-d$ smallest eigenvalues of $M + N_{\tau}$ are those of
$M_{00} + \tau I_{q-d}$. By Fact~\ref{fact:weyl}(b), 
$\lambda_{d+j}(M) \le \lambda_j(M_{00}) + \tau$, for $j = 1, \dots, q - d$.  
Letting $\tau \downarrow 0$ yields $\lambda_{d+j}(M) \le \lambda_j(M_{00})$.

Next, apply a similar argument to $M - N'_{\tau}$. 
The eigenvalues of its top-left block are at least $\lambda_{\min}(M_{11}) - \tau$,
while those of the bottom-right block are at most $\lambda_{\max}(M_{00})$.
Thus, if $0< \tau < \delta$, the $q-d$ smallest eigenvalues of $M - N'_{\tau}$ are exactly those of $M_{00} - \tau^{-1}M_{01}M_{10}$. Hence, for $j = 1, \dots, q - d$, 
\begin{align*}
    \lambda_{d+j}(M)
    \ge
    \lambda_j(M_{00} - \tau^{-1}M_{01}M_{10})
    \ge
    \lambda_j(M_{00}) - \tau^{-1}\|M_{01}\|_{op}^2,  
\end{align*}
where the second inequality follows from Fact~\ref{fact:weyl}(a). 
Taking $\tau = \delta/2$ completes the proof.  
\end{enumerate}
\end{proof}

\begin{lemma}[Size of the fluctuation]\label{lem:eps}
Let $\varepsilon_n = \|\bB - \gamma_n I_q\|_F$, where $\gamma_n = s/m$. Then 
\begin{align*}
    \bE(\varepsilon_n^2)
    \le
    \frac{2q^2s(m-s)}{m^2(m-1)}
    \le
    \frac{2q^2p}{m(m-1)}.
\end{align*} 
Hence, if $H$ and $d$ are fixed, $\varepsilon_n = O_p(\sqrt p/n)$.
\end{lemma}

\begin{proof}
By Lemma~\ref{lem:rep}, $\bB \sim \cB_q(s/2, (m-s)/2)$. 
Applying Lemma~\ref{lem:proj}(b) with $t = q$ yields the first inequality for $\bE(\varepsilon_n^2)$,
while the second inequality follows from $s \leq p$ and $m-s < m$. 

If $H$ and $d$ are fixed, the right-hand side of the inequality is of order $O(p/n^2)$.
Hence, by Markov's inequality, $\varepsilon_n = O_p(\sqrt p/n)$.
\end{proof}

Let $M^*$ denote the $q\times q$ diagonal matrix defined by   
\begin{align*}
   M^* 
   = \Lambda + \gamma_n D^2 
   = \diag\{
   \gamma_n + (1 - \gamma_n)\ta_1, \ldots, 
   \gamma_n + (1 - \gamma_n)\ta_d, 
   \gamma_n, \ldots, \gamma_n\}.
\end{align*}
Since $M - M^* = D(\bB-\gamma_nI_q)D$ and $\|D\|_{op} = \sqrt{\lambda_{\max}(D^TD)}  \le 1$, 
we have 
\begin{align}
   \|M - M^*\|_{F} 
   \leq \|D\|_{op}^2 \|\bB - \gamma_n I_{q}\|_F
   \leq \varepsilon_n,  
\label{eq:MMbar}
\end{align}
where $\varepsilon_n = \|\bB - \gamma_n I_{q}\|_F$ is defined in Lemma~\ref{lem:eps}.

By Fact~\ref{fact:weyl}(a), 
\begin{align*}
 |\lambda_{\min}(M_{11}) - \lambda_{\min}(M^*_{11})| 
 \leq \|M_{11} - M_{11}^*\|_{F}, \\
 |\lambda_{\max}(M_{00}) - \lambda_{\max}(M^*_{00})| 
 \leq \|M_{00} - M_{00}^*\|_{F}. 
\end{align*}
Both $M_{11} - M_{11}^*$ and $M_{00} - M_{00}^*$ are sub-blocks of $M - M^*$. 
Hence, the right-hand sides of both inequalities are bounded by $\varepsilon_n$ due to equation (\ref{eq:MMbar}). Therefore,
\begin{align*}
  \lambda_{\min}(M_{11}) 
  &\geq \lambda_{\min}(M^*_{11}) - \varepsilon_n 
   = \gamma_n + (1 - \gamma_n)\ta_d - \varepsilon_n, \\
  \lambda_{\max}(M_{00}) 
  &\leq \lambda_{\max}(M^*_{00}) + \varepsilon_n 
   = \gamma_n + \varepsilon_n,  
\end{align*}
which implies that 
\begin{align*}
  \delta 
  = \lambda_{\min}(M_{11}) - \lambda_{\max}(M_{00})
  \geq (1 - \gamma_n)\ta_d - 2\varepsilon_n. 
\end{align*}

Consider the event defined below
\begin{align*}
    \cE_n
    = \{\ta_d \ge \underline a\}
    \cap\{(1-\gamma_n)\underline a - 2\varepsilon_n \ge (1-\gamma)\underline a/2\}.
\end{align*}
Under Assumption~\ref{ass:signal}, 
$\bP(\ta_d \ge \underline a) \to 1$. 
Moreover, if $p/n \to \gamma < 1$, then $\gamma_n\to\gamma$ and $\bP(\cE_n) \to 1$.
On the event $\cE_n$, we have 
\begin{equation}\label{eq:delta}
  \delta
  = \lambda_{\min}(M_{11}) - \lambda_{\max}(M_{00})
  \ge (1-\gamma_n) \ta_d - 2\varepsilon_n
  \ge (1-\gamma)\underline a/2,
  \qquad
  \|M_{01}\|_F \le \varepsilon_n.
\end{equation}

\begin{proof}[Proof of Theorem~\ref{thm:values1}]
By Fact~\ref{fact:weyl}(a) and \eqref{eq:MMbar},
$|\hat\lambda_k - \lambda_k(M^*)| \le \varepsilon_n$ for every $k$.
The ordered eigenvalues of $M^*$ are $\gamma_n + (1-\gamma_n)\ta_k$ for $k\le d$ and
$\gamma_n$ for $k>d$.
Lemma~\ref{lem:eps} establishes the first claim without any signal assumption.
If $p/n\to\gamma$, then $\gamma_n \to \gamma$.
Moreover, $\sqrt{p}/n \le n^{-1/2} \to 0$, which establishes the stated limits.
\end{proof}

\begin{lemma}[Signal eigenvectors]\label{lem:eigvec}
Suppose that Assumption~\ref{ass:null} holds, $H$ and $d$ are fixed,
$p/n\to\gamma\in[0,1)$, and $\ta_i\toP\rho_i^2$ for $i\le d$.
Fix $k\le d$ such that $\rho_k^2>0$ and $\rho_k^2\neq\rho_i^2$ for all $i\neq k$,
and set $\rho_i^2 = 0$ for $i>d$. Define 
\begin{align*}
    g = (1-\gamma)\min_{i\neq k}|\rho_k^2-\rho_i^2| > 0 .
\end{align*}
Then, with probability tending to one, $\hat\lambda_k = \lambda_k(M)$ is a simple
eigenvalue of $M$. Moreover, a corresponding unit eigenvector $w_k$,
with its sign chosen such that $w_k^Te_k\ge 0$, satisfies
\begin{align*}
    \|w_k - e_k\| \le \frac{2^{5/2}\varepsilon_n}{g} = O_p(\sqrt p/n).
\end{align*}
\end{lemma}

\begin{proof}
The matrix $M^*$ is diagonal, with eigenvalues 
$\gamma_n + (1-\gamma_n)\ta_i$ for $i\le d$ and $\gamma_n$ for $i>d$.
Thus, $e_k$ is a unit eigenvector of $M^*$ corresponding to $\lambda_k(M^*)$, 
and the eigengap of $M^*$ at $k$ is
$g_n = (1-\gamma_n)\min_{i\ne k}|\ta_k-\ta_i|$, 
where $\ta_i := 0$ for $i>d$.
Since $\gamma_n\to\gamma$ and $\ta_i\toP\rho_i^2$, we have $g_n\toP g$.

By Lemma~\ref{lem:eps}, $\varepsilon_n = O_p(\sqrt p/n)\toP 0$.
Consequently, the event $\{g_n\ge g/2\}\cap\{\varepsilon_n < g/4\}$ has probability tending to one.
On this event, Fact~\ref{fact:weyl}(a) and \eqref{eq:MMbar} imply that 
$|\lambda_i(M)-\lambda_i(M^*)|\le\varepsilon_n$ for every $i$.
It follows that 
\begin{align*}
    |\lambda_k(M) - \lambda_i(M)| 
    &\ge |\lambda_k(M^*) - \lambda_i(M^*)| - |\lambda_k(M^*) - \lambda_k(M)| - |\lambda_i(M) - \lambda_i(M^*)| \\
    &\ge g_n - 2\varepsilon_n > 0, \qquad i\ne k,
\end{align*}
and hence $\lambda_k(M)$ is simple. 
Furthermore, Fact~\ref{fact:dk} and \eqref{eq:MMbar} yield
\begin{align*}
    \|w_k-e_k\|
    \le \frac{2^{3/2}\|M-M^*\|_{op}}{g_n}
    \le \frac{2^{5/2}\varepsilon_n}{g}.
\end{align*}
This completes the proof. 
\end{proof}

\begin{proof}[Proof of Theorem~\ref{thm:values2}]
Let $\bG_n = (m/\sqrt{s})(\bB - \gamma_n I_q)$. 
By Lemma~\ref{lem:rep}, $\bG_n$ is independent of $\cF$.
Moreover, Lemma~\ref{lem:proj}(c), applied to a $q$-frame,
yields $\bG_n \tod (1-\gamma)^{1/2}\cG_q$, 
since $s \to \infty$ and $s/m = \gamma_n \to \gamma$. 
Finally, $(n/\sqrt p)/(m/\sqrt s) \to 1$.

\begin{enumerate}
\item[(a)] 
By Lemma~\ref{lem:rep}, $M_{00} = \bB_{00}$. 
On the event $\cE_n$, Lemma~\ref{lem:block}(b) and \eqref{eq:delta} imply that 
\begin{align*}
  0
  \le
  \lambda_j(\bB_{00}) - \hat\lambda_{d+j}
  \le
  \frac{2\varepsilon_n^2}{\delta}
  = O_p(p/n^2). 
\end{align*}
Since $\lambda_j(\bB_{00} - \gamma_n I_t) = \lambda_j(\bB_{00}) - \gamma_n$, we have 
\begin{align*}
  \frac{n}{\sqrt p} \bigl(\hat\lambda_{d+j} - \gamma_n\bigr)
  &= \frac{n}{\sqrt p}\lambda_j\bigl(\bB_{00} - \gamma_nI_t\bigr) 
  + \frac{n}{\sqrt{p}} \bigl( \hat\lambda_{d+j} - \lambda_j(\bB_{00}) \bigr) \\
  &= \frac{n}{\sqrt p}\lambda_j\bigl(\bB_{00} - \gamma_nI_t\bigr) + O_p(\sqrt p/n) \\
  &=\lambda_j\bigl((\bG_n)_{00}\bigr) + o_p(1).
\end{align*}
The result follows because $\bG_n \tod (1-\gamma)^{1/2}\cG_q$
and the continuity of the eigenvalues of a symmetric matrix as functions of its entries.

\item[(b)] 
Fix $k\le d$. By Lemma~\ref{lem:eigvec}, with probability tending to one,
$\hat\lambda_k$ is a simple eigenvalue of $M$. Moreover, a corresponding unit eigenvector $u$
of $M$ can be chosen such that $\|u - e_k\| = O_p(\sqrt{p}/n)$.
Since $(M - \hat\lambda_k I_q)u = 0$, we have
\begin{align*}
  M_{kk} - \hat\lambda_k
  = e_k^T (M - \hat\lambda_k I_q) e_k
  = (e_k-u)^T (M - \hat\lambda_k I_q) (e_k-u).
\end{align*}
Because $0\preceq M\preceq I$ and $\hat\lambda_k\in[0,1]$, it follows that 
\begin{align*}
    |M_{kk} - \hat\lambda_k| \le \|e_k - u\|^2 = O_p(p/n^2). 
\end{align*}
From $M = \Lambda + D\bB D$, we have  
$M_{kk} = \ta_k + (1 - \ta_k)\bB_{kk}$.
Therefore, 
\begin{align*}
  \frac{n}{\sqrt p}\bigl\{\hat\lambda_k - \gamma_n - (1-\gamma_n)\ta_k\bigr\}
  &= (1-\ta_k) \frac{n}{\sqrt p}\bigl(\bB_{kk} - \gamma_n\bigr) 
  + \frac{n}{\sqrt p}\bigl(\hat\lambda_k - M_{kk}\bigr) \\
  &=(1-\ta_k)\frac n{\sqrt p}(\bB_{kk}-\gamma_n)+O_p(\sqrt p/n) \\
  &=(1-\ta_k)(\bG_n)_{kk} + o_p(1).
\end{align*}
The result follows because $\bG_n \tod (1-\gamma)^{1/2}\cG_q$ 
and $\ta_k \toP \rho_k^2$. 

\item[(c)] 
Let $\Xi_n$ be $\cF$-measurable and suppose that $\Xi_n \tod \Xi$. 
Since $\bG_n$ is independent of $\cF$, 
the joint characteristic function of $(\bG_n, \Xi_n)$ factorizes.
Hence, $(\bG_n, \Xi_n) \tod ((1-\gamma)^{1/2}\cG_q, \Xi)$, 
where the two limiting components are independent.
Moreover, the coefficients $1 - \ta_k$ are $\cF$-measurable and converge in probability to $1 - \rho_k^2$. 
The joint statement then follows from Slutsky's theorem and the continuous mapping theorem.
\end{enumerate}
\end{proof}

\begin{proof}[Proof of Corollary~\ref{coro:values}]
\begin{enumerate}
\item[(a)] 
By \eqref{eq:value-rate}, for each $k \leq d$, 
\begin{align*}
  \hat\rho_k^2 - \ta_k
  = \frac{1}{1-\gamma_n}
    \{\hat\lambda_k - \gamma_n - (1-\gamma_n)\ta_k\} 
  = O_p(\sqrt p/n),
\end{align*}
which implies that $\hat\rho_k^2 \toP \rho_k^2$. 

\item[(b)] 
Observe that 
\begin{align*}
\sqrt{n}(\hat\rho_k^2-\ta_k)
= (p/n)^{1/2}(1-\gamma_n)^{-1}\cdot(n/\sqrt p)\{\hat\lambda_k-\gamma_n-(1-\gamma_n)\ta_k\}. 
\end{align*}
This converges in distribution to $\gamma^{1/2}(1-\gamma)^{-1/2}(1-\rho_k^2)(\cG_{11})_{kk}$
by Theorem~\ref{thm:values2}(b).
The result follows from the fact that the diagonal entries of $\cG_{11}$ are independent $\cN(0,2)$ random variables.

\item[(c)] 
The result follows from Theorem~\ref{thm:values2}(c), 
by taking $\Xi_n = \sqrt{n}(\ta_k - \rho_k^2)_{k\le d}$ and 
using the independence of the limiting components.
\end{enumerate}
\end{proof}

\section{Proofs for Section~\ref{sec:test}}
\label{sec:proofs-sec5}

This section provides proofs of the results in Section~\ref{sec:test}.

\begin{proof}[Proof of Theorem~\ref{thm:Td}]
\begin{enumerate}
\item[(a)] 
By Lemma~\ref{lem:rep}, 
$T_d^*/n \sim \Pi(s,t,m)$ and is independent of $\cF$. 
By Lemma~\ref{lem:block}(a),
\begin{align*}
 T_d 
 = n\sum_{i>d} \lambda_i(M)
 \le n\, \tr (M_{00}) 
 = T_d^*. 
\end{align*}
When $d = 0$, then $M = \bB$, so $T_0 = n\, \tr(\bB) = T_0^*$. 

\item[(b)] 
The result follows by applying Lemma~\ref{lem:proj}(b) with a $t$-frame
and substituting 
$m-s = n-1-p$, 
$m-t = n-1-q$, 
$m+2 = n+1-d$, and 
$m-1 = n-2-d$.

\item[(c)] 
By part (a), $T_d \leq T_d^*$, so 
$\{T_d > c_{n,\alpha}(d)\} \subset \{T_d^*>c_{n,\alpha}(d)\}$.
The result follows.
\end{enumerate}
\end{proof}

\begin{proof}[Proof of Theorem~\ref{thm:test}]
\begin{enumerate}
\item[(a)] 
On the event $\cE_n$, Lemma~\ref{lem:block}(a) and equation \eqref{eq:delta} give
\begin{align*}
   0
   \le T_d^* - T_d
   \le \frac{2n\varepsilon_n^2}{\delta}
   \le \frac{4n\varepsilon_n^2}{(1-\gamma) \underline a}
   = O_p(p/n). 
\end{align*}
By Lemma~\ref{lem:proj}(d) with $t=q-d$ fixed,
$s\to\infty$ and $m-s = n-p-1\to\infty$,
\begin{align*}
\frac{T_d^*-nts/m}{n\{2ts(m-s)/m^3\}^{1/2}}
\tod \cN(0,1), 
\end{align*}
and recall that $f_d = ts$, 
\begin{align*}
 \frac{n^2\cdot2ts(m-s)/m^3}{2f_d(1-p/n)}
 = \frac{n^3(n-p-1)}{(n-d-1)^3(n-p)}
 \to 1.
\end{align*}
By Slutsky's Theorem, 
\begin{align*}
  \frac{T_d^* - nf_d/m}{\sqrt{2f_d(1 - p/n)}} 
  \tod \cN(0, 1).
\end{align*}
Finally, write 
\begin{align*}
  \frac{T_d - f_d}{\sqrt{2f_d(1 - p/n)}} 
  =
  \frac{T_d^* - nf_d/m}{\sqrt{2f_d(1 - p/n)}} 
  + 
  \frac{nf_d/m - f_d}{\sqrt{2f_d(1 - p/n)}} 
  - 
  \frac{T_d^* - T_d}{\sqrt{2f_d(1 - p/n)}}.
\end{align*}
The first term converges in distribution to $\cN(0, 1)$. 
The second and third terms are both $o_p(1)$ because 
the denominator $\sqrt{2f_d(1 - p/n)} \to \infty$ as $f_d \to\infty$ and $p/n\to\gamma$, while 
the corresponding numerators satisfy 
$nf_d/m - f_d = f_d(d+1)/m = O(p/n) = O(1)$ and $T_d^* - T_d = O_p(p/n) = O_p(1)$. 
This theorem then follows from Slutsky's Theorem.

\item[(b)] 
Let $d'<d$ and $f_{d'}=(p-d')(q-d')$. By \eqref{eq:value-rate}, for $d'<i\le d$,
\begin{align*}
  n\hat\lambda_i
  = n\gamma_n + n(1-\gamma_n)\ta_i
   + O_p(\sqrt p). 
\end{align*}
We can decompose $T_{d'}$ as follows.
\begin{align*}
    T_{d'} = n \sum_{i = d' + 1}^{H-1} \hat\lambda_i = n \sum_{i=d'+1}^d \hat\lambda_i + T_d. 
\end{align*}
Summing over the fixed number $d - d'$ of indices yields
\begin{align*}
  T_{d'} - f_{d'}
  = n(1-\gamma_n) \sum_{i=d'+1}^d \ta_i
  + (T_d - f_d) 
  + \bigl\{n\gamma_n(d-d')+f_d-f_{d'}\bigr\}
  + O_p(\sqrt p).
\end{align*}
We first analyze the second and third terms on the right-hand side separately. 
\begin{itemize}
\item 
{\em The second term}. 
By part (a), 
$(T_d - f_d)/\sqrt{2f_d(1 - p/n)} \tod \cN(0, 1)$. 
Therefore, $T_d - f_d = O_p(\sqrt{f_d}) = O_p(\sqrt{p})$. 

\item 
{\em The third term}. Notice that 
\begin{align*}
    f_{d'} - f_d 
    &= (d - d')(p + q - d - d'), \\
    n\gamma_n 
    &= \frac{n(p-d)}{n - 1 - d}
     = (p-d) + \frac{(p-d)(d+1)}{n-1-d}.
\end{align*}
We have 
\begin{align*}
    n\gamma_n (d - d') + f_d - f_{d'}
    &= (d - d')
    \left\{ \frac{(p-d)(d+1)}{n-1-d} - (q - d') \right\}
    = O(1). 
\end{align*}
\end{itemize}

The event $\{\ta_d \geq \underline a\}$ occurs with probability tending to one by Assumption~\ref{ass:signal}.
On this event, $\ta_i \geq \ta_d \geq \underline a$ for $i \leq d$. Then, as $n \to \infty$,
\begin{align*}
    T_{d'} - f_{d'}
    \geq n(1 - \gamma_n)(d-d')\underline a + O_p(\sqrt{p}) 
    \geq \frac{1}{2} n(1 - \gamma)(d - d')\underline a + O_p(\sqrt{p}). 
\end{align*}
The normalizer satisfies $\sqrt{2f_{d'}(1 - p/n)} \leq \sqrt{2p(H-1)} = O(\sqrt{p})$.
Therefore,
\begin{align*}
   \frac{T_{d'} - f_{d'}}{\sqrt{2f_{d'}(1 - p/n)}}
   \geq \frac{n(1 - \gamma)(d - d')\underline a/2}{O(\sqrt{p})} + O_p(1)
   \toP \infty,
\end{align*}
because $n/\sqrt{p} \geq \sqrt{n} \to \infty$. This completes the proof.
\end{enumerate}
\end{proof}

\begin{proof}[Proof of Corollary~\ref{coro:regimes}]
\begin{enumerate}
\item[(a)]
This can be shown using the same argument as in part (b), with $\gamma = 0$.

\item[(b)]
The upper-$\alpha$ quantile of $\chi^2_{f}$ satisfies 
$f + z_\alpha\sqrt{2f}\{1 + o(1)\}$ as $f\to\infty$.
By the asymptotic normality of $T_d$ established in Theorem~\ref{thm:test}(a), 
the size of Li's $\chi^2$-test satisfies 
\begin{align*}
  \bP\left[\frac{T_d-f_d}{\sqrt{2f_d(1-p/n)}} 
  > \frac{z_\alpha+o(1)}{\sqrt{1-p/n}}\right]
  \to 1-\Phi(z_\alpha/\sqrt{1-\gamma}).
\end{align*}

\item[(c)]
By the asymptotic normality of $T_d$ established in Theorem~\ref{thm:test}(a),
the corrected test has asymptotic size $\alpha$. 
For the exact test, Lemma~\ref{lem:proj}(d),
together with the computation in the proof of Theorem~\ref{thm:test},
yields 
\begin{align*}
 \frac{c_{n,\alpha}(d) - f_d}{\sqrt{2f_d(1-p/n)}}
 \to z_\alpha.
\end{align*}
Consequently, the exact test also has asymptotic size $\alpha$.
The finite-sample bound is given in Theorem~\ref{thm:Td}(c).
\end{enumerate}
\end{proof}

\begin{proof}[Proof of Corollary~\ref{coro:test}]
At $d = d_0$, both tests fail to reject $H_0$ with probability tending to $1 - \alpha$. 
By Theorem~\ref{thm:test}(b), for either calibration, 
the test rejects $H_0$ with probability tending to one at every step $d' < d_0$.
The event $\{\hat d_0 > d_0\}$ requires rejection at step $d_0$, 
which occurs with probability at most $\alpha$ for every $n$ under the exact test.
This completes the proof. 
\end{proof}

\begin{proof}[Proof of Corollary~\ref{coro:lmax}]
The limit follows from Theorem~\ref{thm:values2}(a) with $j=1$, 
since $\sqrt{p(1-p/n)}/\sqrt{p} \to \sqrt{1-\gamma}$. 
Under Assumption~\ref{ass:signal} with $d_0>d$, 
applying Theorem~\ref{thm:values1} at step $d_0$ yields 
\begin{align*}
   \hat\lambda_{d+1} - \gamma'_n
   \ge
   (1-\gamma'_n)\underline a - O_p(\sqrt{p}/n), 
\end{align*}
where $\gamma'_n = (p-d_0)/(n-d_0-1)$ is evaluated at step $d_0$.
The normalized test statistic at step $d$ can be decomposed as 
\begin{align*}
    L_d 
    &= \frac{n}{\sqrt{p(1 - p/n)}} (\hat\lambda_{d+1} - \gamma_n) \\
    &= \frac{n}{\sqrt{p(1 - p/n)}} (\hat\lambda_{d+1} - \gamma'_n)
     + \frac{n}{\sqrt{p(1 - p/n)}} (\gamma'_n - \gamma_n),
\end{align*}
where $\gamma_n = (p-d)/(n-d-1)$ is evaluated at step $d$. 
We analyze the two terms separately. 
\begin{align*}
    \frac{n}{\sqrt{p(1 - p/n)}} (\hat\lambda_{d+1} - \gamma'_n)
    &\geq \frac{n}{\sqrt{p(1 - p/n)}}(1-\gamma'_n)\underline a  
     - O_p(1) \toP \infty, \\
    \frac{n}{\sqrt{p(1 - p/n)}} (\gamma'_n - \gamma_n)
    &= \frac{n}{\sqrt{p(1 - p/n)}} O(n^{-1}) \to 0. 
\end{align*}
Consequently, $L_d\toP\infty$. 
Therefore, by an argument analogous to that used 
in the proof of Corollary~\ref{coro:test}, 
we obtain $\bP(\hat d_0 = d_0) \to 1 - \alpha$.
\end{proof}

\section{Proofs for Section~\ref{sec:vectors}}
\label{sec:proofs-sec6}

This section provides proofs of the results in Section~\ref{sec:vectors}.
We first state the following lemma, which will be used in the proof of Theorem~\ref{thm:vectors-align}. 

\begin{lemma}\label{lemma:Zplus}
Let $Z \in \bR^{m\times s}$ have iid\ $\cN(0,1)$ entries,
and define $Z^+ = (Z^TZ)^{-1}Z^T$. 
For a fixed unit vector $\psi\in\bR^m$, 
if $m\to\infty$, $s/m \to \gamma \in [0,1)$, 
then 
\begin{align*}  
    \|Z^+\|_{op} = O_p(m^{-1/2}), \qquad 
    m\|Z^+\psi\|^2 \toP \frac{\gamma}{1-\gamma}.
\end{align*}    
\end{lemma}

\begin{proof}[Proof of Lemma~\ref{lemma:Zplus}]
By Fact~\ref{fact:rmt}, with probability tending to one,
$\sigma_{\min}(Z)^2/m \ge (1-\sqrt\gamma)^2/2$.
Consequently, $\|Z^+\|_{op}=1/\sigma_{\min}(Z) = O_p(m^{-1/2})$.  

Let $Z^TZ = V_Z\Lambda_ZV_Z^T$ be the spectral decomposition,
where $\Lambda_Z = \diag(\lambda_1,\dots,\lambda_s)$.
By the preceding bound, with probability tending to one, 
\begin{align*}
\max_{i\le s}\lambda_i^{-1} = \lambda_{\min}^{-1} \le \frac{2}{m(1-\sqrt\gamma)^2}.
\end{align*}
Consequently, $\max_i\lambda_i^{-1} = O_p(m^{-1})$.
Define $O_Z=ZV_Z\Lambda_Z^{-1/2} \in \bR^{m\times s}$. Then $O_Z^TO_Z = I_s$ and
\begin{align*}
    \|Z^+ \psi\|^2 
    = \|V_Z\Lambda_Z^{-1/2} O_Z^T \psi\|^2
    = \sum_{i = 1}^s \frac{(O_Z^T\psi)_i^2}{\lambda_i}. 
\end{align*}
Let $Q_0$ be a Haar-distributed orthogonal matrix independent of $Z$.
Since $(Q_0Z)^T(Q_0Z) = Z^TZ$, replacing $Z$ by $Q_0Z$ leaves $\Lambda_Z$ and $V_Z$ unchanged
and replaces $O_Z$ by $Q_0O_Z$.
Because $Q_0Z \eqd Z$, it follows that $(\Lambda_Z, O_Z^T\psi) \eqd (\Lambda_Z, O_Z^TQ_0^T\psi)$.
Conditional on $Z$, $Q_0^T\psi$ is uniformly distributed on the unit sphere.
Since $O_Z$ has orthonormal columns, $O_Z^TQ_0^T\psi$ has the same distribution 
as the first $s$ coordinates of a uniformly distributed unit vector.
Since this conditional distribution does not depend on $\Lambda_Z$,
$O_Z^TQ_0^T\psi$ is independent of $\Lambda_Z$, and we may write 
$O_Z^TQ_0^T\psi \eqd (z_1, \dots, z_s)^T/\|z\|$ with $z \sim \cN_m(0,I_m)$, 
where $z$ is independent of $\Lambda_Z$.
Therefore,
\begin{align*}
  m\|Z^+\psi\|^2
  \eqd
  \frac m{\|z\|^2}\sum_{i=1}^s\frac{z_i^2}{\lambda_i}.
\end{align*}
Since $m/\|z\|^2 \toP 1$, it remains to study the sum on the right. 

Conditional on $\Lambda_Z$, the mean and variance of the sum are 
\begin{align*}
    \bE\left(\sum_{i=1}^s\frac{z_i^2}{\lambda_i} \Bigm| \Lambda_Z\right) 
    &= \sum_{i=1}^s \frac{1}{\lambda_i}  
    = \tr\{(Z^TZ)^{-1}\}, \\
    var\left(\sum_{i=1}^s\frac{z_i^2}{\lambda_i} \Bigm| \Lambda_Z\right) 
    &= \sum_{i=1}^s \frac{2}{\lambda_i^2}  
    = O_p(s/m^2) \to 0.    
\end{align*} 
If $\gamma = 0$, then
\begin{align*}
\tr\{(Z^TZ)^{-1}\}\le \frac{s}{\lambda_{\min}}  \le \frac{2s}{m(1-\sqrt\gamma)^2} \toP 0. 
\end{align*}
If $\gamma > 0$, Fact~\ref{fact:mp} yields 
\begin{align*}
\tr\{(Z^TZ)^{-1}\}
= \frac{s}{m} \cdot s^{-1}\sum_i\left(\frac{\lambda_i}{m}\right)^{-1}
\toP
\gamma\int x^{-1}dF_\gamma(x)
= \frac{\gamma}{1-\gamma},
\end{align*}
where we use the fact that, with probability tending to one, 
all $\lambda_i/m$ lie in a compact subset of $(0,\infty)$.
This completes the proof.     
\end{proof}

\begin{proof}[Proof of Theorem~\ref{thm:vectors-align}]
Work in the coordinates provided by Lemma~\ref{lem:reduction}. 
Then $\gamma_n\to\gamma$ and 
$\varepsilon_n = O_p(\sqrt p/n) \toP 0$.
Note that $p\to\infty$ is not needed. 
Let $\cJ$ be the set of indices $j\le d$ for which $\rho_j^2>0$ is a simple eigenvalue. Under the assumptions of Theorem~\ref{thm:vectors-align}, $\cJ= \{1,\dots,d\}$. 

\emph{Step 1: least-squares form.} 
For $j\in\cJ$,  
$\hat\lambda_j \toP \gamma + (1-\gamma) \rho_j^2 > 0$ 
follows from Theorem~\ref{thm:values1}. 
By \eqref{eq:betab}, 
$\hat\beta_j = \sqrt{n}\,b^{(j)}/r_j$,
where $b^{(j)} = (\bar X^T\bar X)^{-1}\bar X^TF_{\bar G}u_j$.

\emph{Step 2: eigenvectors.} 
Since $u_j$ is the unit eigenvector of $K$ corresponding to $r_j^2$, we have $Ku_j = r_j^2u_j$.
Recall that the orthogonal matrix $U$ is defined in Lemma~\ref{lem:rep} and $K = UMU^T$.
Hence, $MU^Tu_j = r_j^2U^Tu_j$, which implies that $U^Tu_j$ is a unit eigenvector of $M$ for $r_j^2 = \hat\lambda_j$. 
By Lemma~\ref{lem:eigvec}, for $j\in\cJ$ and a suitable sign of $u_j$,
\begin{equation}\label{eq:xi}
    \xi_j := U^Tu_j-e_j, \qquad \|\xi_j\| = O_p(\sqrt{p}/n).
\end{equation}

\emph{Step 3: decomposition.} 
Recall $C_n = P_{\bar X_S} + P_{\cL}$ and $\check X_N = P_{\cL}X_N$ from Lemma~\ref{lem:geometry}(b).
Write 
\begin{align*}
    \bar X_N = C_nX_N = \bar X_S\Gamma + \check X_N,
\end{align*}
where $\Gamma = (\bar X_S^T\bar X_S)^{-1}\bar X_S^TX_N$. 
Then for any vector $b$, 
\begin{align*}
    \bar Xb
    = \bar X_Sb_S + \bar X_Nb_N
    = \bar X_S(b_S+\Gamma b_N) + \check X_Nb_N,
\end{align*} 
while for $b^{(j)}$ defined in Step 1, 
by $P_{\bar X} = P_{\bar X_S} + P_{\cV}$ from Lemma~\ref{lem:geometry}(b), 
\begin{align*}
    \bar Xb^{(j)} 
    = P_{\bar X}F_{\bar G}u_j
    =P_{\bar X_S}F_{\bar G}u_j + P_{\cV}F_{\bar G}u_j.
\end{align*}
Matching the components in the orthogonal subspaces $\col(\bar X_S)$ and $\cV$,
\begin{align*}
  b_N^{(j)} = \check X_N^+ F_{\bar G}u_j = \check X_N^+Wu_j, \qquad 
  b_S^{(j)} = \check b^{(j)} - \Gamma b_N^{(j)}, \qquad
  \check b^{(j)} = (\bar X_S^T\bar X_S)^{-1}\bar X_S^TF_{\bar G}u_j,
\end{align*}
where $\check X_N^+ = (\check X_N^T\check X_N)^{-1} \check X_N^T$,
and we used $\check X_N^+ = \check X_N^+ P_{\cL}$ and $P_{\cL}F_{\bar G} = W$.
Additionally, by Lemma~\ref{lem:geometry}(c), $\check X_N = F_{\cL}Z$ and we can write 
$\check X_N^+ = Z^+F_{\cL}^T$, where $Z^+ = (Z^TZ)^{-1}Z^T$ is defined in Lemma~\ref{lemma:Zplus}. 

\emph{Step 4: noise part.} 
By Lemma~\ref{lem:rep} and \eqref{eq:xi},
\begin{align*}
Wu_j = WU(U^Tu_j) = WU(e_j+\xi_j) = (1-\ta_j)^{1/2} \phi_j + F_{\phi} D\xi_j, 
\end{align*}
so
\begin{align*}
  \sqrt{n}\,b_N^{(j)}
  = (1-\ta_j)^{1/2} \sqrt{n}\,Z^+\psi_j+\Delta_j,
\end{align*}
where   
$\psi_j = F_{\cL}^T \phi_j$ and 
$\Delta_j = \sqrt{n} Z^+F_{\cL}^T F_{\phi} D\xi_j$. 
We discuss these two terms separately.
Consider the second term $\Delta_j$ first.
Lemma~\ref{lemma:Zplus} and $m/n\to 1$ give $\|Z^+\|_{op} = O_p(n^{-1/2})$.
$F_{\cL}^TF_{\phi}$ has orthonormal columns. 
Therefore, 
\begin{align*}
\|\Delta_j\|
\le \sqrt n\|Z^+\|_{op}\|\xi_j\|
 = O_p(\sqrt p/n) \toP 0. 
\end{align*}
Now, consider the first term.
The $\psi_j$ are $\cF$-measurable and orthonormal. 
Given $\cF$, $Z$ has iid\ $\cN(0,1)$ entries by Lemma~\ref{lem:geometry}(c). 
By rotation invariance, for any $\cF$-measurable $w\in\bR^m$, 
the conditional distribution of $\|Z^+w\|$ is that of $\|w\|\|Z^+e_1\|$, so
Lemma~\ref{lemma:Zplus} and $m/n\to 1$ give $n\|Z^+w\|^2-\|w\|^2\gamma/(1-\gamma) \toP 0$. 

Now, consider 
\begin{align*}
    n\psi_j^TZ^{+T}Z^+\psi_l
    &= \frac{n}{4}\left\{
    (\psi_j + \psi_l)^T(Z^{+T}Z^+)(\psi_j + \psi_l)
    - 
    (\psi_j - \psi_l)^T(Z^{+T}Z^+)(\psi_j - \psi_l)
    \right\} \\
    &= \frac{1}{4} \left\{
    n\|Z^+(\psi_j + \psi_l)\|^2 - n\|Z^+(\psi_j - \psi_l)\|^2
    \right\} \\
    &\toP \frac{\gamma}{4(1 - \gamma)} \left\{
    \|\psi_j + \psi_l\|^2 - \|\psi_j - \psi_l\|^2
    \right\}
    = \delta_{jl}\frac{\gamma}{1 - \gamma}.
\end{align*}
where the last equation holds because $\psi_j$ and $\psi_l$ are orthonormal. 
Therefore, with $(1 - \ta_j) \toP (1 - \rho_j^2)$ and $\Delta_j \toP 0$, 
\begin{align*}
  n\,b_N^{(j)T}b_N^{(l)}
  \toP
  \delta_{jl}(1-\rho_j^2) \frac{\gamma}{1-\gamma},
  \qquad j,l\in\cJ.
\end{align*}

\emph{Step 5: signal part.} 
With $u_j=U(e_j+\xi_j)$,
\begin{align*}
 \sqrt{n}\,\check b^{(j)}
 = \sqrt n\,b_{\rm or}^{(j)}
 +
 \sqrt n(\bar X_S^T\bar X_S)^{-1}\bar X_S^TF_{\bar G}U\xi_j, 
\end{align*}
where 
$b_{\rm or}^{(j)} = (\bar X_S^T\bar X_S)^{-1}\bar X_S^TF_{\bar G}\tu_j$. 
Since $F_{\bar G}U$ is orthonormal, 
the second term has norm at most $\sqrt{n}\,\sigma_{\min}(\bar X_S)^{-1}\|\xi_j\|$,
which has order $O_p(\sqrt p/n)$
because $\bar X_S^T\bar X_S/n \toP I_d$ by the law of large numbers
and $\sigma_{\min}(\bar X_S)^2/n = \lambda_{\min}(\bar X_S^T\bar X_S/n)\toP 1$.

By applying Proposition~\ref{prop:cca} to the $d$-dimensional data $\bar X_S$, 
$b_{\rm or}^{(j)}$ is the unnormalized oracle SIR direction:
$
   b_{\rm or}^{(j)T}\bar X_S^T\bar X_Sb_{\rm or}^{(l)}
   = \tu_j^T A \tu_l
   = \ta_j \delta_{jl}, 
$
where the last equality holds because $\tu_j$ and $\tu_l$ are eigenvectors of $A$ by definition. 
Then 
\begin{align*}
   \sqrt{n}\,b_{\rm or}^{(j)} = \ta_j^{1/2}\hat v_j,
\end{align*}
where  
$\hat v_j$ is the normalized oracle SIR direction defined in Section~\ref{subsec:oracle}
and $\hat v_j^T(\bar X_S^T\bar X_S/n)\hat v_l = \delta_{jl}$. 
Let $\hat S_X = \bar X_S^T\bar X_S/n$, so that $\hat S_X \toP I_d$.
By the normalization $\hat v_j^T\hat S_X\hat v_j = 1$, we have 
$\|\hat v_j\|^2 \leq \lambda_{\min}(\hat S_X)^{-1} = O_p(1)$. Consequently,
\begin{align*}
    \hat v_j^T\hat v_l
    = \delta_{jl} + \hat v_j^T(I_d - \hat S_X)\hat v_l
    \toP \delta_{jl}.
\end{align*}
Since $\ta_j \toP \rho_j^2$ and the second term in the expansion of
$\sqrt{n}\,\check b^{(j)}$ is $o_p(1)$, it follows that  
\begin{align*}
  n\,\check b^{(j)T}\check b^{(l)}
  \toP \rho_j^2 \delta_{jl}.
\end{align*}

\emph{Step 6: the term $\Gamma b_N$.} 
Let $F_S$ be an $\cF$-measurable frame spanning $\col(\bar X_S)$. 
Since $F_SF_S^T = P_{\bar X_S}$ and $\bar X_S^T = \bar X_S^T P_{\bar X_S}$, we have 
\begin{align*}
    \Gamma b_N^{(j)} 
    = (\bar X_S^T\bar X_S)^{-1}\bar X_S^TF_S (F_S^TX_Nb_N^{(j)}).
\end{align*} 
Notice that $(F_S, F_{\cL})$ has orthonormal columns by Lemma~\ref{lem:geometry}. 
Given $\cF$, the matrices $F_S^TX_N$ and $Z = F_{\cL}^TX_N$ are independent with iid\ $\cN(0,1)$ entries.
Therefore, $K$, hence $u_j$ and $b_N^{(j)}$, 
depend on $X_N$ only through $Z$. 
Given $(\cF, Z)$, we may treat $\bar X_S$, $F_S$, and $b_N^{(j)}$ as fixed and then 
\begin{align*}
 \Gamma b_N^{(j)} 
 \sim 
 \cN\{0,\|b_N^{(j)}\|^2 (\bar X_S^T\bar X_S)^{-1}\}. 
\end{align*}
Moreover, 
since $n\|b_N^{(j)}\|^2 = O_p(1)$ by step 4 and 
\begin{align*}
  \bE\bigl(n \|\Gamma b_N^{(j)}\|^2 \mid \cF, Z\bigr)
  = n\|b_N^{(j)}\|^2\ \tr{(\bar X_S^T\bar X_S)^{-1}}
  = \|b_N^{(j)}\|^2\ \tr{(\bar X_S^T\bar X_S/n)^{-1}}, 
\end{align*}
we have $n\|\Gamma b_N^{(j)}\|^2 = O_p(1/n)$. 
Hence $n\,b_S^{(j)T}b_S^{(l)} \toP \rho_j^2\delta_{jl}$.

\emph{Step 7: conclusion.} 
By Lemma~\ref{lem:reduction}, take $\mu = 0$, $\Sigma = I_p$, $B = (I_d,0)^T$ 
and $\beta_k = e_k$, so that $V_S$ is diagonal and $v_k = e_k$.
The oracle consistency reads $\hat v_k \toP e_k$ up to sign. 
Fix the sign of $\tu_k$ so that $\hat v_k\toP e_k$, 
and that of $u_k$ as in \eqref{eq:xi}. 
Recall that $\hat\beta_j = \sqrt{n}\,b^{(j)}/r_j$ by Step 1,
and $\sqrt n\,b_S^{(j)} = \ta_j^{1/2}\hat v_j + o_p(1) \toP \rho_je_j$
by Steps 5 and 6. 
We have 
\begin{align*}
(\hat\beta_{1:d}^T\Sigma\beta_{1:d})_{jl}
= \hat\beta_j^Te_l
= \frac{\sqrt{n}\,b^{(j)}_{S,l}}{r_j} \toP \frac{\rho_j\delta_{jl}}{\sqrt{\gamma+(1-\gamma)\rho_j^2}}
= \kappa_j\delta_{jl}. 
\end{align*}
In particular $\hat\beta_k^T\Sigma\beta_k > 0$ with probability tending to one,
so the sign conventions agree.
This establishes $\hat\beta_{1:d}^T\Sigma\beta_{1:d} \toP \sK$. 
Similarly, 
\begin{align*}
  \hat\beta_j^T\hat\beta_l
  = \frac{n\,b_S^{(j)T}b_S^{(l)} + n\,b_N^{(j)T} b_N^{(l)}}{r_jr_l}
  \toP
  \delta_{jl}
  \frac{\rho_j^2+(1-\rho_j^2)\gamma/(1-\gamma)}{\gamma+(1-\gamma)\rho_j^2}
  = \delta_{jl}(\kappa_j^2+\tau_j^2),
\end{align*}
which completes the proof.  
\end{proof}

\begin{proof}[Proof of Theorem~\ref{thm:accuracy}]
Work in the coordinates provided by Lemma~\ref{lem:reduction}. 
We use only Steps 1--6 in the proof of Theorem~\ref{thm:vectors-align}.
These steps do not require the consistency of the oracle directions, which is used only in Step 7.
Therefore, Theorem~\ref{thm:accuracy} does not require consistency of the oracle directions, unlike Theorem~\ref{thm:vectors-align}.
  
For part (a), in the coordinates provided by Lemma~\ref{lem:reduction}, 
for any $b\in\bR^p$, $b=(b_S^T,b_N^T)^T$ and 
\begin{equation*}
  R^2(b) = \frac{\|b_S\|^2}{\|b_S\|^2+\|b_N\|^2}.
\end{equation*}
Next, the argument in Theorem~\ref{thm:vectors-align} only needs to apply to the case $\cJ = \{k\}$. 
Steps 4--6 in the proof of Theorem~\ref{thm:vectors-align} give  
\begin{align*}
  R^2(\hat\beta_k)
  = \frac{n\|b_S^{(k)}\|^2}{n\|b_S^{(k)}\|^2+n\|b_N^{(k)}\|^2}
  \toP
  \frac{\rho_k^2}{\rho_k^2+(1-\rho_k^2)\gamma/(1-\gamma)}
  = \varphi_\gamma(\rho_k^2).
\end{align*}

For part (b), take $\cJ = \{1,\dots,d\}$.
Define the $d\times d$ matrices $Q_S$ and $Q_N$ by
\begin{align*}
    (Q_S)_{jl} = n\,b_S^{(j)T}b_S^{(l)}, \qquad
    (Q_N)_{jl} = n\,b_N^{(j)T}b_N^{(l)}, \qquad j,l\leq d,
\end{align*}
and let $\Lambda_\rho = \diag(\rho_1^2,\dots,\rho_d^2)$.
Steps 4--6 imply that 
\begin{align*}
    Q_S \toP \Lambda_\rho, \qquad
    Q_N \toP \frac{\gamma}{1-\gamma}(I_d - \Lambda_\rho).
\end{align*}
Let $\hat\beta_{1:d,S}$ denote the first $d$ rows of $\hat\beta_{1:d}$, and
let $D_r = \diag(r_1,\dots,r_d)$, which is invertible with probability tending to one by Step 1.
Since $\hat\beta_j = \sqrt{n}\,b^{(j)}/r_j$, we have
$\hat\beta_{1:d,S}^T\hat\beta_{1:d,S} = D_r^{-1}Q_SD_r^{-1}$ and
$\hat\beta_{1:d}^T\hat\beta_{1:d} = D_r^{-1}(Q_S+Q_N)D_r^{-1}$.
Because $\Sigma = I_p$ and $B = (I_d,0)^T$, it follows that
\begin{align*}
    R^2(\hat\beta_{1:d})
    =
    \frac{1}{d}\tr\bigl\{
    (\hat\beta_{1:d}^T \hat\beta_{1:d})^{-1} \hat\beta_{1:d,S}^T \hat\beta_{1:d,S}
    \bigr\}
    =
    \frac{1}{d}\tr\bigl\{(Q_S+Q_N)^{-1}Q_S\bigr\}.
\end{align*}
The limit of $Q_S+Q_N$ is a diagonal matrix with strictly positive diagonal entries,
since $\rho_d^2 > 0$. Thus, by the continuous mapping theorem,
\begin{align*}
    R^2(\hat\beta_{1:d})
    \toP
    \frac{1}{d}\sum_{k=1}^d
    \frac{\rho_k^2}{\rho_k^2+(1-\rho_k^2)\gamma/(1-\gamma)}
    =
    \frac{1}{d}\sum_{k=1}^d \varphi_\gamma(\rho_k^2).
\end{align*}
This completes the proof. 
\end{proof}

\begin{proof}[Proof of Theorem~\ref{thm:pred}]
Use the coordinates and sign conventions of the proof of
Theorem~\ref{thm:vectors-align}, and we have $w_S = x_{0, S}$.

We first establish the following decomposition. 
For every $n$,
\begin{equation}
  \hat\beta_{1:d}^T (x_0 - \bar x)
  = \hatsK\,\beta_{1:d}^T (x_0 - \mu) + \zeta_n
  -\hat\beta_{1:d}^T (\bar x - \mu), \qquad
  \hatsK:=\hat\beta_{1:d}^T \Sigma \beta_{1:d},
\label{eq:dec}
\end{equation}
where, conditionally on the sample, $\zeta_n\sim \cN_d(0, \Omega_n)$ and
$\Omega_n = \hat\beta_{1:d}^T \Sigma \hat\beta_{1:d} - \hatsK\hatsK^T$. 
Write $x_0 - \mu = \Sigma\beta_{1:d}w_S + r$, where  
$r = (I_p - \Sigma\beta_{1:d} \beta_{1:d}^T)(x_0 - \mu)$. 
Then 
\begin{align*}
  cov(w_S, r) = \beta_{1:d}^T\Sigma (I_p - \beta_{1:d}\beta_{1:d}^T\Sigma) = 0,
\end{align*}
where we use the normalization condition $\beta_{1:d}^T\Sigma\beta_{1:d} = I_d$. 
Since $w_S$ and $r$ are jointly normally distributed, this implies that $r\indep w_S$. 
Furthermore, as in Lemma~\ref{lem:geometry}(a), 
model \eqref{eq:model-cs} implies that $r\indep(w_S, y_0)$. 
Since the new observation is independent of the sample, conditional on the sample, 
$\zeta_n = \hat\beta_{1:d}^Tr$ follows $\cN_d(0,\Omega_n)$, 
where 
$\Omega_n = \hat\beta_{1:d}^T(\Sigma - \Sigma\beta_{1:d}\beta_{1:d}^T\Sigma)\hat\beta_{1:d}$. 
Moreover, $\zeta_n$ is independent of $(w_S, y_0)$. 
This establishes \eqref{eq:dec}. 

By Theorem~\ref{thm:vectors-align}, we have $\hatsK \toP \sK$. 
Furthermore, Theorem~\ref{thm:vectors-align} implies that 
$\Omega_n \toP (\sK^2+\sT^2) - \sK^2 = \sT^2$. 

Next, we show that $\hat\beta_j^T(\bar x - \mu) \toP 0$ for $1 \leq j \leq d$.
In the present coordinates, $\mu = 0$. Partition
$\hat\beta_j = (\hat\beta_{j,S}^T, \hat\beta_{j,N}^T)^T$ and
$\bar x = (\bar x_S^T, \bar x_N^T)^T$ into signal and noise components. Then 
\begin{align*}
    \hat\beta_j^T\bar x = \hat\beta_{j,S}^T\bar x_S + \hat\beta_{j,N}^T\bar x_N.
\end{align*}
First, $\hat\beta_j = O_p(1)$, because $\hat\beta_j^T\hat\Sigma\hat\beta_j = 1$ 
and $\lambda_{\min}(\hat\Sigma)$ is bounded away from zero with probability tending to one. 
For the signal component, since $\bar x_S \sim \cN_d(0, I_d/n)$ and $d$ is fixed,
$\|\bar x_S\|= O_p(n^{-1/2})$.
It follows that $|\hat\beta_{j,S}^T\bar x_S| \leq \|\hat\beta_j\| \|\bar x_S\| = O_p(n^{-1/2})$. 
For the noise component, recall that the columns of $X_N$ are iid $\cN_n(0, I_n)$. 
Thus, $\bar x_N = X_N^T1_n/n$ is independent of $\bar X_N = C_nX_N$.
By Lemma~\ref{lem:geometry}(a), $\bar x_N$ is also independent of $\cF$. 
Since $\hat\beta_j$ is $\sigma(\cF, \bar X_N)$-measurable,
conditional on $\bar X_N$ and $\cF$, we have 
$\hat\beta_{j, N}^T\bar x_N \sim \cN(0, \|\hat\beta_{j, N}\|^2/n)$. 
Consequently, since $\hat\beta_j = O_p(1)$, we have $\hat\beta_{j, N}^T\bar x_N = O_p(n^{-1/2})$. 
Combining the bounds for the signal and noise components yields $\hat\beta_j^T(\bar x - \mu) \toP 0$. 

We are now in a position to prove the theorem.
\begin{enumerate}
\item[(a)] 
By the preceding argument, 
$\hat\beta_{1:d}^T(x_0 - \bar x) = \sK w_S + \zeta_n + o_p(1)$.
Conditionally on the sample, $\zeta_n \sim \cN_d(0, \Omega_n)$, where $\Omega_n \toP \sT^2$.
Moreover, by bounded convergence, for any vector $a$, the characteristic function is 
\begin{align*}
    \bE[\exp(\imath a^T\zeta_n)] = \bE[\exp(-\tfrac12a^T\Omega_na)] \to 
    \exp(-\tfrac{1}{2} a^T\sT^2a),
\end{align*}
which implies that $\zeta_n \tod \sT w_N$, where $w_N \sim \cN_d(0, I_d)$.
Since $\zeta_n$ is independent of $(w_S, y_0)$, whose joint distribution does not depend on $n$, 
we further obtain $(\zeta_n, w_S, y_0) \tod (\sT w_N, w_S, y_0)$,
where $w_N$ is independent of $(w_S, y_0)$. 
The desired result then follows from the continuous mapping theorem and Slutsky's theorem.

\item[(b)] 
By (a), 
\begin{align*}
    \hat\beta_k^T(x_0 - \bar x)-\beta_k^T(x_0 - \mu)
    \tod
    (\kappa_k-1)w_{S, k} + \tau_k w_{N, k}
    \sim \cN\{0,(1-\kappa_k)^2 + \tau_k^2\}.
\end{align*} 
If $\gamma = 0$, then $\kappa_k=1$ and $\tau_k=0$. 
If $\gamma > 0$, then $\tau_k>0$ since $\rho_k^2 < 1$ by Proposition~\ref{prop:cca-pop}(c). 
\end{enumerate}
\end{proof}

\end{document}